\documentclass[lettersize,journal]{IEEEtran}
\usepackage{amsmath,amssymb,amsfonts}
\usepackage[linesnumbered,ruled]{algorithm2e}
\usepackage{array}
\usepackage{caption}
\usepackage{textcomp}
\usepackage{tabularx}
\usepackage{hyperref}
\usepackage{booktabs}
\usepackage{diagbox}
\usepackage{makecell}
\usepackage{stfloats}
\usepackage{url}
\usepackage{verbatim}
\usepackage{graphicx}
\usepackage{cite}
\usepackage{subfig}
\usepackage{svg}
\usepackage{pifont} % 引入pifont宏包
\newtheorem{remark}{Remark}
\newtheorem{assumption}{Assumption}

\newtheorem{lemma}{Lemma}
\newtheorem{theorem}{Theorem}
\newtheorem{proof}{Proof}
\begin{document}
\allowdisplaybreaks[4]
\title{Optimal Decentralized Dynamic Energy Management over Asynchronous Peer-to-Peer Transactive Networks via Operator Splitting}
        % <-this % stops a space
\author{Xi Zhang, Huqiang Cheng, Guo Chen,~\IEEEmembership{Member,~IEEE}, Huaqing Li,~\IEEEmembership{Senior Member,~IEEE}, Liang Ran, Jinhui Hu, and Tingwen Huang,~\IEEEmembership{Fellow,~IEEE}
\thanks{This work is supported in part by National Natural Science Foundation of China under Grant 62573364, in part by Australian Research Council under Grants FT190100156 and DP230100801, and in part by the Fundamental Research Funds for the Central Universities under Grant SWU-XDJH202312.
}
\thanks{X. Zhang is with the School of Electrical Engineering and Automation, Hubei Normal University, Hubei, China (e-mail: xizhang@hbnu.edu.cn); H. Cheng is with the College of Computer Science, Chongqing University, Chongqing 400044, China (e-mail:
huqiangcheng@126.com); G. Chen is with the School of Electrical Engineering and Telecommunications, University of New South Wales, Sydney, NSW 2052, Australia (e-mail: guo.chen@unsw.edu.au); H. Li and L. Ran are with the Chongqing Key Laboratory of Nonlinear Circuits and Intelligent Information Processing, College of Electronic and Information Engineering, Southwest University, Chongqing 400715, China (e-mail: huaqingli@swu.edu.cn; ranliang\_rl@163.com); J. Hu is with the School of Artificial Intelligence and Computer Science, Hubei Normal University, Hubei, China, and also with the Department of Mechanical Engineering, City University of Hong Kong, Kowloon Tong, Kowloon, Hong Kong SAR, China (e-mail: jinhuihu3-c@my.cityu.edu.hk); T. Huang is with the School of Computer Science and Control Engineering, Shenzhen University of Advanced Technology, Shenzhen 518055, China (email: huangtingwen@suat-sz.edu.cn). X. Zhang and H. Cheng contributed equally to this work.
}
}

\maketitle

\begin{abstract}
Peer-to-peer (P2P) energy management facilitates decentralized resource allocation among prosumers, improving local hosting capacity for renewables and minimizing energy expenditures while ensuring data privacy through distributed coordination. However, conventional P2P energy management methods are confined to synchronous scheduling paradigms, creating synchronization bottlenecks that fundamentally conflict with the dynamic and decentralized nature of P2P energy management tasks. To bridge this gap, this paper focuses on resolving a class of dynamic energy management problems over asynchronous P2P (Asyn-P2P) transactive networks. We first recast the dynamic energy management problems into a saddle-point problem, and then propose a \underline{syn}chronous \underline{d}ecentralized d\underline{y}namic e\underline{n}ergy m\underline{a}nagement algorithm, dubbed \textit{Syn-DYNA}, based on operator splitting theory. To eliminate the global synchronization clock in \textit{Syn-DYNA}, we introduce a random activation scheme, together with local buffers for latest state tracking, to develop an \underline{a}synchronous variant of \textit{Syn-DYNA}, namely \textit{Asyn-DYNA}. Based on monotone operator theory, theoretical analysis proves a non-asymptotic linear convergence rate for \textit{Syn-DYNA} and establishes the almost sure convergence of \textit{Asyn-DYNA}. Numerical experiments validate effectiveness of \textit{Syn-DYNA} and \textit{Asyn-DYNA} algorithms by tackling a dynamic energy management task over P2P transactive networks.
\end{abstract}


\section{Introduction}
%1.概述P2P研究领域的重要性（去中心化、透明性、灵活性）；2.明确研究问题的重要性；3.简要总结前人研究的不足
\IEEEPARstart{P}{2P} energy management is a key enabler for the modernization of power systems, facilitating efficient and decentralized energy distribution \cite{Tariq2025}. Driven by this decentralized nature, decentralized optimization algorithms (DOAs) are increasingly being applied in P2P trading systems  \cite{Scutari2010}. These algorithms enable producers to participate in dynamic energy transactions with neighboring producers, which not only alleviates the communication bottleneck of centralized architectures but also protects individual privacy. However, practical implementation of DOAs in P2P energy trading systems faces significant challenges arising from prosumers' heterogeneous computational capabilities and stochastic communication delays. These factors inevitably lead to substantial synchronization overheads, for instance idle waiting times, thereby severely degrading the overall convergence efficiency of DOAs.

\subsection{Literature Review}\label{LR}
%因为很多能源管理的需要，所以p2p应运而生，p2p用在能源管理、金融、共享经济、碳交易等多个领域被应用
The widespread adoption of renewable energy has accelerated a structural transition from centralized to decentralized energy systems \cite{Howell2017}. This shift urgently necessitates novel energy management solutions based on prosumer or agent interaction. Such solutions are essential to efficiently harness energy and harmonize the intermittency of renewables with dynamic demand patterns.
%p2p的重要性和好处，其核心逻辑是“去中心化的直接交易模式”。透明性、灵活性
\begin{figure}[!htp]
  \centering
  \includegraphics[width=3.3in,height=1.7in]{{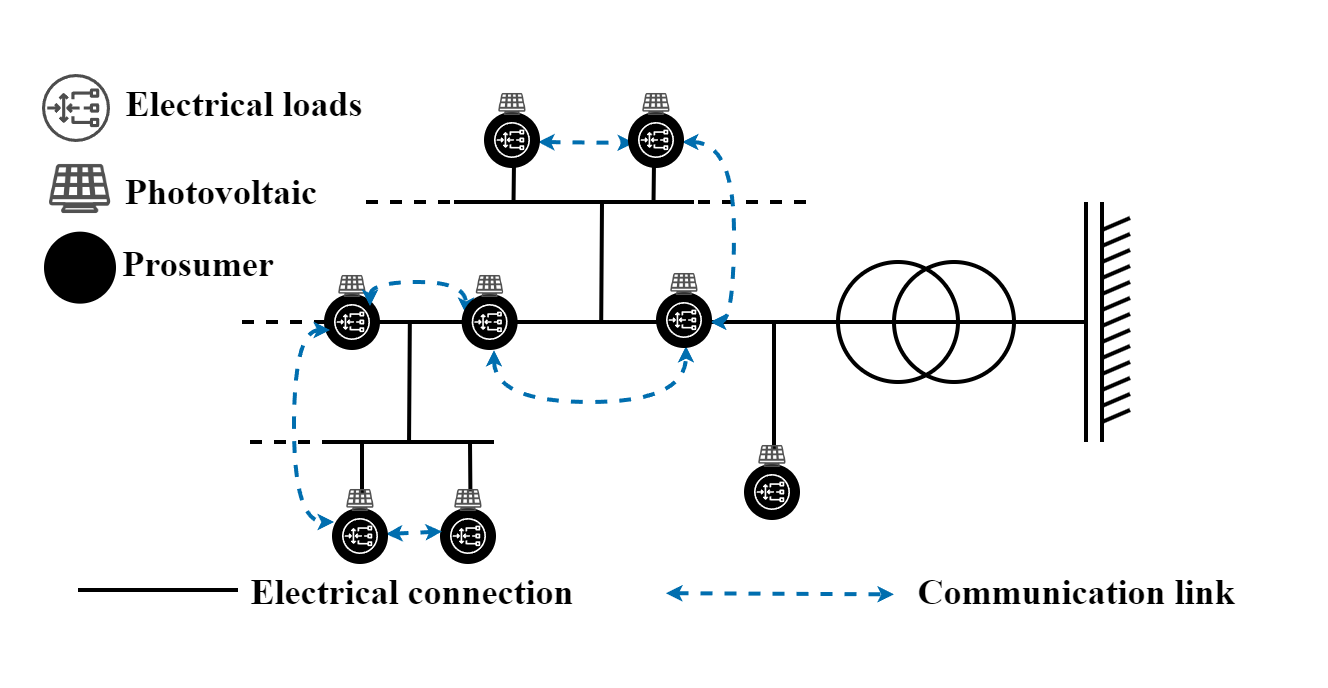}}
  \caption{A conceptual P2P trading system example}\label{Fig.0}
\end{figure}
In this context, P2P energy management has attracted significant attention as a viable mechanism to optimize energy management \cite{Chang2022, Chien2023}. Characterized by an open, decentralized direct trading model and high transparency, this system fundamentally shifts the transaction paradigm. Its core value lies in dynamically balancing local supply and demand, thereby enhancing the economic viability for prosumers and promoting regional renewable energy consumption \cite{Khorasany2021}. To illustrate the structural evolution compared to traditional systems, Fig. \ref{Fig.0} presents a conceptual P2P transformation network model. Within this framework, the P2P trading system integrates prosumers' strategic interactions with energy carrier, while explicitly investigating resource flexibility \cite{Wang2024d}.

%最近对P2P的研究：由集中式--》分布式（列举）

To operationalize the P2P trading systems, extensive research has been conducted on developing trading algorithms. Initially, centralized approaches, such as auction-based mechanisms \cite{Wu2018c} and coalition game theory models \cite{Zhang2019a}, were employed to facilitate transactions in community microgrids. However, the dependence of these methods on a central coordinator introduces significant challenges regarding user privacy and system scalability. To address these limitations and eliminate the need for a central authority, the research focus has shifted towards decentralized techniques. For examples, analytic target cascading \cite{Xiao2024} and enhanced benders decomposition \cite{Xia2022}, which allow prosumers to operate independently based on decomposition principles. Furthermore, recent studies have advanced this field through consensus-based algorithms \cite{Zhang2011,Hug2015} and the alternating direction method of multipliers (\textit{ADMM}) \cite{Zhang2017,Morstyn2019}, enabling multi-class energy trading without compromising privacy.

%针对分布式算法，很多不同的算法和使用场景都被考虑到，例如\cite{Chang2023}解决了隐私方面的风险问题，\cite{Chen2019,Wang2019}提出a Q-cube framework。这些都是将优化问题视为一个同步并行分布问题，所有产消者同步解决子问题，所有耦合变量在每次迭代中同时更新。但是，使用此类方法不适合异步产消者。同步的方法会导致一些计算能力良好的产消费者需要等待来自计算能力较差的产消费者的交易信息。在这种情况下，每次迭代所耗费的时间是由计算能力最不发达的产消费者决定的，这进而限制了解决方案的整体效率。此外，这种方法还会导致闲置的产消费者所拥有的计算资源得不到充分利用。为此，异步优化算法在众多现有研究中陆续得到应用，例如异步的交替方向乘子法，是在电力分配系统和传输系统的能源优化方面[19]、[20]、[21]被较多次的提到。同时，异步优化算法也被用于点对点能源交易[22]、[23]和机器学习领域[\cite{Liu2025}]。针对无向网络上的共识优化问题，文献\cite{Wu2023a}设计了分散式梯度下降（DGD）和先适应后组合技术（DGD-ATC）算法在部分异步和完全异步情况下建立了收敛保证。这使得拥有良好计算能力的产消费者能够利用其他计算能力较差的产消费者的先前交易信息来进行优而不是等待最新的更新信息。虽然上述的异步算法提高了计算效率，但是计算能力良好的消费者可能会在错误的方向更新自己的交易信息，从而导致最优结果向次优结果偏移。因此，提出一种异步DOEM来应对上述挑战，同时在数值证明下保证收敛，进一步考虑通信延迟。

Despite the privacy and scalability advantages of the aforementioned decentralized methods, the majority algorithms, including the \textit{ADMM} and consensus algorithms, address the optimization problems in a synchronous manner. In such framework, agents with superior computing capabilities are forced to idle and wait for the slowest participants to complete their updates before proceeding to the next iteration  \cite{Guo2017}. This synchronization barrier not only limits overall system efficiency but also results in the significant underutilization of the computing resources \cite{Li2024}. To overcome these bottlenecks, asynchronous DOAs have increasingly attracted attention. By allowing prosumers to update variables using currently available local information without waiting for global synchronization, these methods significantly enhance the speed of iterative calculations. Recent studies have successfully applied asynchronous DOAs to power distribution systems \cite{Zhai2022a}, particularly in P2P trading systems \cite{Distribution2022}. To name a few, Ullah et al. \cite{Distribution2022} leveraged \textit{ADMM} with asynchronous communications to design a decentralized voltage management method for P2P energy trading, which improves the efficiency and robustness of the communications among prosumers and  shows superiorities in short-term market clearings. Jin et al. \cite{Jin2025} considered the heterogeneous computation and communication abilities of prosumers in P2P multi-energy trading of smart buildings and designed an asynchronous \textit{ADMM}, which promotes the flexibility of building's heating loads via introducing multi-energy trading. Wang et al. \cite{Wang2024e} explored the flexibilities of heating loads of low-carbon buildings in P2P energy trading and developed an asynchronous \textit{ADMM} to adapt prosumers to various communication network infrastructures. While these \textit{ADMM}-based methods \cite{Distribution2022,Jin2025} attempt to incorporate asynchrony into decentralized P2P energy trading, they fail to provide efficient mechanisms to handle communication delays. Consequently, they remain susceptible to significant synchronization overheads, for instance prolonged idle waiting times, in real-world scenarios. Furthermore, as extensions of the \textit{ADMM} framework, these asynchronous studies \cite{Distribution2022,Jin2025,Wang2024e} can only provide close-loop solution of the P2P energy management problems and suffer from an absence of rigorous theoretical results on complexity and convergence, as well as a lack of theoretical basis for parameter tuning.

\subsection{Motivation and Challenge}\label{MC}
The aforementioned gaps motivate us to devise a DOA for a category of dynamic energy management tasks over Asyn-P2P transactive networks. This approach obviates the need for a global synchronization clock among prosumers and ensures robustness against time-varying communication delays. Furthermore, to bridge the theoretical gaps in existing P2P energy management studies \cite{Distribution2022,Jin2025,Wang2024e}, we provide a comprehensive analysis of complexity and convergence properties for the proposed algorithms. Achieving these goals is non-trivial, as it requires establishing a technical framework that fundamentally departs from conventional \textit{ADMM}-based paradigms, necessitating a novel analytical approach.

\subsection{Contributions}\label{Contri}
The main contributions of this work are stated blow.
\begin{enumerate}
  \item We propose a novel technical line based on operator-splitting theory to design two decentralized algorithms, namely \textit{Syn-DYNA} and \textit{Asyn-DYNA}, for a class of dynamic P2P energy management problems. Both \textit{Syn-DYNA} and \textit{Asyn-DYNA} are proved to converge to the optimal solution of the dynamic P2P energy management problems. \textit{Asyn-DYNA} extends \textit{Syn-DYNA} to asynchronous settings, which not only improves the efficiency of P2P energy management compared to synchronous P2P (Syn-P2P) methods \cite{Chang2023,Morstyn2018,Zhou2023,Shah2024,Liu2023a,Ullah2021} by eliminating the global synchronization clock, but also enhances the robustness of existing asynchronous studies \cite{Distribution2022,Jin2025,Wang2024e} against communication delays.
  \item We leverage monotone operator theory to provide a rigorous convergence analysis for both algorithms. Specifically, a faster linear convergence rate is established for the synchronous setting and an asymptotic convergence is proved for the asynchronous counterpart. This offers a novel technical trajectory for analyzing P2P energy management problems, distinct from the methods employed in existing studies \cite{Distribution2022,Jin2025,Wang2024e,Chang2023,Morstyn2018,Zhou2023,Shah2024,Liu2023a,Ullah2021}.
  \item Different to the operator-splitting studies \cite{Li2024,Li2021}, this paper focuses on studying the dynamic P2P energy management problems, which is a more challenging problem formulation comprising of coupling constraints and diverse dynamic objective functions. We propose a novel framework to tackle these challenges by employing mathematical mappings and transformations.
\end{enumerate}

\subsection{Organization}\label{Organ}
We provide the remainder of this work in this subsection. The basic notation, communication network model, and the dynamic energy management problem formulation are presented in Section \ref{Section 2}. Section \ref{Section 3} describes the connection of the proposed algorithms development and detailed updates. In Section \ref{Section 4}, detailing the convergence results of the proposed algorithms. Section \ref{Section 5} presents the numerical experiments and analysis on proposed algorithms, with Section \ref{Section 6} following to conclude the paper. Some mathematical derivations and simulation results can be found in the supplementary material.

\section{Preliminaries}\label{Section 2}

\subsection{Basic Notation}\label{Section 2-1}
For arbitrary two vectors $ x,y \in {\mathbb{R}^n}$, the conjugate function of $h:{\mathbb{R}^n} \to \mathbb{R} $ is denoted by $h^ \otimes $, which is defined by ${h^ \otimes }\left( x \right) := {\max _y}\left\{ {{x^{\top} }y - h\left( y \right)} \right\}$; considering a closed, proper, and convex function, $f:{\mathbb{R}^n} \to \mathbb{R} $, together with a symmetric positive definite matrix $Z \in {\mathbb{R}^{n \times n}}$, we denote the proximal operator (resolvent) of its subdifferential as ${\mathbf{prox}}_f^Z\{ x \} = \arg {\min _y}\{ {f\left( y \right) + \left\| {y - x} \right\|_Z^2} \}$, while for a scalar $a>0$, the proximal operator is defined as: ${\mathbf{pro}}{{\mathbf{x}}_{af}}\{ x \} = \arg {\min _y}\{ {f\left( y \right) + \frac{1}{{2a}}{{\left\| {y - x} \right\|}^2}} \}$. The definition of the indicator function is given by ${\mathcal{I}_\mathcal{X}}\left( x \right) = \left\{ \begin{gathered}
  0, \quad \,\, {\text{         }}x \in \mathcal{X} \hfill \\
   + \infty, {\text{     }}x \notin \mathcal{X} \hfill \\
\end{gathered}  \right.$ and ${\mathcal{P}_\mathcal{X}}\left\{ x \right\}$ is denoted as a projection of $x$ into the set $\mathcal{X}$. Since some operators are also considered as mapping matrices, they are interchangeable throughout the paper. We have summarized the remaining basic symbols and notations of this article in TABLE \ref{Table 1}.

\begin{table}[!h]
\centering
\renewcommand{\arraystretch}{1.5}
\begin{tabularx}{8.8cm}{lX}  % 10cm 減去前兩個欄位寬度後，剩下的通通給
\hline                      % 第三欄位使用，文字超出的部份會自動折行
\bf{Symbols}  & \bf{Definitions}  \\
\hline
${\mathbb R}$, ${{\mathbb R}^n}$, ${{\mathbb R}^{m \times n}}$ & the sets of real numbers, $n$-dimensional column real vectors, $m \times n$ real matrices, respectively\\
${\mathbb{N}_0}$ & the set of natural numbers including 0 \\
$\mathbb{E} {\cdot} $ & the expectation of any random variable \\
:= & the definition symbol\\
%${\rm{diag}}\left\{ \nu \right\}$ & a diagonal matrix with all the elements of vector $\nu \in {\mathbb{R}^n}$ laying on its main diagonal\\
%$X \le Y$ & each element in $Y - X$  is nonnegative, where $X$ and $Y$ are two vectors or matrices with same dimensions\\
% $\tilde x \otimes \tilde y$ & the Kronecker product of vectors $\tilde x$ and $\tilde y$\\
$\left| \cdot \right|$ & an operator to represent the absolute value of a scalar or the cardinality of a set\\
${\cdot ^ \top }$ & the transpose of any matrices or vectors \\
${\mathbf{0}}$   & a zero matrix with appropriate dimensions\\
${\mathbf{I}}$   & an identity matrix with appropriate dimensions\\
% ${0_{n}}$   & an $n$-dimensional column vector with all-zero elements\\
${\mathbf{1}}$   & an all-one matrix with appropriate dimensions\\
\hline
\end{tabularx}
\caption{Basic notations.}
\label{Table 1}
\end{table}

\subsection{Communication Network Model}\label{Section 2-2}
Every prosumer is assigned a unique agent, which is responsible for communication, computation, and decision making. To simplify terminology, the terms `prosumer' and `agent' are used interchangeably throughout this paper. As shown in Fig. \ref{Fig.0}, all agent forms an communication network.

\begin{assumption}\label{A2-2-1}
The undirected communication network is bidirectionally connected.
\end{assumption}

Note that the conditions on communication delays are entailed by convergence analysis and thus are deferred to Section \ref{Section 4}.

\subsection{Dynamic Energy Management Problem}\label{Section 2-3}
This paper considers three kinds of objective functions of prosumers in the P2P trading systems, which are introduced below.
\begin{enumerate}
\item {\bf{Seller-Buyer Traded Cost:}} Define the traded power between prosumers $i$ and $j$ by ${p_{i,j,t}}$, which is positive when $i$ is a seller and $j$ is a buyer and when $j$ is a seller and $i$ is a buyer otherwise, and the sum of traded power of the prosumer $i$ by ${p_{i,t}} := {\text{col}}{\left\{ {{p_{i,j,t}}} \right\}_{j \in {\mathcal{N}_i}}} \in {\mathbb{R}^{\left| {{\mathcal{N}_i}} \right|}}$. Without loss of generality, the cost function for the seller $i$ or buyer $i$ at time $t$, $\forall t \in \mathcal{T}$,  is characterized by the following time-varying quadratic function
\begin{equation}\label{E2-3-1}
{s_{i,t}}\left( {{p_{i,t}}} \right) := \sum\limits_{j \in {\mathcal{N}_i}} {{a_{i,j,t}}p_{i,j,t}^2}  + \sum\limits_{j \in {\mathcal{N}_i}} {{\tilde b _{i,j,t}}{p_{i,j,t}}} + {c_{i,j,t}},
\end{equation}
where ${{a_{i,j,t}}}>0$ is time-varying cost coefﬁcients associated with the  coupling traded energy between two prosumers $i$ and $j$, $i \in \mathcal{V}$, $j \in {\mathcal{N}_i}$. Note that these time-varying coupling cost coefficients can better characterize the traded situations among prosumers in contrast to the existing works  \cite{Ullah2021,Ullah2020,Chang2023,Zhou2023}.

\item {\bf{Network Utilization Fee:}} The network utilization fee is a charge applied to P2P transactive networks participants for using the power network infrastructure, which is calculated based on the net electrical distance between trading parties to cover costs such as grid modernization, taxes, polices, and maintenance. Inspired by  \cite{Liu2023a,Ullah2021,Chang2023}, the network utilization fee for prosumer $i$ at time $t$, is characterized by the following linear function
\begin{equation}\label{E2-3-2}
{n_{i,t}}\left( {{p_{i,t}}} \right) := \sum\limits_{j \in {\mathcal{N}_i}} {{\hat b_{i,j,t}}{p_{i,j,t}}}, i \in \mathcal{V}, t \in \mathcal{T},
\end{equation}
where ${\hat b_{i,j,t}} \ge 0$ is a time-varying coefficient associated with the electrical transport distance between the prosumer $i$ and its trader $j$, $j \in {\mathcal{N}_i}$.

\item {\bf{Reputation Income:}} The reputation income quantifies the benefit brought by each participant based on the bilateral trading reputation, which is determined by traders’ previous relation and experience. In reality, prosumers are inclined to negotiate with the traders of good transaction performance and reputation in past transactive records. Therefore, it is inspired from  \cite{Morstyn2018, Ullah2021} that the reputation income is characterized by the following time-varying linear function
\begin{equation}\label{E2-3-3}
{r_{i,t}}\left( {{p_{i,t}}} \right) := \sum\limits_{j \in {\mathcal{N}_i}} {{\bar b_{i,j,t}}{p_{i,j,t}}}, i \in \mathcal{V}, t \in \mathcal{T},
\end{equation}
where ${\bar b_{i,j,t}} \ge 0$ is a time-varying bilateral trading reputation factor associated with the prosumer $i$ and its trader $j$, $j \in {\mathcal{N}_i}$.
\end{enumerate}

Different to most existing works, e.g., \cite{Chang2023,Morstyn2018,Zhou2023,Shah2024,Liu2023a,Ullah2021}, all coefﬁcients of the above objective functions are assumed to be bilateral and time-varying, which can better
capture the dynamic shifting of system states. Apart from the above objective functions, some physical constraints are considered as follows:
\begin{enumerate}
\item {\bf{Power Set-Point Constraint:}} Let ${\hat p_{i,t}} = \sum\nolimits_{j \in {\mathcal{N}_i}} {{p_{i,j,t}}} $ denote the power set point of prosumer $i$, $\forall i \in \mathcal{V}$, at time $t$, $\forall t \in \mathcal{T}$. The limit of prosumers' power-set point is defined by the following box constraint
\begin{equation}\label{E2-3-4}
\hat p_{i,t}^{{\text{min}}} \le {\hat p_{i,t}} \le \hat p_{i,t}^{{\text{max}}}, i \in \mathcal{V}, t \in \mathcal{T}.
\end{equation}
where $\hat p_{i,t}^{{\text{min}}}$ and $\hat p_{i,t}^{{\text{max}}}$ are the upper and lower bounds on the power set point of prosumer $i$ at time $t$, respectively.

\item {\bf{Bilateral Reciprocity Constraint:}} We define the sets of energy sellers and buyers at time $t$ as $\mathcal{V}_t^{\text{s}}$ and $\mathcal{V}_t^{\text{s}}$ , respectively, such that $\mathcal{V} = \mathcal{V}_t^{\text{s}} \cup \mathcal{V}_t^{\text{c}}$. Here, the time index $t$ is adopted to denote a possible alteration of prosumers' identities over time. To avoid the energy transaction happens between two producers or two buyers, we assume that for any two traders $i$ and $j$, $j \in {\mathcal{N}_i}$, if $i$ is an energy seller, i.e., $p_{i,j,t} \ge 0$, then $j$ must be an energy buyer, i.e., $p_{i,j,t} \le 0$, and vice versa. If there is no energy transactions happening between $i$ and $j$, then $p_{i,j,t} = p_{i,j,t} = 0$. Different with \cite{Chang2023,Morstyn2018,Zhou2023,Shah2024,Liu2023a,Ullah2021}, we consider the energy loss in transportation $loss_{i,j,t}$, which is always a nonnegative value since the selling energy must be no less than the received energy. Therefore, the bilateral reciprocity constraints are given by
\begin{subequations}\label{E2-3-5}
\begin{align}
 \label{E2-3-5-1} & {p_{i,j,t}} + {p_{j,i,t}} = l_{i,j,t}^{\text{p}}, i \in \mathcal{V}, j \in {\mathcal{N}_i}, t \in \mathcal{T}, \\
 \label{E2-3-5-2} & {p_{i,j,t}} \ge 0, i \in \mathcal{V}_t^{\text{s}}, j \in \mathcal{N}_i, t \in \mathcal{T}, \\
 \label{E2-3-5-3} & {p_{i,j,t}} \le 0, i \in \mathcal{V}_t^{\text{c}}, j \in \mathcal{N}_i, t \in \mathcal{T}.
\end{align}
\end{subequations}
%\item 3
\end{enumerate}
To recap, this paper aims to minimize the total cost (maximize the overall social welfare) of all prosumers, which can be modelled as an dynamic P2P energy management problem as follows:
\begin{equation}\label{E2-3-6}
\begin{aligned}
\mathop {\min }\limits_{{p_{i,t}}} &\sum\limits_{t = 1}^T {\sum\limits_{i = 1}^m {{s_{i,t}}\left( {{p_{i,t}}} \right) + {n_{i,t}}\left( {{p_{i,t}}} \right) - {r_{i,t}}\left( {{p_{i,t}}} \right)} }\\
{\text{s.t.}}  \;\; & \hat p_{i,t}^{{\text{min}}} \le {\hat p_{i,t}} \le \hat p_{i,t}^{{\text{max}}}, i \in \mathcal{V}, t \in \mathcal{T},\\
&{p_{i,j,t}} + {p_{j,i,t}} = l_{i,j,t}^{\text{p}}, i \in \mathcal{V}, j \in {\mathcal{N}_i}, t \in \mathcal{T},\\
&{p_{i,j,t}} \ge 0, i \in \mathcal{V}_t^{\text{s}}, j \in \mathcal{N}_i, t \in \mathcal{T},\\
&{p_{i,j,t}} \le 0, i \in \mathcal{V}_t^{\text{c}}, j \in \mathcal{N}_i, t \in \mathcal{T},
\end{aligned}
\end{equation}
which is strongly-convex and smooth but with complicated coupling constraints.

\subsection{Problem Reformulation}\label{Section 2-4}
To make an equivalent transformation of the P2P energy management problem (\ref{E2-3-1}) into a more solvable formulation, we first define a scalar ${b_{i,j,t}} = {{\tilde b}_{i,j,t}} + {{\hat b_{i,j,t}}} + {{\bar b_{i,j,t}}}$, vectors ${a_{i,t}}: = {\text{col}}{\left\{ {{a_{i,j,t}}} \right\}_{j \in {\mathcal{N}_i}}} \in {\mathbb{R}^{\left| {{\mathcal{N}_i}} \right|}}$, ${{b}_{i,t}}: = {\text{col}}{\left\{ {{b_{i,j,t}}} \right\}_{j \in {\mathcal{N}_i}}} \in {\mathbb{R}^{\left| {{\mathcal{N}_i}} \right|}} $, ${b_i}: = {\text{col}}{\left\{ {{b_{i,t}}} \right\}_{t \in \mathcal{T}}} \in {\mathbb{R}^{T\left| {{\mathcal{N}_i}} \right|}}$, ${p_i}: = {\text{col}}\left\{ {{p_{i,t}}} \right\}_{t = 1}^T\in {\mathbb{R}^{T\left| {{\mathcal{N}_i}} \right|}}$, $p: = {\text{col}}\left\{ {{p_i}} \right\}_{i = 1}^m \in {\mathbb{R}^{T\sum\nolimits_{i = 1}^m {\left| {{\mathcal{N}_i}} \right|} }}$, $\hat p_i^{{\text{min}}}: = {\text{col}}{\left\{ {\hat p_{i,t}^{{\text{min}}}} \right\}_{t \in \mathcal{T}}} \in {\mathbb{R}^T}$, $\hat p_i^{{\text{max}}}: = {\text{col}}{\left\{ {\hat p_{i,t}^{{\text{max}}}} \right\}_{t \in \mathcal{T}}} \in {\mathbb{R}^T}$, and $l_{ij}^{\text{p}}: = {\text{col}}{\left\{ {l_{i,j,t}^{\text{p}}} \right\}_{t \in \mathcal{T}}} \in {\mathbb{R}^T}$, together with matrices ${\tilde A_{i,t}} := {\text{diag}}\left\{ {{a_{i,t}}} \right\} \in {\mathbb{R}^{\left| {{\mathcal{N}_i}} \right| \times \left| {{\mathcal{N}_i}} \right|}}$, ${A_i} = \left[ {\begin{array}{*{20}{c}}
  {{{\tilde A}_{i,1}}}&{\mathbf{0}}&{\mathbf{0}}&{\mathbf{0}} \\
  {\mathbf{0}}&{{{\tilde A}_{i,2}}}&{\mathbf{0}}&{\mathbf{0}} \\
  {\mathbf{0}}&{\mathbf{0}}& \ddots &{\mathbf{0}} \\
  {\mathbf{0}}&{\mathbf{0}}&{\mathbf{0}}&{{{\tilde A}_{i,T}}}
\end{array}} \right] \in {\mathbb{R}^{T\left| {{\mathcal{N}_i}} \right| \times T\left| {{\mathcal{N}_i}} \right|}}$,
%${\tilde \Xi _{i,t}}: = \left[ {{\mathbf{0}} \ldots ,{\mathbf{0}},{\mathbf{I}},{\mathbf{0}}, \ldots {\mathbf{0}}} \right] \in {\mathbb{R}^{\left| {{\mathcal{N}_i}} \right| \times T\left| {{\mathcal{N}_i}} \right|}}$, ${\Xi _{i,t}}: = {\mathbf{1}}_{\left| {{\mathcal{N}_i}} \right|}^ \top {\tilde \Xi _{i,t}}$,
and ${\Xi _i}: = {\text{col}}{\left\{ {{\Xi _{i,t}}} \right\}_{t \in \mathcal{T}}} \in {\mathbb{R}^{T \times T\left| {{\mathcal{N}_i}} \right|}}$ with ${\Xi _{i,t}}$ being a vector with all-zero elements except from ${\left( {t - 1} \right)\left| {{\mathcal{N}_i}} \right|}$ to ${ {t} \left| {{\mathcal{N}_i}} \right|}$-th elements being 1. Based on these definitions, the P2P energy management problem (\ref{E2-3-1}) can be equivalently written as the following constrained optimization problem
\begin{equation}\label{E2-4-1}
\begin{aligned}
\mathop {\min }\limits_{{p}} & \sum\limits_{i = 1}^m {p_i^ \top {A_i}{p_i} + b_i^ \top {p_i}}\\
{\text{s.t.}}  \;\; & \hat p_i^{{\text{min}}} \le {\Xi _i}{p_i} \le \hat p_i^{{\text{max}}}, i \in \mathcal{V}, \\
&{\Omega _{ij}}{p_i} + {\Omega _{ji}}{p_j} = l_{ij}^{\text{p}}, i \in \mathcal{V}, j \in {\mathcal{N}_i}, \\
&{\Omega _{ij, t}}{p_i} \ge 0,i \in \mathcal{V}_t^{\text{s}}, j \in {\mathcal{N}_i}, t \in \mathcal{T},\\
&{\Omega _{ij, t}}{p_i} \le 0,i \in \mathcal{V}_t^{\text{c}}, j \in {\mathcal{N}_i}, t \in \mathcal{T},
\end{aligned}
\end{equation}
where the projection operators ${\Omega _{ij, t}} \in {\mathbb{R}^{1 \times T\left| {{\mathcal{N}_i}} \right|}}$ are vectors with all-zero elements except $\left( {\left( {t - 1} \right)\left| {{\mathcal{N}_i}} \right| + {\text{index}}\left\{ j \right\}} \right)$\footnote{Here ${\text{index}}\left\{ j \right\}$ denotes the index position of the prosumer $j$ in the neighboring set of the prosumer $i$.}-th element being $1$ and ${\Omega _{ij}}: = {\text{col}}{\left\{ {{\Omega _{ij, t}}} \right\}_{t \in \mathcal{T}}} \in {\mathbb{R}^{T \times T\left| {{\mathcal{N}_i}} \right|}}$. To handle the box constraint, we define a set $\mathcal{S}_i^{\text{p}}: = \left\{ {{p_i}|\hat p_i^{{\text{min}}} \le {\Xi _i}{p_i} \le \hat p_i^{{\text{max}}}} \right\}$ such that the original box constraint (\ref{E2-3-4}) reduces to
\begin{equation}\label{E2-4-2}
{p_i} \in \mathcal{S}_i^{\text{p}},i \in \mathcal{V},
\end{equation}
where $\mathcal{S}_i^{\text{p}}$ is a convex polyhedra owing to the affine property of the constraint. We then define an another projection operator ${\Psi _{\left( {i,j} \right)}}:{\text{ }}p \to {\left[ {{{\left( {{\Omega _{ij}}{p_i}} \right)}^ \top },{\text{ }}{{\left( {{\Omega _{ji}}{p_j}} \right)}^ \top }} \right]^ \top } \in {\mathbb{R}^{2{T} \times T\sum\nolimits_{i = 1}^m {\left| {{\mathcal{N}_i}} \right|} }}$ and a set $\mathcal{S}_{\left( {i,j} \right)}^{\text{r}} := \left\{ {{{\left[ {z_1^ \top ,z_2^ \top } \right]}^ \top } \in {\mathbb{R}^{2{T}}}\left| {{z_1} + {z_2} = l_{ij}^{\text{p}}} \right.} \right\}$ such that the coupled equality constraint in (\ref{E2-4-1}) reduces to
\begin{equation}\label{E2-4-3}
{\mathcal{I}_{{\mathcal{S}^{\text{r}}}}}\left( {\Psi p} \right): = \sum\limits_{i = 1}^m {\sum\limits_{\left( {i,j} \right) \in \mathcal{E}} {{\mathcal{I}_{\mathcal{S}_{\left( {i,j} \right)}^{\text{r}}}}\left( {{\Psi _{\left( {i,j} \right)}}p} \right)} },
\end{equation}
where ${\mathcal{S}^{\text{r}}}: = \prod\nolimits_{\left( {i,j} \right) \in \mathcal{E}} {\mathcal{S}_{\left( {i,j} \right)}^{\text{r}}} $ with $\prod\nolimits_{\left( {i,j} \right) \in \mathcal{E}} {\mathcal{S}_{\left( {i,j} \right)}^{\text{r}}} $ denoting the Cartesian product of the set ${\mathcal{S}_{\left( {i,j} \right)}^{\text{r}}}$ over all edges ${\left( {i,j} \right) \in \mathcal{E}}$. Note that the sets $\mathcal{S}_{\left( {i,j} \right)}^{\text{r}}$ and ${\mathcal{S}^{\text{r}}}$ are convex since all constraints are affine. To proceed, we define the following sets to tackle the inequality constraints in (\ref{E2-4-1}).
\begin{subequations}\label{E2-4-4}
\begin{align}
\label{E2-4-4-1}\mathcal{S}_i^{\text{s}} := \left\{ {{p_i}|{\Omega _{i,j,t}}{p_i} \ge 0, i \in \mathcal{V}_t^{\text{s}}, j \in \mathcal{N}_i, t \in \mathcal{T}} \right\},\\
\label{E2-4-4-2}\mathcal{S}_i^{\text{c}} := \left\{ {{p_i}|{\Omega _{i,j,t}}{p_i} \le 0, i \in {\mathcal{V}_t^{\text{c}}}, j \in {\mathcal{N}_i}, t \in \mathcal{T}} \right\},
\end{align}
\end{subequations}
where $\mathcal{S}_i^{\text{s}}$ and $\mathcal{S}_i^{\text{c}}$ are also convex since all constraints are affine. We continue to define a set ${\mathcal{S}_i}: = \mathcal{S}_i^{\text{p}} \cup \mathcal{S}_i^{\text{s}} \cup \mathcal{S}_i^{\text{c}}$ to conclude (\ref{E2-4-2}) and (\ref{E2-4-4}) and the corresponding indicator function
\begin{equation}\label{E2-4-5}
{\mathcal{I}_{{\mathcal{S}_i}}}\left( {{p_i}} \right) = \left\{ \begin{gathered}
  0, \quad \,\, {\text{         }} {p_i} \in {\mathcal{S}_i} \hfill \\
   + \infty, {\text{     }} {p_i} \notin {\mathcal{S}_i} \hfill \\
\end{gathered}  \right..
\end{equation}
In view of (\ref{E2-4-2})-(\ref{E2-4-5}), the constrained optimization problem (\ref{E2-4-1}) can be further transformed into the following unconstrained problem
\begin{equation}\label{E2-4-6}
\mathop {\min }\limits_{{p_i}} \sum\limits_{i = 1}^m (p_i^ \top {A_i}{p_i} \!+\! b_i^ \top {p_i} \!+\! {\mathcal{I}_{{\mathcal{S}_i}}}\left( {{p_i}} \right)) \!+\!  \sum\limits_{i = 1}^m {\sum\limits_{\left( {i,j} \right) \in \mathcal{E}} {{\mathcal{I}_{\mathcal{S}_{\left( {i,j} \right)}^{\text{r}}}}\left( {{\Psi _{\left( {i,j} \right)}}p} \right)} }.
\end{equation}
Via introducing an edge-based dual variable (price signal) ${w_{\left( {i,j} \right)}}: = {\left[ {w_{\left( {i,j} \right),i}^ \top ,w_{\left( {i,j} \right),j}^ \top } \right]^ \top } \in {\mathbb{R}^{2{T}}}$ and its collected form $w: = {\text{col}}{\left\{ {{w_{\left( {i,j} \right)}}} \right\}_{\left( {i,j} \right) \in \mathcal{E}}} \in {\mathbb{R}^{{2T\left| \mathcal{E} \right|} }}$, the saddle-point problem of (\ref{E2-4-6}) is given by
\begin{equation}\label{E2-4-7}
\begin{aligned}
& \mathop {\max }\limits_w \mathop {\min }\limits_p \mathcal{L}\left( {p,w} \right): =\sum\limits_{i = 1}^m {\left(p_i^ \top {A_i}{p_i} + b_i^ \top {p_i} + {\mathcal{I}_{{\mathcal{S}_i}}}\left( {{p_i}} \right)\right)}\\
&+\sum\limits_{i = 1}^m {\sum\limits_{\left( {i,j} \right) \in \mathcal{E}} {{{\left( {{\Psi _{\left( {i,j} \right)}}p} \right)}^ \top }{w_{\left( {i,j} \right)}} - \mathcal{I}_{\mathcal{S}_{\left( {i,j} \right)}^{\text{r}}}^ \otimes \left( {{w_{\left( {i,j} \right)}}} \right)} },
\end{aligned}
\end{equation}
where ${\mathcal{I}_{\mathcal{S}_{\left( {i,j} \right)}^{\text{r}}}^ \otimes }$ is the Fenchel conjugate of ${\mathcal{I}_{\mathcal{S}_{\left( {i,j} \right)}^{\text{r}}} }$.
%Recall the definition of the projection operator ${\Psi _{\left( {i,j} \right)}}$ such that the edge-based saddle-point problem (\ref{E2-4-7}) is equivalent to the following node-based saddle-point problem
%\begin{equation}\label{E2-4-8}
%\begin{aligned}
%& \mathop {\max }\limits_w \mathop {\min }\limits_p \mathcal{L}\left( {p,w} \right): = \sum\limits_{i = 1}^m {\left(p_i^ \top {A_i}{p_i} + b_i^ \top {p_i} + {\mathcal{I}_{{\mathcal{S}_i}}}\left( {{p_i}} \right)\right)} \\
%&+ \sum\limits_{i = 1}^m {\sum\limits_{j \in {\mathcal{N}_i}} {\left( {{{\left( {{\Omega _{ij}}{p_i}} \right)}^ \top }{w_{\left( {i,j} \right),i}} - \mathcal{I}_{{{\mathcal{S}_{\left( {i,j} \right)}^{\text{r}}}}}^ \otimes \left( {{w_{\left( {i,j} \right),j}}} \right)} \right)} }.
%\end{aligned}
%\end{equation}

\section{Algorithm Development}\label{Section 3}
Let ${p^*}: = {\text{col}}{\left\{ {p_i^*} \right\}_{i \in \mathcal{V}}}$ and ${w^*}: = {\text{col}}{\left\{ {w_{\left( {i,j} \right)}^*} \right\}_{\left( {i,j} \right) \in \mathcal{E}}}$ ($w_{\left( {i,j} \right)}^*: = {\left[ {{{\left( {w_{\left( {i,j} \right),i}^*} \right)}^ \top },{{\left( {w_{\left( {i,j} \right),j}^*} \right)}^ \top }} \right]^ \top }$) be the primal and dual solutions to the saddle-point problem (\ref{E2-4-7}), respectively. According to the Karush-Kuhn-Tucker (KKT) condition  \cite{Bertsekas2016}, the first-order primal-dual stationary point  $\left( {{p^*},{w^*}} \right) $ should satisfy
\begin{subequations}\label{E3}
\begin{align}
\label{E3-1}&{\mathbf{0}} \in {A_i}p_i^* + {b_i} + \partial {\mathcal{I}_{{\mathcal{S}_i}}}\left( {p_i^*} \right) + \sum\limits_{j \in {\mathcal{N}_i}} {\Omega _{ij}^ \top w_{\left( {i,j} \right),i}^*},\\
\label{E3-2}&{\mathbf{0}} \in \sum\limits_{\left( {i,j} \right) \in \mathcal{E}} {\left( {{\Psi _{\left( {i,j} \right)}}{p^*} - \partial \mathcal{I}_{\mathcal{S}_{\left( {i,j} \right)}^r}^ \otimes \left( {w_{\left( {i,j} \right)}^*} \right)} \right)}
\end{align}
\end{subequations}
\subsection{Syn-DYNA Development}\label{Section 3-1}
Based on (\ref{E3-1})-(\ref{E3-2}), we resort to an operator splitting method  \cite{Latafat2017} to design an iterative synchronous algorithm, which is briefly presented as follows: for each prosumer $i$, $\forall i \in \mathcal{V}$ and its trader $j$, $\forall j \in {\mathcal{N}_i}$,
\begin{subequations}\label{E3-1-1}
\begin{align}
\label{E3-1-1-1}\bar w_{\left( {i,j} \right)}^k = & {\mathbf{prox}}_{{\beta _{\left( {i,j} \right)}}{\mathcal{I}}_{\mathcal{S}_{\left( {i,j} \right)}^{\text{r}}}^ \otimes }\left\{ {w_{\left( {i,j} \right)}^k + {\beta _{\left( {i,j} \right)}}\left( {{\Psi _{\left( {i,j} \right)}}{p^k}} \right)} \right\},\\
\label{E3-1-1-2}p_i^{k + 1} = & {\mathbf{pro}}{{\mathbf{x}}_{{\alpha _i}{\mathcal{I}_{{\mathcal{S}_i}}}}}\{ {p_i^k - {\alpha _i}( {{A_i}p_i^k + b_i^ \top  \! + \! \sum\limits_{j \in {\mathcal{N}_i}} {\Omega _{ij}^ \top \bar w_{\left( {i,j} \right),i}^k} })} \},\\
\label{E3-1-1-3}w_{\left( {i,j} \right)}^{k + 1} = & \bar w_{\left( {i,j} \right)}^k + {\beta _{\left( {i,j} \right)}}\left( {{\Psi _{\left( {i,j} \right)}}{p^{k + 1}} - {\Psi _{\left( {i,j} \right)}}{p^k}} \right).
\end{align}
\end{subequations}
Note that (\ref{E3-1-1-1})-(\ref{E3-1-1-3}) cannot be implemented in a decentralized manner due to the existence of the global primal variable. Therefore, we apply the relationship $x = {\mathbf{pro}}{{\mathbf{x}}_{\delta {\mathcal{I}_\mathcal{X}}}}\left( x \right) + \delta {\mathbf{pro}}{{\mathbf{x}}_{{\delta ^{ - 1}} \mathcal{I}_\mathcal{X}^ \otimes }}\left( {{\delta ^{ - 1}}x} \right)$ for any constant $\delta > 0$, convex set $\mathcal{X}$, and $\forall x \in {\mathbb{R}^n}$, according to Moreau decomposition  \cite[Theorem 14.3]{Combettes}, together with the fact that ${\mathbf{pro}}{{\mathbf{x}}_{ {\mathcal{I}_\mathcal{X}}}}\left\{  \cdot  \right\} = {\mathcal{P}_\mathcal{X}}\left\{  \cdot  \right\}$ to design a decentralized variant of  (\ref{E3-1-1-1})-(\ref{E3-1-1-3}) as follows: for each prosumer $i$, $ \forall i \in \mathcal{V}$ and its trader $j$, $\forall j \in {\mathcal{N}_i}$,
\begin{subequations}\label{E3-1-2}
\begin{align}
\label{E3-1-2-1}\bar w_{\left( {i,j} \right), i}^k = & \frac{{{\beta _{\left( {i,j} \right)}}}}{2}( {{\Omega _{ij}}p_i^k + {\Omega _{ji}}p_j^k - l_{ {i,j}}^{\text{p}}}) + \frac{ {w_{\left( {i,j} \right), i}^k + w_{\left( {i,j} \right), j}^k}}{2},\\
\label{E3-1-2-2}p_i^{k + 1} = & {\mathcal{P}_{{\mathcal{S}_i}}}\{ {p_i^k - {\alpha _i}( {\left( {{A_i}p_i^k + {b_i}} \right) + \sum\limits_{j \in {\mathcal{N}_i}} {\Omega _{ij}^ \top \bar w_{\left( {i,j} \right),i}^k} } )} \},\\
\label{E3-1-2-3}w_{\left( {i,j} \right),i}^{k + 1} =& \bar w_{\left( {i,j} \right),i}^k + {\beta _{\left( {i,j} \right)}}\left( {{\Omega _{ij}}p_i^{k + 1} - {\Omega _{ij}}p_i^k} \right),
\end{align}
\end{subequations}
which can be implemented in a fully decentralized manner under a global synchronization clock. A more detailed update rule is presented in Algorithm \ref{Algorithm 1}.
\begin{algorithm}[!h]
\small
    \SetKwBlock{Repeat}{Repeat $ k = 0,1,2, \dots$}{}
    \KwIn{proper positive uncoordinated step-sizes $\alpha_i$ and ${{\beta _{\left( {i,j} \right)}}}$, $i \in \mathcal{V}$, $j \in {\mathcal{N}_i}$.}
    \SetKwBlock{Initialize}{Initialize:}{}
    \Initialize{arbitrary primal and dual variables $p_i^0 \in {\mathbb{R}^{T\left| {{\mathcal{N}_i}} \right|}}$ and $w_{\left( {i,j} \right),i}^0 \in {\mathbb{R}^{{T}}}$.}
    \Repeat{
    \For{\rm{prosumers} $i = 1,2, \ldots ,m$ with a global clock}{
     Each prosumer $i$ receives ${{\Omega _{ji}}p_j^k}$ and $w_{\left( {i,j} \right), j}^{k}$ from its traders $j$, $\forall j \in {\mathcal{N}_i}$.\\
     Each prosumer $i$ updates locally the intermediate dual variable according to (\ref{E3-1-2-1}).\\
     Each prosumer $i$ updates locally the primal variable according to (\ref{E3-1-2-2}).\\
     Each prosumer $i$ updates locally the dual variable according to (\ref{E3-1-2-3}).\\
     Each prosumer $i$ sends ${{\Omega _{ij}}p_i^{k+1}}$ and $w_{\left( {i,j} \right), i}^{k + 1}$ to its traders $j$, $\forall j \in {\mathcal{N}_i}$.
        }
    }
    \KwOut{$\left( {{p^k},{w^k}} \right)$ until a predefined criterion is satisfied.}
    \caption{\textit{Syn-DYNA}.}
    \label{Algorithm 1}
\end{algorithm}
\begin{remark}\label{R3-1-1}
Algorithm \ref{Algorithm 1} can be implemented in a decentralized block-coordinate manner, where each coordinate of the local primal decision variable $p_i^{k}$ is $p_{i,j,t}^{k}$, which is the traded power between prosumer $i$ and its trader $j$ at time $t$, $ i \in \mathcal{V}$, $ j \in {\mathcal{N}_i}$, $ t \in \mathcal{T}$. The dual variable $w_{\left( {i,j} \right), i}^{k}$ is served as a price signal associated with each transaction between prosumer $i$ and its trader $j$. Compared with the ADMM-based or consensus-based synchronous DOAs  \cite{Liu2023a,Ullah2021,Zhou2023,Ullah2020,Shah2024,Morstyn2018}, Algorithm \ref{Algorithm 1} is a novel decentralized energy management algorithm inspired from operator splitting, which will be proved to have superiorities of faster convergence, amendable to Asyn-P2P networks, and a rigorous mathematical guarantee.
\end{remark}
\begin{figure}[!h]
\vspace{-0.4cm} % 局部缩小与上文的距离
\subfloat[Synchronous updates.]{\includegraphics[width=1.70in,height=0.9in]{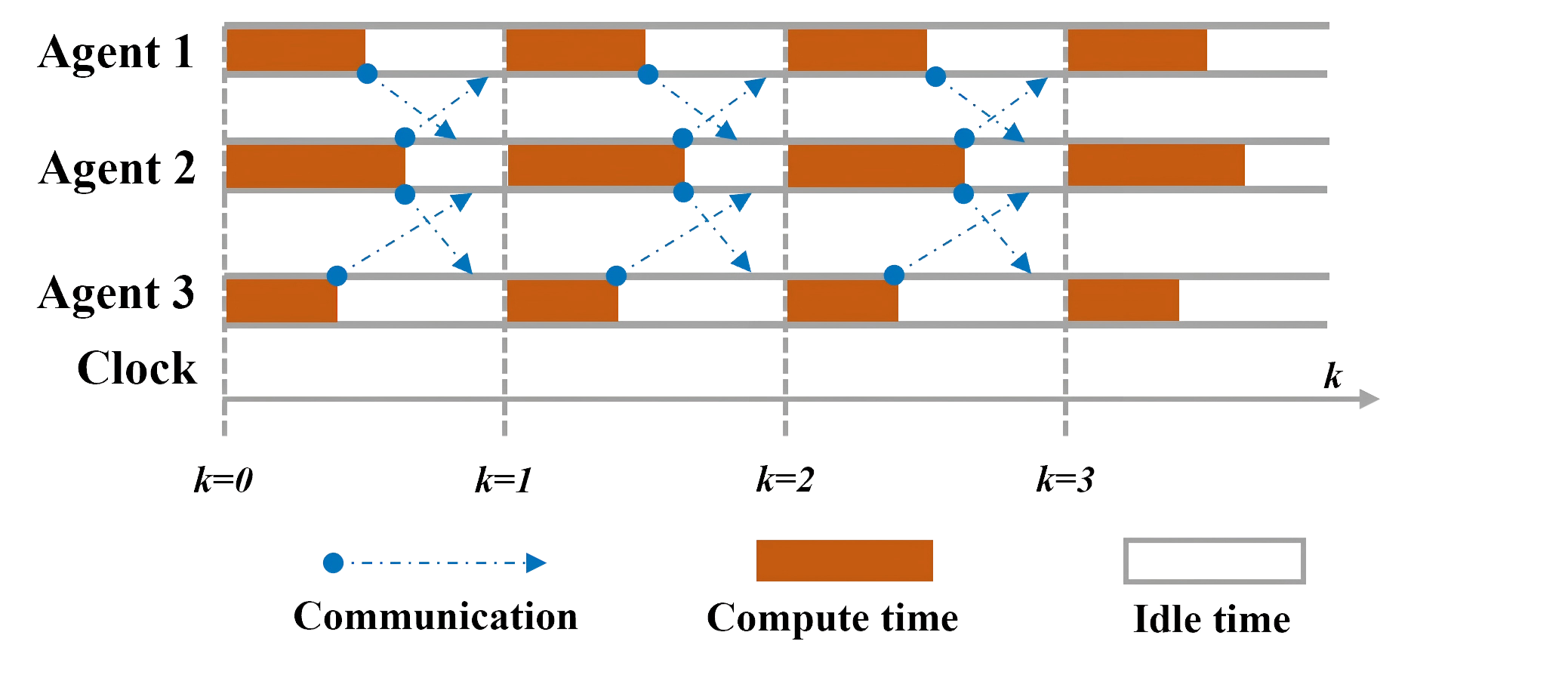}\label{Fig. 1-1}} \hfill
\subfloat[Asynchronous updates.]{\includegraphics[width=1.70in,height=0.9in]{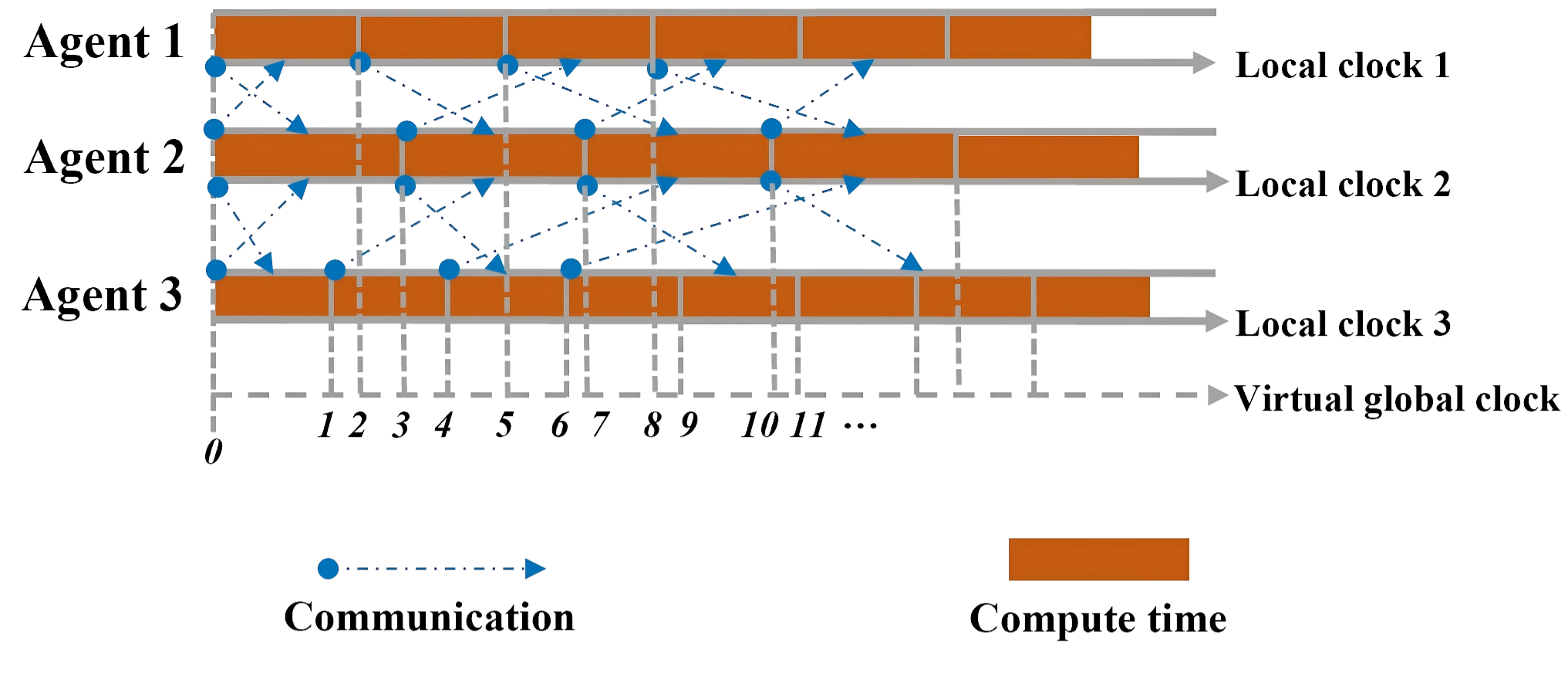}\label{Fig. 1-2}} \hfill
\caption{Synchronous versus asynchronous updates}
\label{Fig. 1}
\vspace{-0.4cm}
\end{figure}
\subsection{Asyn-DYNA Development}\label{Section 3-2}
The implementation of Algorithm \ref{Algorithm 1} relies on a global clock, which makes it only available to Syn-P2P transactive networks. However, in practice, there are unavoidable factors such as communication delays and idle time due to heterogenous computational capabilities of agents (prosumers' computation units)  \cite{Feyzmahdavian2023, Wu2023a, Tian2020}. In fact, running a synchronous algorithm in such a asynchronous environment can be inefficient and unreliable in large-scale systems, as it is often hindered by the slowest agent in the network, leading to idle times for faster agents. Asynchronous updates allow agents to update independently, leveraging their available data without waiting for others, thus improving computational efficiency and robustness against network delays or failures. Fig. \ref{Fig. 1} shows that the asynchronous updates eliminate the idle time and is more resilient to the communication delay in contrast to the synchronous updates. Therefore, it is imperative to extend Algorithm \ref{Algorithm 1} to an asynchronous version. Each agent $i$ over networks is assumed to hold a local clock denoted by ${\mathcal{K}_i}$, $\forall i \in \mathcal{V}$, and the joint set ${\mathcal{K}_1} \cup {\mathcal{K}_2} \cup  \cdots  \cup {\mathcal{K}_m}$ of all local clocks represents the virtual global clock $k$ \footnote{Inspired by  \cite{Wu2023a,Li2024}, we employ a virtual global clock $k$ to index iterations, which is only tailored for mathematical expression of the proposed asynchronous algorithm. This index is unknown by all agents and increases by $1$ whenever there is an update of any agent.}, $k \in {\mathbb{N}_0}$. At the virtual global clock $k$, we use $i_k$ and ${\theta _{{i_k}}}$ to denote the activated agent and the corresponding relaxation factor, ${i_k}  \in \mathcal{V}$, respectively, and employ ${\tau _{{i}}^k}$ and ${\varsigma _{\left( {{i},j} \right)}^k}$ to denote the (time-varying) communication delays associated with primal and dual variables of agent $i$, $i \in \mathcal{V}$, respectively. We then extend the synchronous Algorithm \ref{Algorithm 1} to an asynchronous version, which is briefly described as follows: for each prosumer $i_k$, $ \forall i_k \in \mathcal{V}$ and its trader $j$, $\forall j \in {\mathcal{N}_{i_k}}$, $\forall k \in {\mathcal{K}_{{i_k}}}$,
\begin{subequations}\label{E3-2-1}
\begin{align}
\label{E3-2-1-1}
\bar w_{\left( {{i_k},j} \right),{i_k}}^k = & \frac{{{\beta _{\left( {{i_k},j} \right)}}}}{2}\left( {{\Omega _{{i_k}j}}p_{{i_k}}^{k - \tau _{{i_k}}^k} + {\Omega _{j{i_k}}}p_j^{k - \tau _j^k} - l_{ij}^{\text{p}}} \right) \nonumber \\
& + \frac{1}{2}\left( {w_{\left( {{i_k},j} \right),{i_k}}^{k - \varsigma _{\left( {{i_k},j} \right)}^k} + w_{\left( {{i_k},j} \right),j}^{k - \varsigma _{\left( {{i_k},j} \right)}^k}} \right),\\
\label{E3-2-1-2}
\bar p_{{i_k}}^k =& {\mathcal{P}_{{\mathcal{S}_i}}}\left\{ {p_{{i_k}}^{k - \tau _{{i_k}}^k} - {\alpha _{{i_k}}}\left( {{A_{{i_k}}}p_{{i_k}}^{k - \tau _{{i_k}}^k} + {b_{{i_k}}}} \right)} \right.\nonumber \\
&\left. { - {\alpha _{{i_k}}}\sum\limits_{j \in {\mathcal{N}_{{i_k}}}} {\Omega _{{i_k}j}^ \top \bar w_{\left( {{i_k},j} \right),{i_k}}^k} } \right\},\\
\label{E3-2-1-3}
p_{{i_k}}^{k + 1} = & p_{{i_k}}^k + {\theta _{{i_k}}}\left( {\bar p_{{i_k}}^k - p_{{i_k}}^{k - \tau _{{i_k}}^k}} \right),\\
\label{E3-2-1-4}
w_{\left( {{i_k},j} \right),{i_k}}^{k + 1} = & w_{\left( {{i_k},j} \right),{i_k}}^k + {\theta _{{i_k}}}\left( {\bar w_{\left( {{i_k},j} \right),{i_k}}^k - w_{\left( {{i_k},j} \right),{i_k}}^{k - \varsigma _{\left( {{i_k},j} \right)}^k}} \right) \nonumber \\
& + {\theta _{{i_k}}}{\beta _{\left( {i{_k},j} \right)}}{\Omega _{{i_k}j}}\left( {\bar p_{{i_k}}^k - p_{{i_k}}^{k - \tau _{{i_k}}^k}} \right),
\end{align}
\end{subequations}
which can be implemented in a fully decentralized manner independent of any global synchronization clock. A more detailed update rule is presented in Algorithm \ref{Algorithm 2}.

\begin{algorithm}[!h]
\small
\SetKwBlock{Repeat}{Repeat $ k = 0,1,2, \dots$}{end}
\KwIn{proper positive uncoordinated step-sizes $\alpha_i$ and ${{\beta _{\left( {i,j} \right)}}}$, $i \in \mathcal{V}$, $j \in {\mathcal{N}_i}$.}
\SetKwBlock{Initialize}{Initialize:}{}
\Initialize{arbitrary primal variables $p_i^0 \in {\mathbb{R}^{T\left| {{\mathcal{N}_i}} \right|}}$, dual variables $w_{\left( {i,j} \right),i}^0 \in {\mathbb{R}^{{T}}}$, and local buffers $B_i$, $i \in \mathcal{V}$.}

\For{\rm{prosumers} $i = 1,2, \ldots ,m$  without a global clock}{
 Each prosumer $i$ keeps receiving the possibly delayed messages ${\Omega _{ji}}p_j^{k - \tau _j^k}$ and $w_{\left( {i,j} \right), j}^{k - \varsigma _{\left( {i,j} \right)}^k}$ from its traders $j$, $\forall j \in {\mathcal{N}_i}$, and then store $( {{\Omega _{ji}}p_j^{k - \tau _j^k},j} )$ and $( {w_{\left( {i,j} \right), j}^{k - \varsigma _{\left( {i,j} \right)}^k},j} )$ in a local buffer ${\text{B}}_i$. \\

  \SetKwBlock{If}{if the prosumer $i$ is activated at $k$ {\bf{do}}}{}
 \If{
     The activated prosumer $i_k$ reads the primal and dual variables (if there are multiple records for a same variable, then pick the most recently received one) from its local buffer ${\text{B}}_{i_k}$. \\
%     The activated prosumer $i_k$ updates locally the intermediate dual variable according to (\ref{E3-2-1-1}).\\
%     The activated prosumer $i_k$ updates locally the intermediate primal variable according to (\ref{E3-2-1-2}).\\
%     The activated prosumer $i_k$ updates locally the primal variable according to (\ref{E3-2-1-3}).\\
%     The activated prosumer $i_k$ updates locally the dual variable according to (\ref{E3-2-1-4}).\\
     The activated prosumer $i_k$ updates locally its intermediate dual and primal variables together with the primal and dual variables according to (\ref{E3-2-1-1})-(\ref{E3-2-1-4}).\\
     Each activated prosumer $i_k$ sends ${{\Omega _{{i_k}j}}p_{i_k}^{k+1}}$ and $w_{\left( {i_k,j} \right),i_k}^{k + 1}$ to its traders $j$, $\forall j \in {\mathcal{N}_{i_k}}$, and stores $( {{\Omega _{{i_k}j}}p_{i_k}^{k+1}},i_k )$ and $( w_{\left( {i_k,j} \right),i_k}^{k + 1}, i_k )$ in $B_{i_k}$.\\
 }
 \Else{The inactivated prosumer $i$ maintains its current primal and dual variables, i.e., $p_{{i}}^{k + 1} = p_{{i}}^k$ and $w_{\left( {{i},j} \right),{i}}^{k + 1} = w_{\left( {{i},j} \right), {i}}^k$.
 }
}
\KwOut{$\left( {{p^k},{w^k}} \right)$ until a predefined criterion is satisfied.}
\caption{\textit{Asyn-DYNA}.}
\label{Algorithm 2}
\end{algorithm}

\section{Convergence Analysis}\label{Section 4}
To ensure the solvability of the original P2P energy management problem (\ref{E2-3-6}), we need to make the following standard assumption.
\begin{assumption}(Feasibility and Strong Duality)\label{A4-1}\newline
i) The set of solutions to the original constrained optimization (\ref{E2-3-6}) is nonempty; \newline
ii) There exists at least one set of feasible points $p_i^*$, $i \in \mathcal{V}$, which not only satisfy the local coupled equality inequality constraint (\ref{E2-3-5-1}), but strictly meet the box and inequality constraints (\ref{E2-3-4}), (\ref{E2-3-5-2})-(\ref{E2-3-5-3}).
\end{assumption}
\begin{remark}\label{R4-1}
Assumption \ref{A4-1}-i) is a not uncommon condition in P2P energy management works  \cite{Ullah2021,Zhou2023,Chang2023}, which guarantees the existence of solutions to the original constrained optimization (\ref{E2-3-6}). Assumption \ref{A4-1}-ii) is also a standard requirement in primal-dual methods  \cite{Li2024,Li2021, Yang2019}, which ensures a zero gap between primal and dual problems and enables the KKT condition as a necessary and sufficient condition for optimality in this paper.
\end{remark}

We then make the following definitions ${f_i}\left( {{x_i}} \right) = x_i^ \top {A_i}{x_i} + b_i^ \top {x_i}$ with ${x_i} = {p_i}$, ${g_i}\left( {{x_i}} \right) = {\mathcal{I}_{\mathcal{S}_i}}\left( {{x_i}} \right)$, and ${x}: = {\text{col}}{\left\{ {x_i} \right\}_{i \in \mathcal{V}}}$ such that the unconstrained optimization problem (\ref{E2-4-7}) can be equivalently transformed into the following formulation
\begin{equation}\label{E4-1}
  \mathop {\min }\limits_{x} f\left( x \right) + g\left( x \right) + {\mathcal{I}_{{\mathcal{S}^{\text{r}}}}}\left( {\Psi x} \right),
\end{equation}
where $f: = \sum\nolimits_{i = 1}^m {{f_i}} $ and $g: = \sum\nolimits_{i = 1}^m {{g_i}} $. It is clear that the local objective function $f_i$ can be $\mu_i$-strongly convex and $L_i$-smooth, where ${\mu _i}: = {\min _{j \in {\mathcal{N}_i},t \in \mathcal{T}}}{a_{i,j,t}}$ and ${L_i}: = {\max _{j \in {\mathcal{N}_i},t \in \mathcal{T}}}{a_{i,j,t}}$, $\forall i \in \mathcal{V}$. For simplicity, we define ${L_f}: = {\max _{i \in \mathcal{V}}}{L_i}$, ${\mu}: = {\min _{i \in \mathcal{V}}}{\mu_i}$, and the condition number ${\kappa _f}: = {L_f}/\mu $ of the function $f$ in the subsequent analysis. Based on (\ref{E4-1}), we define operators ${\mathbb{T}_A}: = \left[ {\begin{array}{*{20}{c}}
  {\partial \mathcal{I}_{{\mathcal{S}^{\text{r}}}}^ \otimes }&{\mathbf{0}} \\
  {\mathbf{0}}&{\partial g}
\end{array}} \right]$,
${\mathbb{T}_C}: = \left[ {\begin{array}{*{20}{c}}
  {\mathbf{0}}&{\mathbf{0}} \\
  {\mathbf{0}}&{\nabla f}
\end{array}} \right]$, and ${\mathbb{T}_M}: = \left[ {\begin{array}{*{20}{c}}
  {\mathbf{0}}&{ - \Psi } \\
  {{\Psi ^ \top }}&{\mathbf{0}}
\end{array}} \right]$, where ${\mathbb{T}_A}$, ${\mathbb{T}_M}$, and ${\mathbb{T}_C}$ are maximally monotone, monotone, and cocoercive operators, respectively according to  \cite{Combettes}. Let $x_i^* = p_i^*$ and ${x^*}: = {\text{col}}{\left\{ {x_i^*} \right\}_{i \in \mathcal{V}}}$, and define $w^*$  and ${\mathcal{S}^*}$ by the dual solution and set of primal-dual solutions to (\ref{E4-1}), respectively. Note that ${\mathcal{S}^*}$ is non-empty under Assumption \ref{A4-1}. Therefore, seeking the primal-dual solutions to (\ref{E4-1}) is equivalently to solving
\begin{equation}\label{E4-2}
{\mathcal{S}^*} = \left\{ {U|{\mathbf{0}} \in {\mathbb{T}_A}U + {\mathbb{T}_M}U + {\mathbb{T}_C}U} \right\},
\end{equation}
where $U: = {\left[ {{w^ \top },{x^ \top }} \right]^ \top }$.

\subsection{Convergence Analysis for \textit{Syn-DYNA}}\label{Section 4-1}
We further define ${x^k}: = {\text{col}}{\left\{ {x^k_i} \right\}_{i \in \mathcal{V}}}$, ${w^k}: = {\text{col}}{\left\{ {w_{\left( {i,j} \right)}^k} \right\}_{\left( {i,j} \right) \in \mathcal{E}}}$, ${\bar w^k}: = {\text{col}}{\left\{ {\bar w_{\left( {i,j} \right)}^k} \right\}_{\left( {i,j} \right) \in \mathcal{E}}}$, ${\Lambda _\alpha }: = {\text{diag}}\left\{ \alpha  \right\}$ with $\alpha : = {\text{col}}{\left\{ {{\alpha _i}} \right\}_{i \in \mathcal{V}}}$, and ${\Lambda _\beta }: = {\text{diag}}\left\{ \beta  \right\}$ with $\beta : = {\text{col}}{\left\{ {{\beta _{\left( {i,j} \right)}}} \right\}_{\left( {i,j} \right) \in \mathcal{E}}}$, such that a matrix-vector formulation of \textit{Syn-DYNA} is given by
\begin{subequations}\label{E4-2}
\begin{align}
\label{E4-2-1} {{\bar w}^k} = & {\mathbf{prox}}_{{\mathcal{I}^ \otimes }}^{\Lambda _\beta ^{ - 1}}\left\{ {{w^k} + {\Lambda _\beta }\Psi {x^k}} \right\} \\
\label{E4-2-2} {x^{k + 1}} =& {\mathbf{prox}}_{G}^{\Lambda _\alpha ^{ - 1}}\left\{ {{x^k} - {\Lambda _\alpha }\nabla f\left( {{x^k}} \right) - {\Lambda _\alpha }{\Psi ^ \top }{{\bar w}^k}} \right\}\\
\label{E4-2-3} {w^{k + 1}} =& {{\bar w}^k} + {\Lambda _\beta }\Psi \left( {{x^{k + 1}} - {x^k}} \right).
\end{align}
\end{subequations}
To proceed, we define two block-coordinate variables ${U^k}: = {\left[ {{{\left( {{w^k}} \right)}^ \top },{{\left( {{x^k}} \right)}^ \top }} \right]^ \top }$ and ${\bar U_k}: = {\left[ {\bar w_k^ \top ,\bar x_k^ \top } \right]^ \top }$, and two intermediate operators ${\mathbb{T}_H}: = \left[ {\begin{array}{*{20}{c}}
  {\Lambda _\beta ^{ - 1}}&{\mathbf{0}} \\
  {{\Psi ^ \top }}&{\Lambda _\alpha ^{ - 1}}
\end{array}} \right]$ and ${\mathbb{T}_s}: = \left[ {\begin{array}{*{20}{c}}
  {{\Lambda^{ - 1} _\beta }}&{\mathbf{0}} \\
  {\mathbf{0}}&{{\Lambda ^{ - 1} _\alpha }}
\end{array}} \right]$ such that (\ref{E4-2-1})-(\ref{E4-2-3}) can be rewritten in a compact formulation as follows:
\begin{subequations}\label{E4-3}
\begin{align}
\label{E4-3-1} {{\bar U}^k} = & {\left( {{\mathbb{T}_H} + {\mathbb{T}_A}} \right)^{ - 1}}\left( {{\mathbb{T}_H} - {\mathbb{T}_M} - {\mathbb{T}_C}} \right){U^k}, \\
\label{E4-3-2} {U^{k + 1}} = & {U^k} + \mathbb{T}_s^{ - 1}\left( {{\mathbb{T}_H} - {\mathbb{T}_M}} \right)\left( {{{\bar U}^k} - {U^k}} \right),
\end{align}
\end{subequations}
which is the operator update of \textit{Syn-DYNA}. If we define $\mathbb{T}: = {\mathbf{Id}} + {\mathbb{T}}_s^{ - 1}\left( {{\mathbb{T}_H} - {\mathbb{T}_M}} \right)\left( {{{\left( {{\mathbb{T}_H} + {\mathbb{T}_A}} \right)}^{ - 1}}\left( {{\mathbb{T}_H} - {\mathbb{T}_M} - {\mathbb{T}_C}} \right) - {\mathbf{Id}}} \right)$ with ${{\mathbf{Id}}}$ being an identity operator, then (\ref{E4-3-1})-(\ref{E4-3-2}) reduces to
\begin{equation}\label{E4-4}
{U^{k + 1}} = \mathbb{T}{U^k}.
\end{equation}
To proved the convergence of the update (\ref{E4-4}), many existing works will show that the operator $\mathbb{T}$ is a nonexpansive or averaged operator under proper conditions. However, this paper seeks a different technical line to prove the convergence, which leverages the (Quasi)-Fej\'{e}r monotonicity  \cite{Combettes2001a}.
\begin{lemma}\label{L4-1}
Suppose that Assumptions \ref{A2-2-1}-\ref{A4-1} hold. The set of primal-dual solutions to (\ref{E4-1}) coincides with the set of fixed points of $\mathbb{T}$, i.e., ${\mathcal{S}^*} = {\mathbf{Fix}}\mathbb{T}$.
\begin{proof}
We first assume $U \in {\text{Fix}}\mathbb{T}$ such that $U = \mathbb{T}U$ and $\bar U = U$. It follows from (\ref{E4-3-1}) that
\begin{equation}\label{E4-5}
{\left( {{\mathbb{T}_H} + {\mathbb{T}_A}} \right)^{ - 1}}\left( {{\mathbb{T}_H} - {\mathbb{T}_M} - {\mathbb{T}_C}} \right)U = U.
\end{equation}
Rearranging both sides of (\ref{E4-5}) obtains $\left( {{\mathbb{T}_H} - {\mathbb{T}_M} - {\mathbb{T}_C}} \right)U \in \left( {{\mathbb{T}_H} + {\mathbb{T}_A}} \right)U$, which implies ${{\mathbf{0}} \in {\mathbb{T}_A}U + {\mathbb{T}_M}U + {\mathbb{T}_C}U}$, i.e., $U \in {\mathcal{S}^*}$. Therefore, for any $U$, if $U \in {\text{Fix}}\mathbb{T}$, then we have $U \in {\mathcal{S}^*}$, and vice versa.
\end{proof}
\end{lemma}
In the subsequent analysis, we denote any optimal solution to the saddle-point problem (\ref{E2-4-7}) by ${U^*} \in {\mathcal{S}^*}$, which coincides with the fixed point of Algorithm \ref{Algorithm 1} according to Lemma \ref{L4-1}.
\begin{lemma}(Fej\'{e}r monotonicity)\label{L4-2}
Suppose that Assumptions \ref{A2-2-1}-\ref{A4-1} hold. If the uncoordinated step-sizes satisfy $0 < {\alpha _i} < 1/\left( {\left( {{L_f}/2} \right) + \left\| {\sum\nolimits_{j \in {\mathcal{N}_i}} {{\beta _{\left( {i,j} \right)}}\Omega _{ij}^ \top {\Omega _{ij}}} } \right\|} \right)$, $\forall i \in \mathcal{V}$, the sequence ${\left\{ {{U^k}} \right\}_{k \ge 0}}$ generated by Algorithm \ref{Algorithm 1} takes a Fej\'{e}r-type relationship as follows:
\begin{equation}\label{E4-6}
\left\| {{U^{k + 1}} - {U^ * }} \right\|_{{\mathbb{T}_s}}^2 \le \left\| {{U^k} - {U^ * }} \right\|_{{\mathbb{T}_s}}^2 - \left\| {{U^{k + 1}} - {U^k}} \right\|_{{\mathbb{T}_s} - 2{\mathbb{T}_{d}}}^2,
\end{equation}
where ${\mathbb{T}_{d}} := \left[ {\begin{array}{*{20}{c}}
  {\Lambda _\beta ^{ - 1}}&{ - \frac{1}{2}\Psi } \\
  { - \frac{1}{2}{\Psi ^ \top }}&{\Lambda _\alpha ^{ - 1} - \frac{{{L_f}}}{4}{\mathbf{I}}}
\end{array}} \right]$.
\begin{proof}
See Section VII-A in the supplementary material.
\end{proof}
\end{lemma}
In the following theorem, we establish a linear convergence rate for Algorithm \ref{Algorithm 1} when there are no communication delays and without consideration of idle time among agents. As far as we concern, this linear convergence is faster than almost all the existing P2P energy management methods, such as  \cite{Zhou2023,Shah2024,Morstyn2018,Liu2023a,Ullah2021,Chang2023}. This superiority originates from adopting the operator-splitting method in the development of Algorithm \ref{Algorithm 1}. Let $\gamma := \left\| {{\mathbb{T}_s}} \right\|\left\| {{{\left( {2{\mathbb{T}_{d}} - {\mathbb{T}_s}} \right)}^{ - 1}}} \right\|{\left\| {{\mathbb{T}_s}{{\left( {{\mathbb{T}_H} - {\mathbb{T}_M}} \right)}^{ - 1}}} \right\|^2}$ ${\left( {1 + \eta \left( {\left\| {{\mathbb{T}_H} - {\mathbb{T}_M}} \right\| + {L_f}} \right)} \right)^2}$, and it can be verified that $\gamma >1$ under the conditions of Lemma \ref{L4-2}.
\begin{theorem}(Linear convergence of Algorithm \ref{Algorithm 1})\label{T4-1}
Under the conditions of Lemma \ref{L4-2}, the sequence ${\left\{ {{U^k}} \right\}_{k \ge 0}}$ generated by Algorithm \ref{Algorithm 1} converges linearly to the optimal solution at a rate of $\mathcal{O}\left( {{{\left( {1 - 1/\gamma } \right)}^k}} \right)$, i.e.,
\begin{equation}\label{E4-7}
\left\| {\mathbb{T}{U^{k + 1}} - {U^*}} \right\|_{{\mathbb{T}_s}}^2 \le \left( {1 - \frac{1}{\gamma }} \right)\left\| {{U^k} - {U^*}} \right\|_{{\mathbb{T}_s}}^2,
\end{equation}
where ${U^*} \in {\mathcal{S}^*}$ is an optimal solution.
\begin{proof}
See Section VII-B in the supplementary material.
\end{proof}
\end{theorem}

\subsection{Convergence Analysis for \textit{Asyn-DYNA}}\label{Section 4-2}
The subsequent theorem targets to prove the asymptotic convergence for Algorithm \ref{Algorithm 2}. To begin with, we need to define ${x^{k - {\tau ^k}}}: = {\text{col}}\left\{ {x_1^{k - \tau _1^k},x_2^{k - \tau _2^k}, \ldots ,x_m^{k - \tau _m^k}} \right\}$, $w_{\left( {i,j} \right)}^{k - \varsigma _{\left( {i,j} \right)}^k}: = {\left[ {{{\left( {w_{\left( {i,j} \right),i}^{k - \varsigma _{\left( {i,j} \right)}^k}} \right)}^ \top }{{\left( {,w_{\left( {i,j} \right),j}^{k - \varsigma _{\left( {i,j} \right)}^k}} \right)}^ \top }} \right]^ \top }$, and ${w^{k - {\varsigma ^k}}}: = {\text{col}}{\left\{ {w_{\left( {i,j} \right)}^{k - \varsigma _{\left( {i,j} \right)}^k}} \right\}_{\left( {i,j} \right) \in \mathcal{E}}}$. Recall the synchronous fixed-point operator defined in (\ref{E4-4}), we define a zero-padding operator ${\mathbb{T}_i}:U \to {\text{col}}\left\{ {{\text{0,}} \ldots {\text{,}}{{\left[ {\mathbb{T}U} \right]}_i}{\text{,}} \ldots {\text{,0}}} \right\}$ and an intermediate operator $\mathbb{Z}: = {\mathbf{Id}} - \mathbb{T}$ together with the associated local zero-padding operator ${\mathbb{Z}_i}: = {\text{col}}\left\{ {{\text{0,}} \ldots {\text{,}}{{\left[ {{\mathbf{Id}} - \mathbb{T}} \right]}_i}{\text{,}} \ldots {\text{,0}}} \right\}$ such that a compact update of Algorithm \ref{Algorithm 2} is given by
\begin{equation}\label{E4-2-1}
{U^{k + 1}} = {U^k} - {\theta _{{i_k}}}{\mathbb{Z}_{{i_k}}}{\hat U^k},
\end{equation}
where ${\hat U^k}: = {\left[ {{{\left( {{x^{k - {\tau ^k}}}} \right)}^ \top },{{\left( {{w^{k - {\varsigma ^k}}}} \right)}^ \top }} \right]^ \top }$. We denote the $\sigma$-algebra measurement of the sequence $\left\{ {{U^0},{{\hat U}^1},{U^1},{{\hat U}^2},{U^2}. \ldots .,{{\hat U}^k},{U^k}} \right\}$ by ${\mathcal{F}^k}$, i.e., ${\mathcal{F}^k}: = \sigma \left\{ {{U^0},{{\hat U}^1},{U^1},{{\hat U}^2},{U^2}. \ldots .{{\hat U}^k},{U^k}} \right\}$. To facilitate the subsequent analysis, we make the following assumption
\begin{assumption}(Bounded delays and independent activation)\label{A4-2-1}\newline
i) At each global virtual clock $k$, $k \in {\mathcal{K}_i}$, there exists a unified constant $d \ge 0$, which upper-bounds the communication delays $\tau _i^k$ and $\varsigma _{\left( {i,j} \right)}^k$, $i \in \mathcal{V}$, $j \in {\mathcal{N}_i}$; \newline
ii) Each agent $i$, $i \in \mathcal{V}, $ following a Poisson process with parameter ${\pi _i}$, is activated with probability ${P_i} \in \left( {0,1} \right)$ at the global virtual clock $k$, i.e., $\mathbb{P}\left( {{i_k} = i} \right) = {P_i}$. Moreover, the activation takes mutually independent increments, i.e., ${i_1},{i_2}, \ldots ,{i_k}$ being independent.
\end{assumption}
\begin{remark}\label{R4-2-1}
Assumption \ref{A4-2-1} is standard in asynchronous analysis  \cite{Wu2018a, Li2024,Peng2016}, which implies ${P_i} = {\pi _i}/\sum\nolimits_{i = 1}^m {{\pi _i}} $ and $\sum\nolimits_{i = 1}^m {{P_i}}  = 1$. It is worthwhile to mention that there are some decentralized optimization methods, such as  \cite{Wu2018a,Ramaswamy2022,Ren2020,Wu2023a}, that are able to tolerate unbounded communication delays. However, these works not only show degraded convergence rates but can only handle monotonous optimization problems with limited applications to practical engineering fields, for instance, P2P energy management.
\end{remark}
\begin{lemma}\label{L4-2-1}
Suppose that Assumption \ref{A4-2-1}-i) holds. For $J\left( k \right) \subseteq \left\{ {k - d,k - d + 1, \ldots ,k - 1} \right\}$, we have
\begin{equation}\label{E4-2-2}
{\hat U^k} = {U^k} + \sum\limits_{l \in J\left( k \right)} {\left( {{U^l} - {U^{l + 1}}} \right)}.
\end{equation}
\begin{proof}
See  \cite[Lemma 2]{Wu2018a} for the proof.
\end{proof}
\end{lemma}
The following lemma proves that the operator (\ref{E4-4}) is nonexpansive under certain conditions, which is a key relationship to establish the convergence for Algorithm \ref{Algorithm 2}. This is different with  \cite{Li2024} that requires the operator to be averaged leading to a more conservative step-size selection.
\begin{lemma}\label{L4-2-2}
Suppose that Assumptions \ref{A2-2-1} and \ref{A4-1} hold. If the step-size satisfies $0 < {\alpha _i} \le 1/\left( {\kappa _f} {\left( {L_f/\left( {2} \right)} \right) + \left\| {\sum\nolimits_{j \in {\mathcal{N}_i}} {{\beta _{\left( {i,j} \right)}}\Omega _{ij}^ \top {\Omega _{ij}}} } \right\|} \right)$, then the operator $\mathbb{T}$ is nonexpansive in $\mathbb{T}_s$-norm space, i.e.,
\begin{equation}\label{E4-2-3}
\left\| {\mathbb{T}U - \mathbb{T}Z} \right\|_{{\mathbb{T}_s}} \le \left\| {U - Z} \right\|_{{\mathbb{T}_s}},
\end{equation}
where $U$ and $Z$ are two arbitrarily vectors.
\begin{proof}
See Section VII-C in the supplementary material.
\end{proof}
\end{lemma}

%\begin{lemma}\label{L4-2-2}
%Suppose that Assumptions \ref{A2-2-1}, \ref{A4-1}, and \ref{A4-2-1} hold. Considering the sequence ${\left\{ {{U^k}} \right\}_{k \in \mathbb{N}}}$ generated by the asynchronous update (\ref{E4-2-1}), there exists a constant $c > 0$ such that
%\begin{equation}\label{E4-2-3}
%\begin{aligned}
%& \mathbb{E}\left[ {\left\| {{U^{k + 1}} - {U^*}} \right\|_{{\mathbb{T}_s}}^2|{\mathcal{F}^k}} \right]\\
%\le & \left\| {{U^k} - {U^*}} \right\|_{{\mathbb{T}_s}}^2 + \frac{c}{m}\sum\limits_{l \in J\left( k \right)} {\left\| {{U^l} - {U^{l + 1}}} \right\|_{{\mathbb{T}_s}}^2} \\
%& - \frac{1}{m}\left( {\frac{1}{{\tilde \theta }} - \frac{{{\kappa _{\text{s}}}}}{{m{p_{\min }}}} - \frac{{\left| {J\left( k \right)} \right|}}{c}} \right)\left\| {{{\tilde U}^{k + 1}} - {U^k}} \right\|_{{\mathbb{T}_s}}^2,
%\end{aligned}
%\end{equation}
%where .
%\begin{proof}
%See  \cite[Lemma 4]{Li2024} for the proof.
%\end{proof}
%\end{lemma}
Considering the asynchronous update defined in (\ref{E4-2-1}),  we define the lower and upper bounds of the activation probability as ${P_{\min }} := {\min _{i \in \mathcal{V}}}{P_i}$ and ${P_{\max }}  := {\max _{i \in \mathcal{V}}}{P_i}$, respectively. The condition number of the activation probabilities is defined by ${\kappa _P}: = {P_{\max }}/{P_{\min }}$ and the condition number of the uncoordinated step-sizes is defined by ${\kappa _s}: = {\lambda _{\max }}\left( {{\mathbb{T}_s}} \right)/{\lambda _{\min }}\left( {{\mathbb{T}_s}} \right)$, where ${\lambda _{\min }}\left( { \cdot} \right)$ and ${\lambda _{\max }}\left( { \cdot} \right)$ denote the maximum and minimum eigenvalues of the candidate matrix, respectively. Additionally, we define a virtual synchronous update as:
\begin{equation}\label{E4-2-4}
{\tilde U^{k + 1}} := {U^k} - {\tilde \theta} \mathbb{Z}{\hat U^k},
\end{equation}
where $\tilde \theta : = m{\theta _i}{P_i} $ is a positive constant. We continue to define a stacked vector $Y: = {\text{col}}\left\{ {{y_l}} \right\}_{l = 0}^d $ and $\tilde Y: = {\text{col}}\left\{ {{{\tilde y}_l}} \right\}_{l = 0}^d $, where $y_l$, $ l = 0,1, \ldots ,d $, are arbitrary vectors with appropriate dimensions, ${{\tilde y}_0} := {\mathbb{T}_s}\left( {{y_0} + d \sqrt {{P_{\min }}/{\kappa _s}} \left( {{y_0} - {y_1}} \right)} \right)$, ${{\tilde y}_l}:= {\mathbb{T}_s}$ $ \sqrt {{P_{\min }}/{\kappa _s}}\left( {\left( {l - d  - 1} \right){y_{l - 1}} + \left( {2d  - 2l + 1} \right){y_l} + \left( {l - d } \right){y_{l + 1}}} \right)$, $l = 1,2, \ldots ,d  - 1$, and ${{\tilde y}_d }: = \sqrt {{P_{\min }}/{\kappa _s}} {\mathbb{T}_s}\left( {{y_d } - {y_{d  - 1}}} \right)$. Note that the above definitions are only used for theoretical analysis and not involved in the algorithm updates. The subsequent theorem establishes asymptotic convergence for Algorithm \ref{Algorithm 2} under certain conditions. Before presenting Theorem \ref{T4-2-1}, we consider a well-defined operator $\Theta:  Y \to \tilde Y$ and two stacked vectors ${V^k}: = {\text{col}}\left\{ {{U^{k - l}}} \right\}_{l = 0}^d$ and ${V^*}: = {{\mathbf{1}}_{\left| {d + 1} \right|}} \otimes {U^*}$. We then briefly use ${\mathbb{E}_k}$ to denote the conditional expectation $\mathbb{E}\left[ \cdot {|{\mathcal{F}^k}} \right]$.
\begin{theorem}\label{T4-2-1}
(Asymptotic convergence for Algorithm 2) Suppose that Assumptions \ref{A2-2-1}, \ref{A4-1}, and \ref{A4-2-1} hold. If $0 < {\theta _i} \le 1/\left( {2d\sqrt {{\kappa _s}{\kappa _P}}  + {\kappa _s}{\kappa _P}} \right)$, we have
\begin{equation}\label{E4-2-5}
\mathbb{E}_k {\left\| {{V^{k + 1}} - {V^*}} \right\|_\Theta ^2} \le \left\| {{V^k} - {V^*}} \right\|_\Theta ^2 - \frac{\varpi }{m^2}\left\| {{{\tilde U}^{k + 1}} - {U^k}} \right\|_{{\mathbb{T}_s}}^2,
\end{equation}
where $\varpi : = 1/\left( {{\theta _i}{P_i}} \right) - 2d\sqrt {{\kappa _s}/{P_{\min }}}  - {\kappa _s}/{P_{\min }}$ \cite{Peypouquet2010}. This implies that the sequence ${\left\{ {{U^k}} \right\}_{k \in \mathbb{N}}}$ generated by the asynchronous update (\ref{E4-2-1}) converges to some ${U^*} \in {\mathcal{S}^*}$.
\begin{proof}
See Section VII-D in the supplementary material.
\end{proof}
\end{theorem}

\section{Numerical Experiments}\label{Section 5}
%\subsection{P2P Trading in a Modified IEEE 6-Bus System}

\begin{figure}[!h]
\centering
\subfloat[ $t=1,3$]{\includegraphics[width=1.20in,height=1.0in]{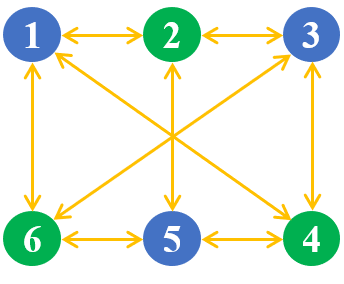}\label{Fig. 5-0-1}} \qquad\qquad
\subfloat[ $t=2, 4$]{\includegraphics[width=1.20in,height=1.0in]{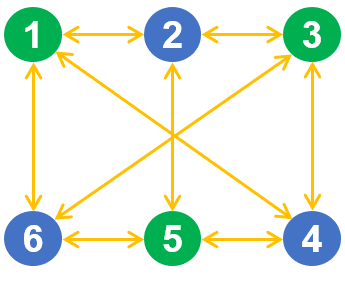}\label{Fig. 5-0-2}}
\caption{P2P communication networks.}
\label{Fig. 5-0}
\end{figure}

In this section, the effectiveness and practicality of the two proposed algorithms are demonstrated via solving a P2P energy management problem, where a network of six prosumers as shown in Fig. \ref{Fig. 5-0} is considered. For each prosumer, the power generation capacity and the desired amount of electricity to be traded vary dynamically over P2P management periods. Consequently, the roles of the prosumers change over different periods. For instance, prosumers 1, 3, and 5 are sellers during periods $t=1, 3$ as shown in Fig. \ref{Fig. 5-0}: (a) and then become buyers during periods $t = 2, 4$ as shown in Fig. \ref{Fig. 5-0}: (b).

\begin{figure*}[!h]
\centering
\subfloat[$t=1$]{\includegraphics[width=1.80in,height=1.50in]{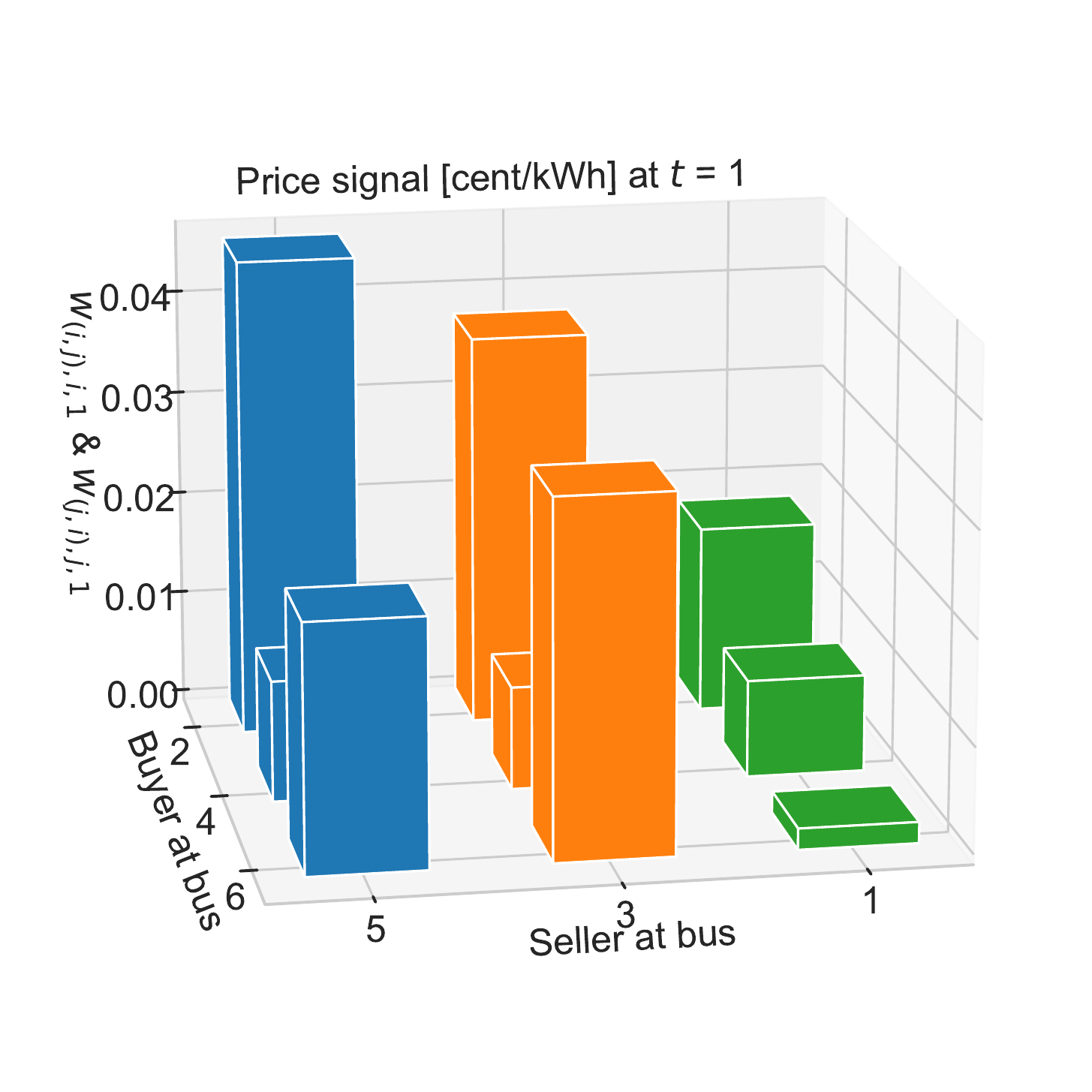}\label{Fig. 5-4-1}}
\subfloat[$t=2$]{\includegraphics[width=1.80in,height=1.50in]{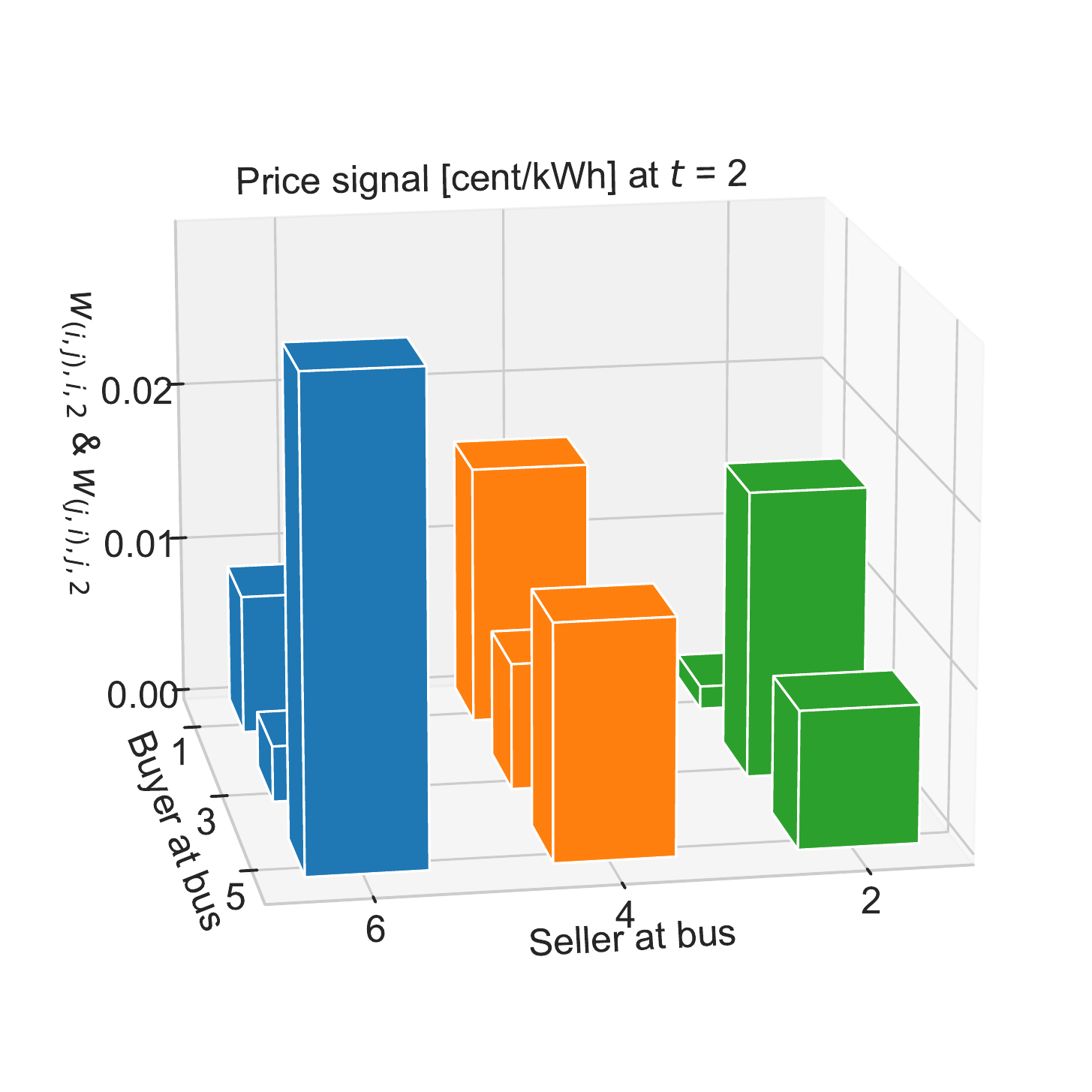}\label{Fig. 5-4-2}}
\subfloat[$t=3$]{\includegraphics[width=1.80in,height=1.50in]{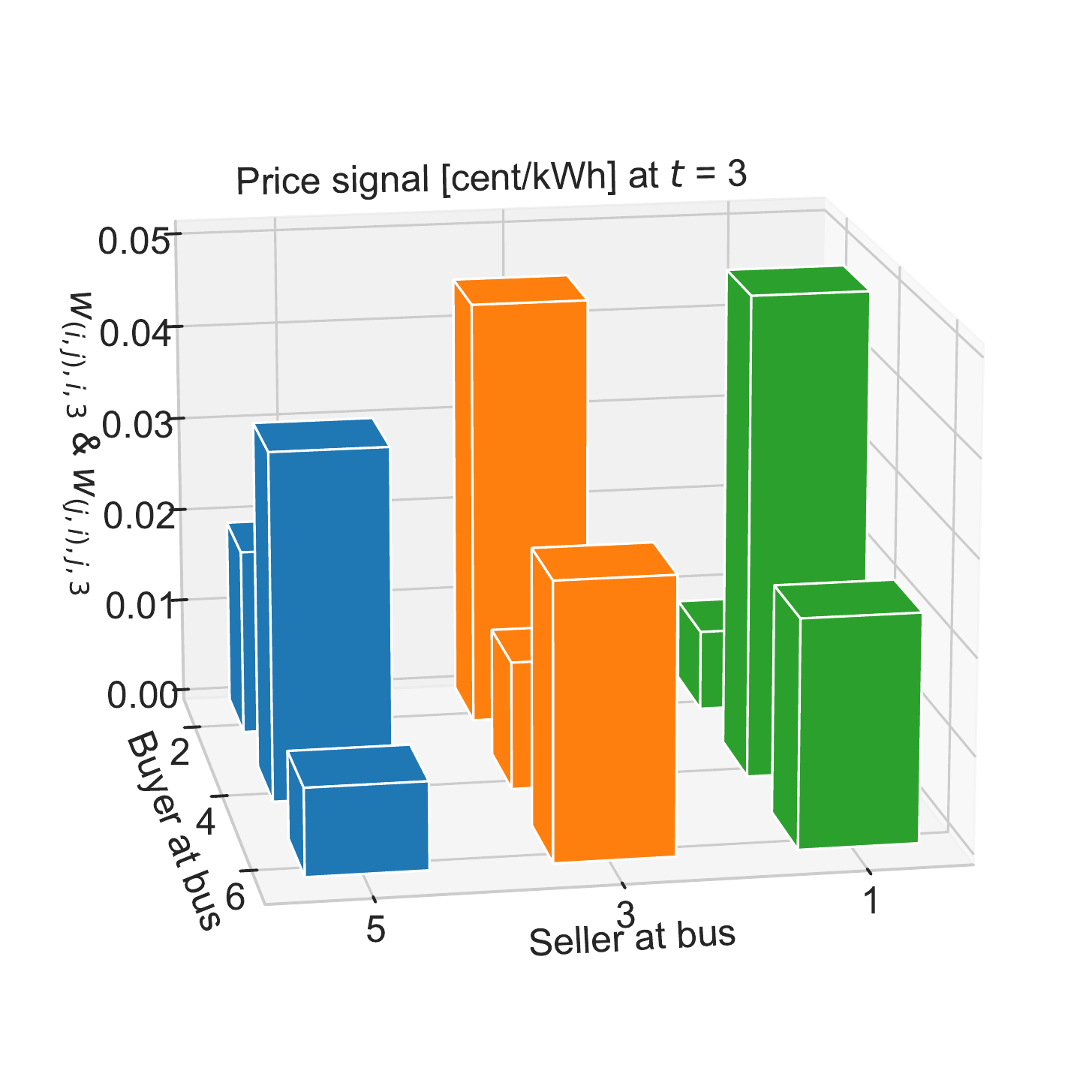}\label{Fig. 5-4-3}}
\subfloat[$t=4$]{\includegraphics[width=1.80in,height=1.50in]{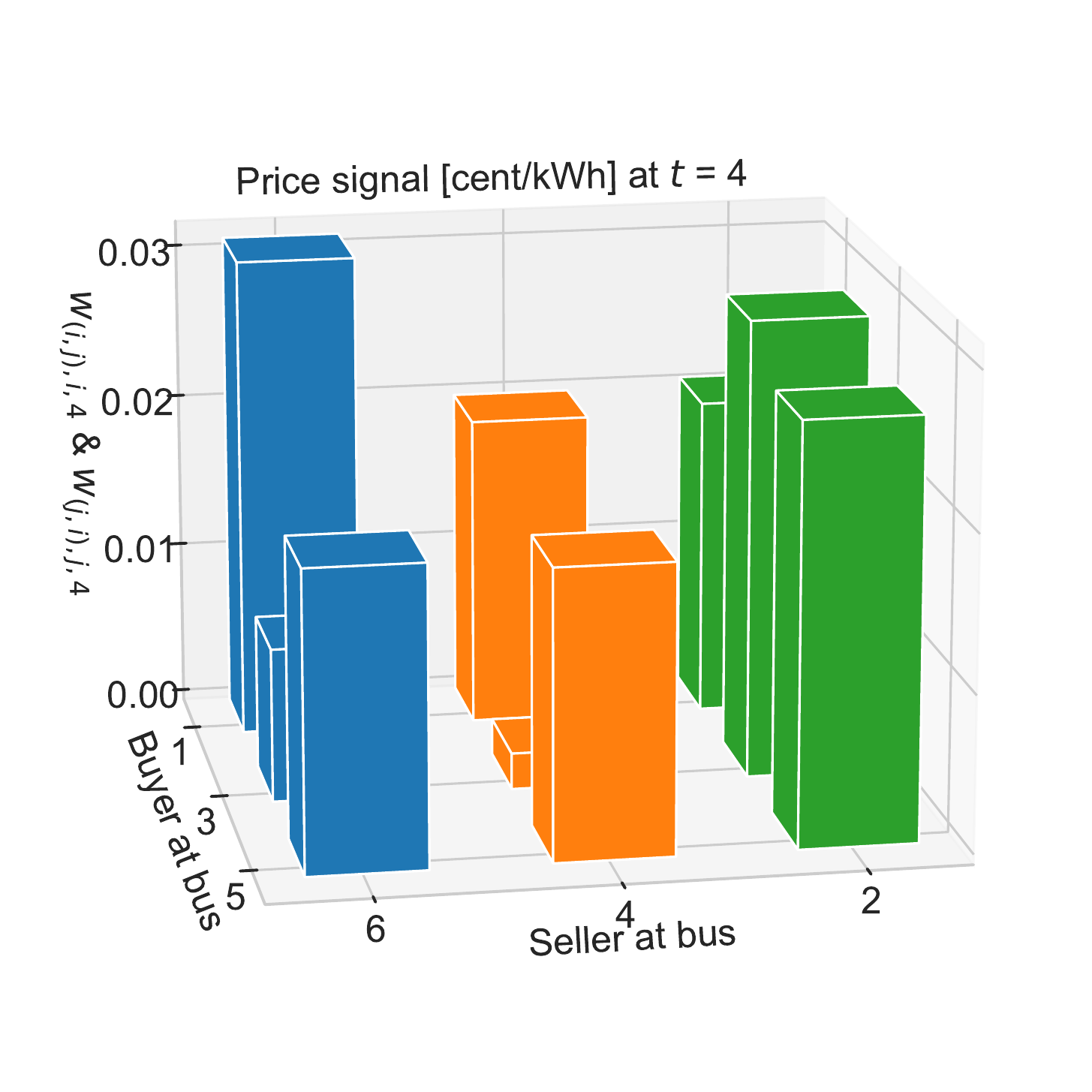}\label{Fig. 5-4-4}}
\caption{Optimal price signal at different time for \textit{Syn-DYNA}.}
\label{Fig. 5-4}
\end{figure*}

\begin{figure*}[!h]
\centering
\subfloat[$t=1$]{\includegraphics[width=1.80in,height=1.50in]{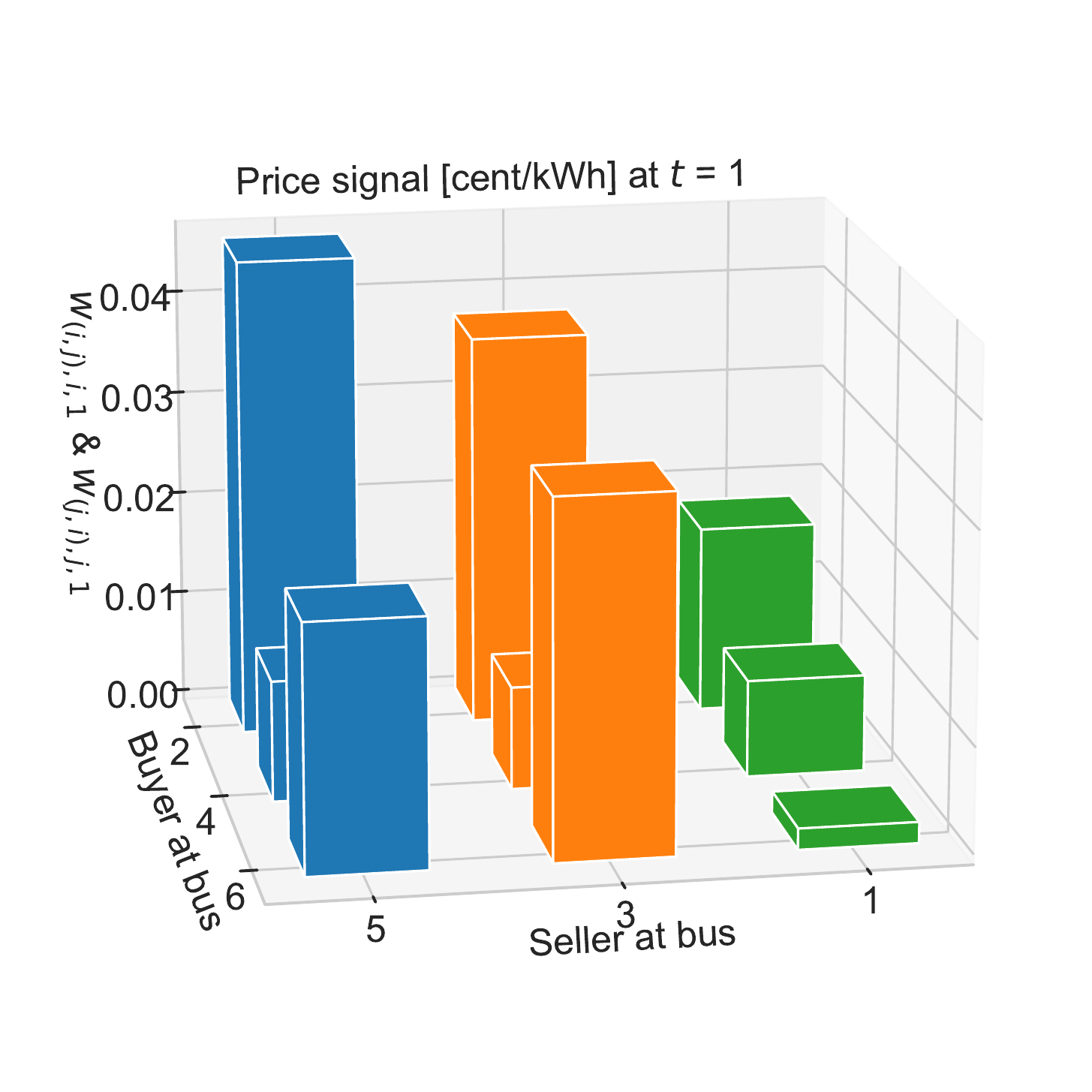}\label{Fig. 5-5-1}}
\subfloat[$t=2$]{\includegraphics[width=1.80in,height=1.50in]{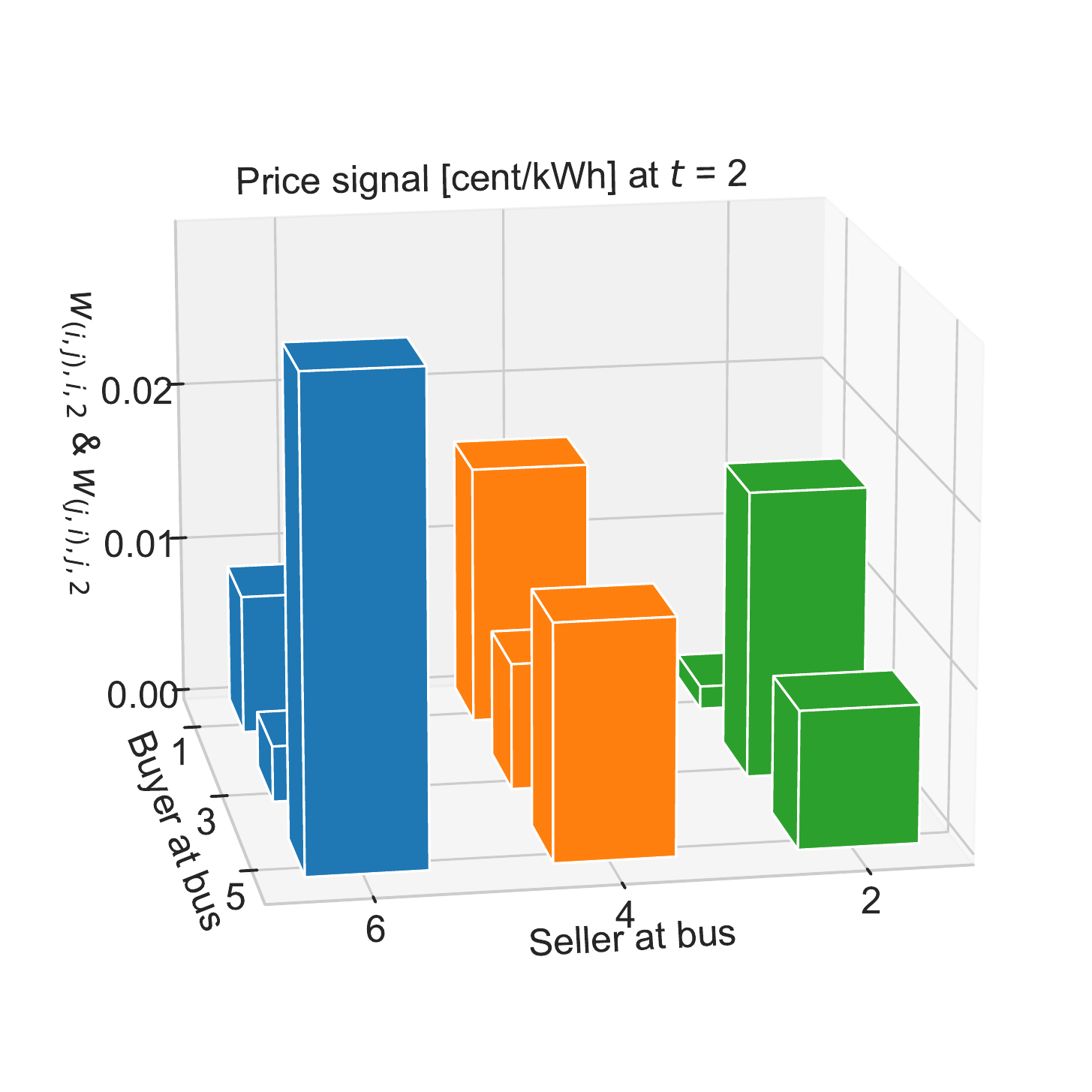}\label{Fig. 5-5-2}}
\subfloat[$t=3$]{\includegraphics[width=1.80in,height=1.50in]{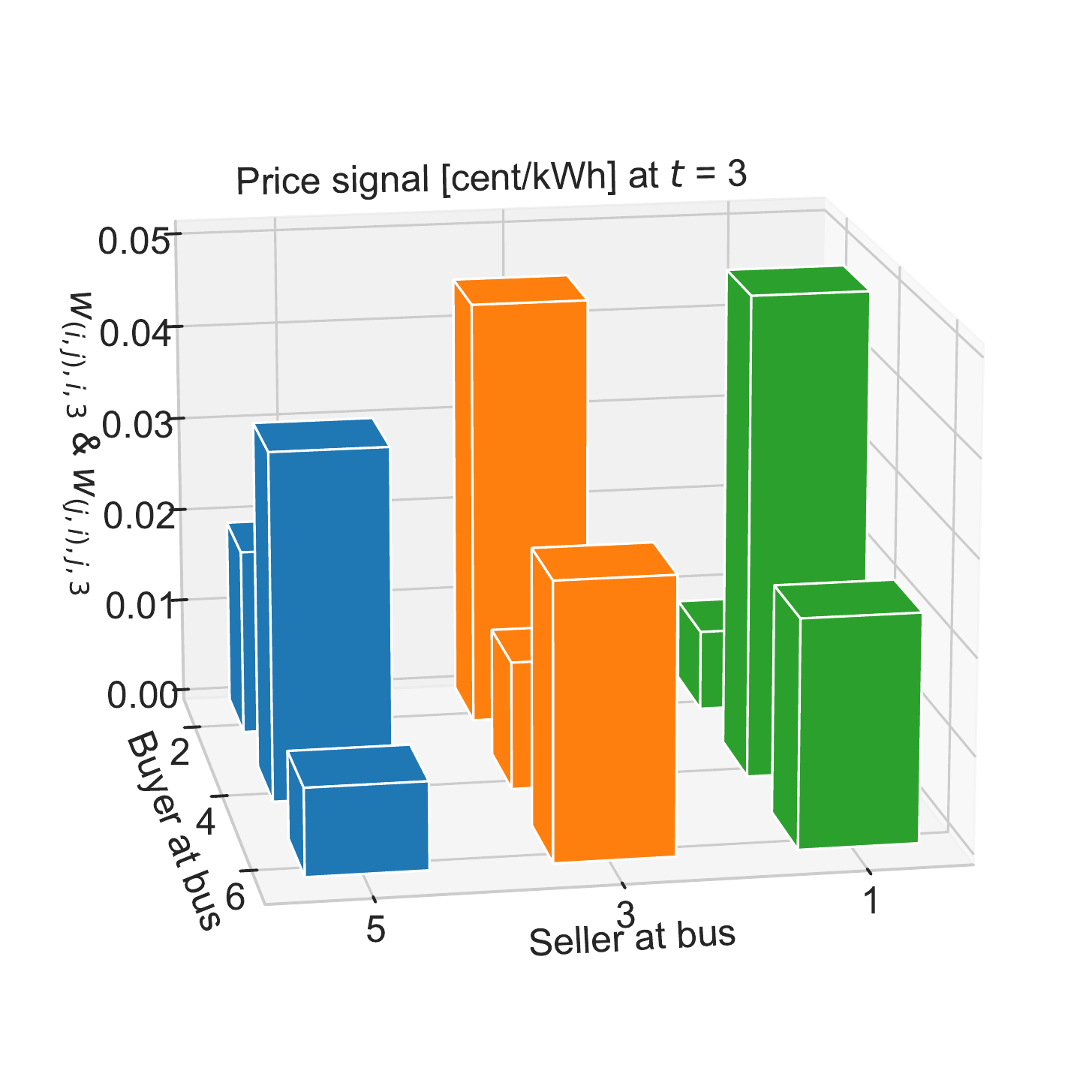}\label{Fig. 5-5-3}}
\subfloat[$t=4$]{\includegraphics[width=1.80in,height=1.50in]{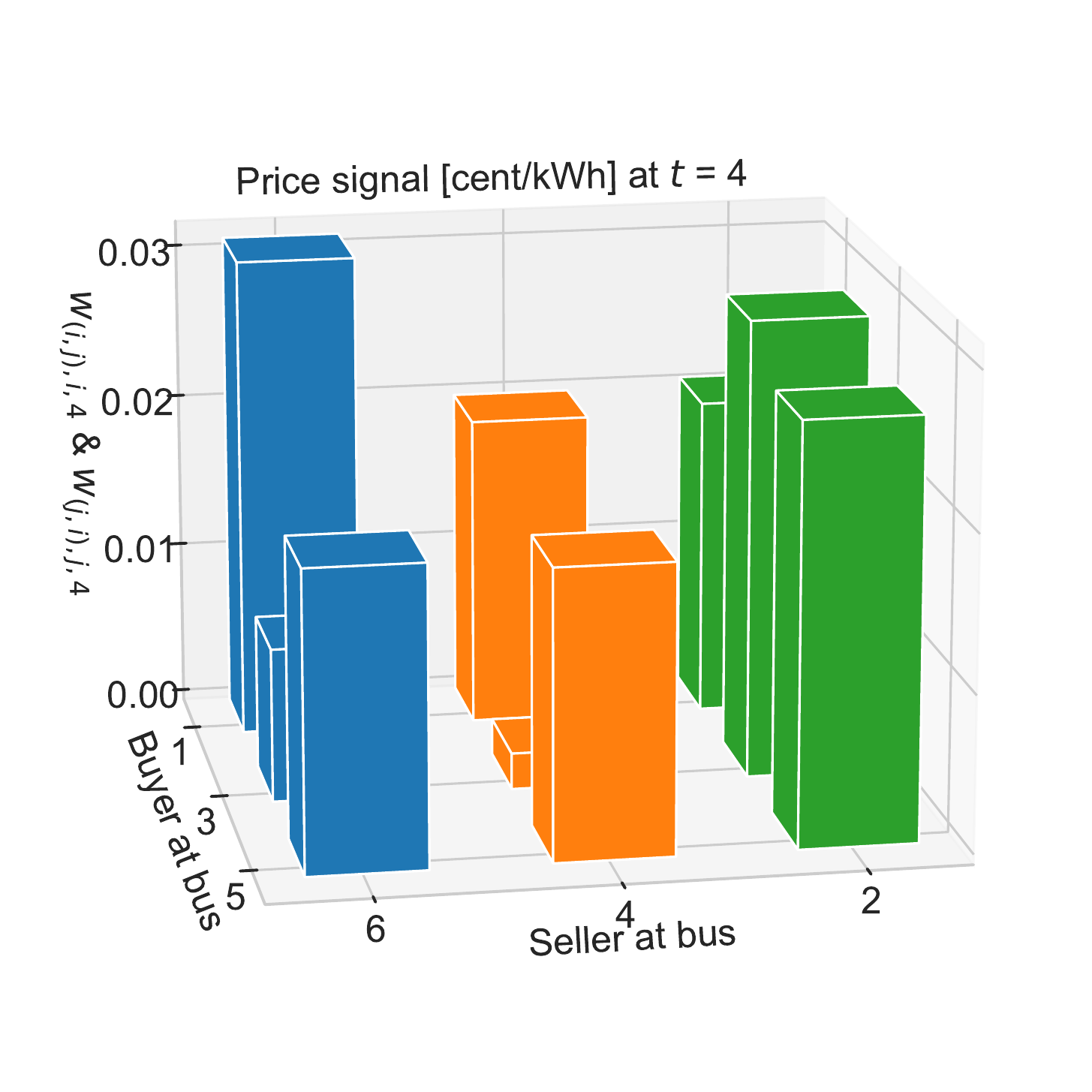}\label{Fig. 5-5-4}}
\caption{Optimal price signal at different time for \textit{Asyn-DYNA} with $d=0$.}
\label{Fig. 5-5}
\end{figure*}

\begin{figure*}[!h]
\centering
\subfloat[$t=1$]{\includegraphics[width=1.80in,height=1.50in]{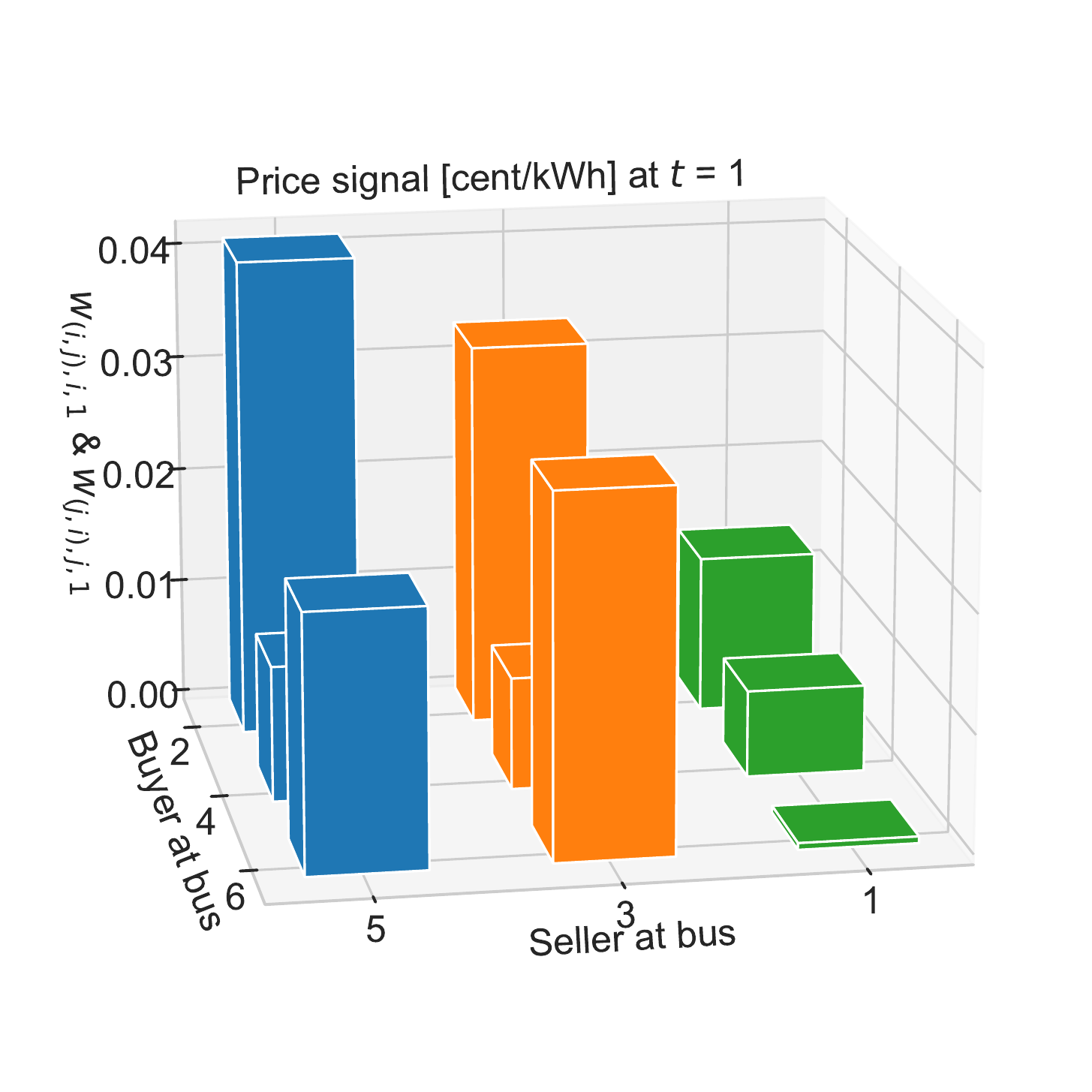}\label{Fig. 5-6-1}}
\subfloat[$t=2$]{\includegraphics[width=1.80in,height=1.50in]{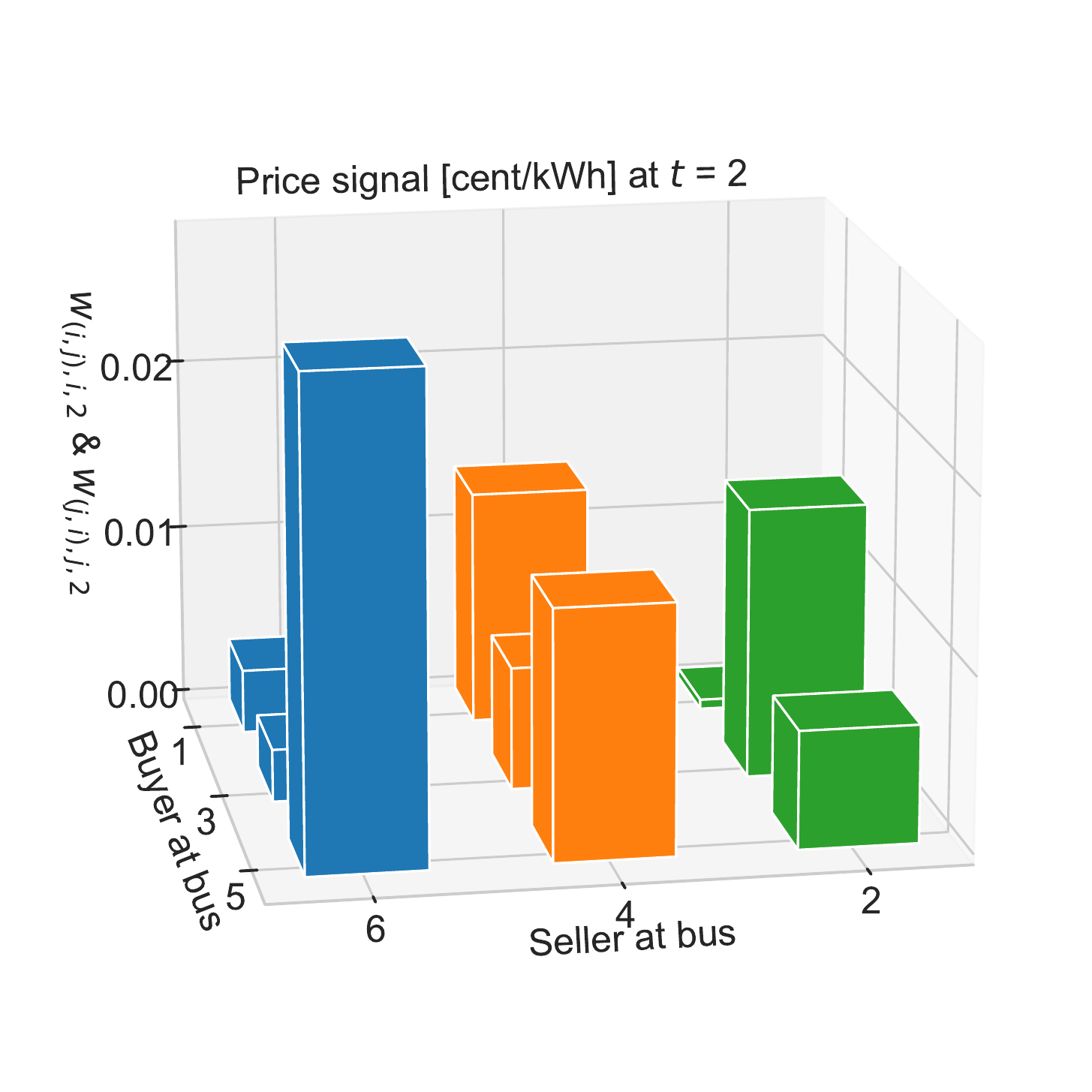}\label{Fig. 5-6-2}}
\subfloat[$t=3$]{\includegraphics[width=1.80in,height=1.50in]{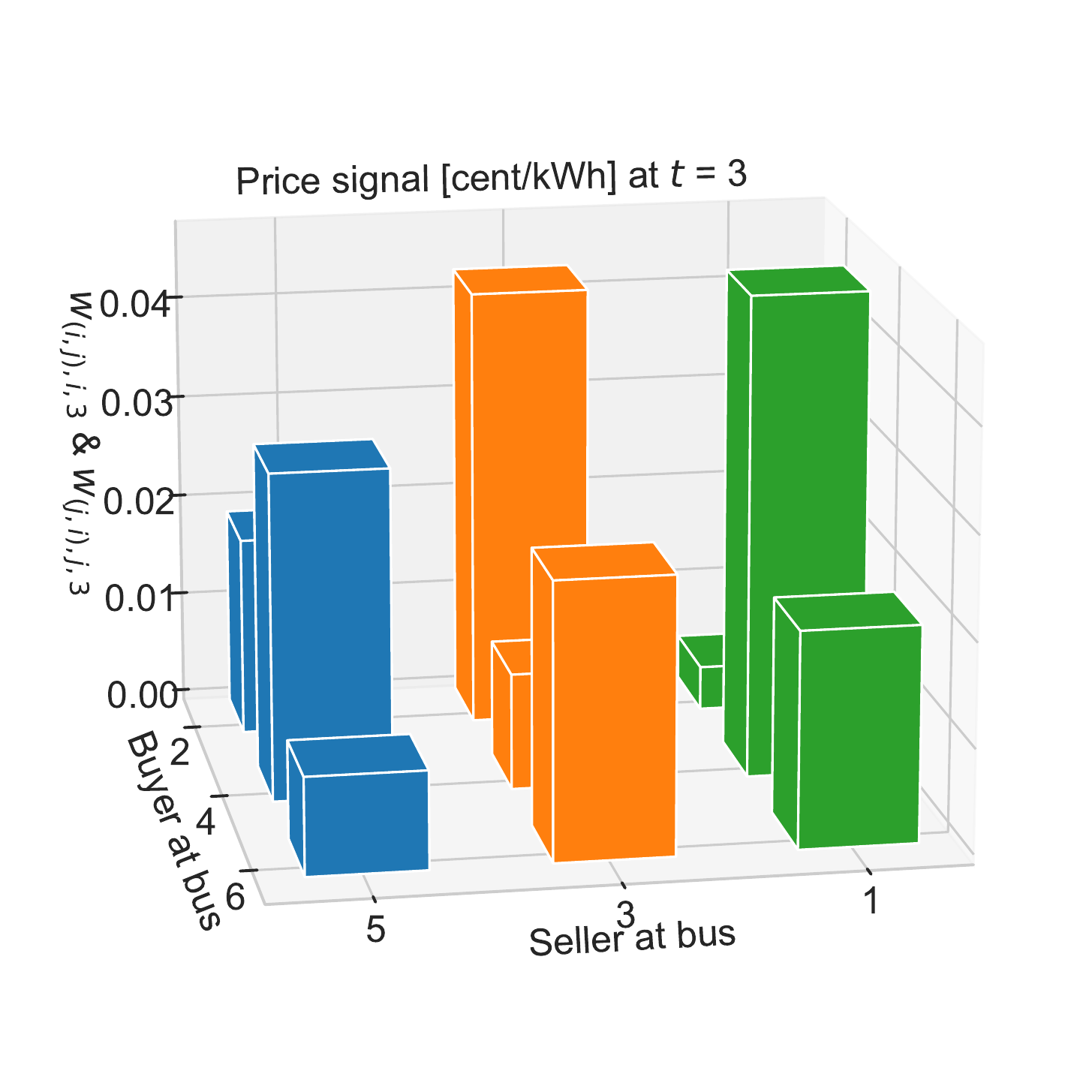}\label{Fig. 5-6-3}}
\subfloat[$t=4$]{\includegraphics[width=1.80in,height=1.50in]{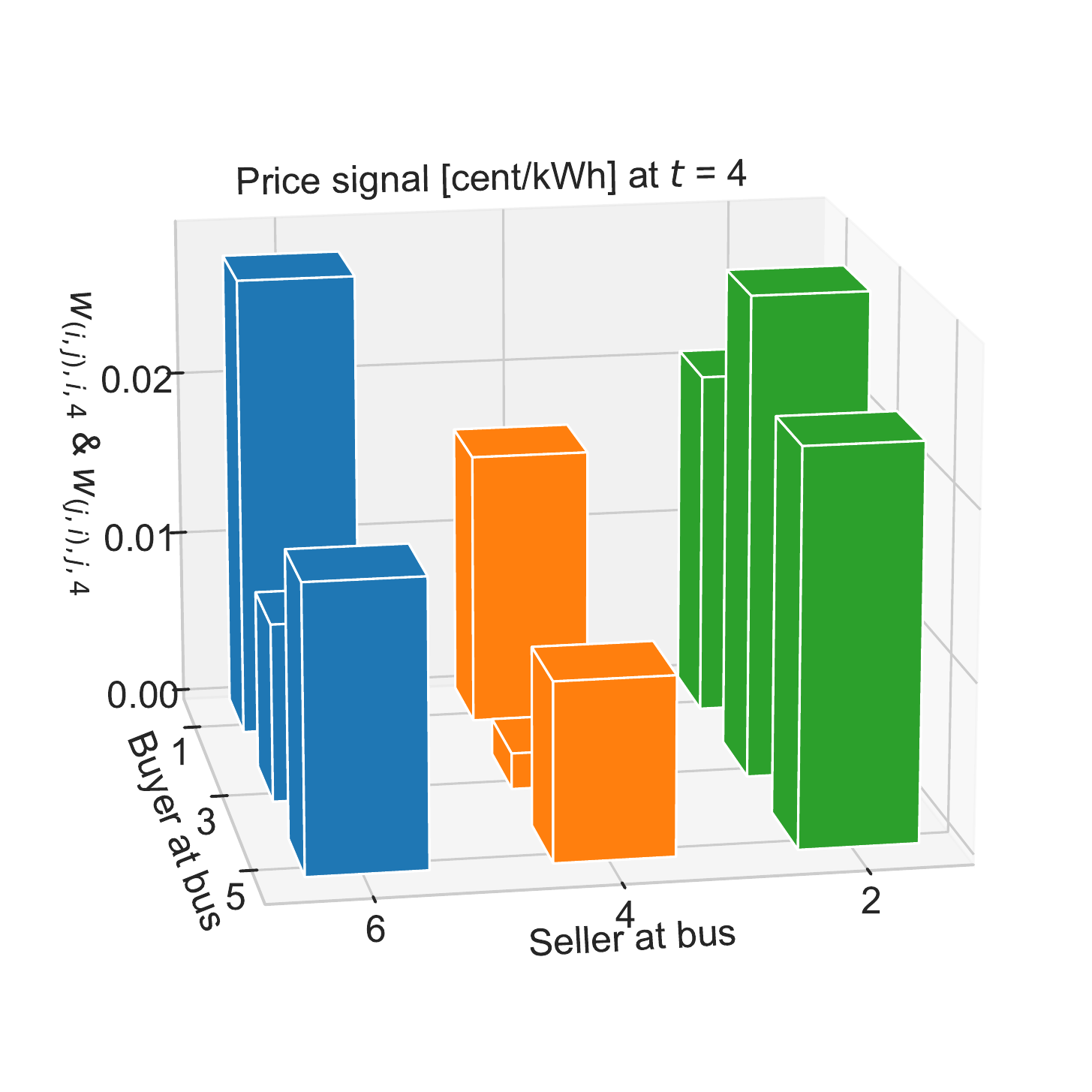}\label{Fig. 5-6-4}}
\caption{Optimal price signal at different time for \textit{Asyn-DYNA} with $d=10$.}
\label{Fig. 5-6}
\end{figure*}

\begin{figure}[!h]
\centering
\subfloat[Convergence comparison over iterations.]{\includegraphics[width=1.15in,height=0.95in]{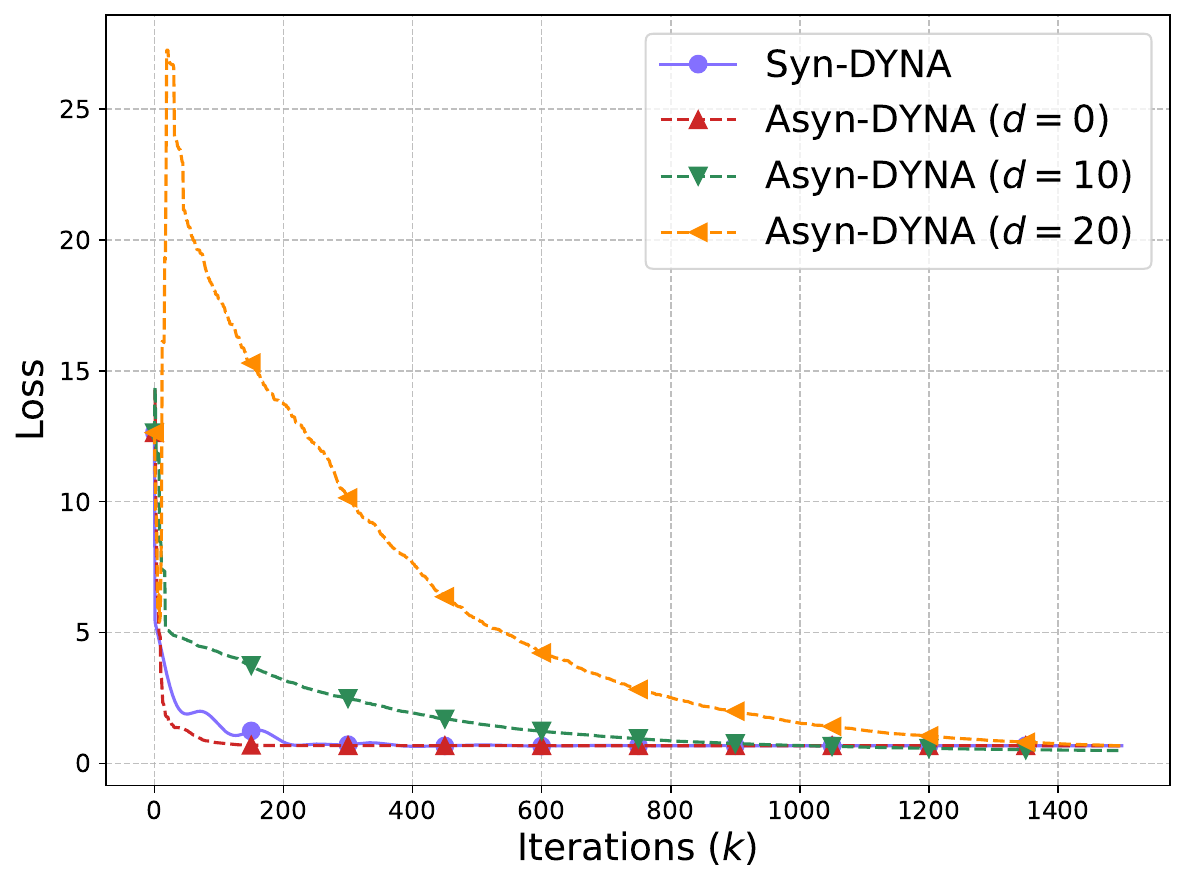}\label{Fig. 5-7-1}} \hfill
\subfloat[Convergence comparison over communication.]{\includegraphics[width=1.15in,height=0.95in]{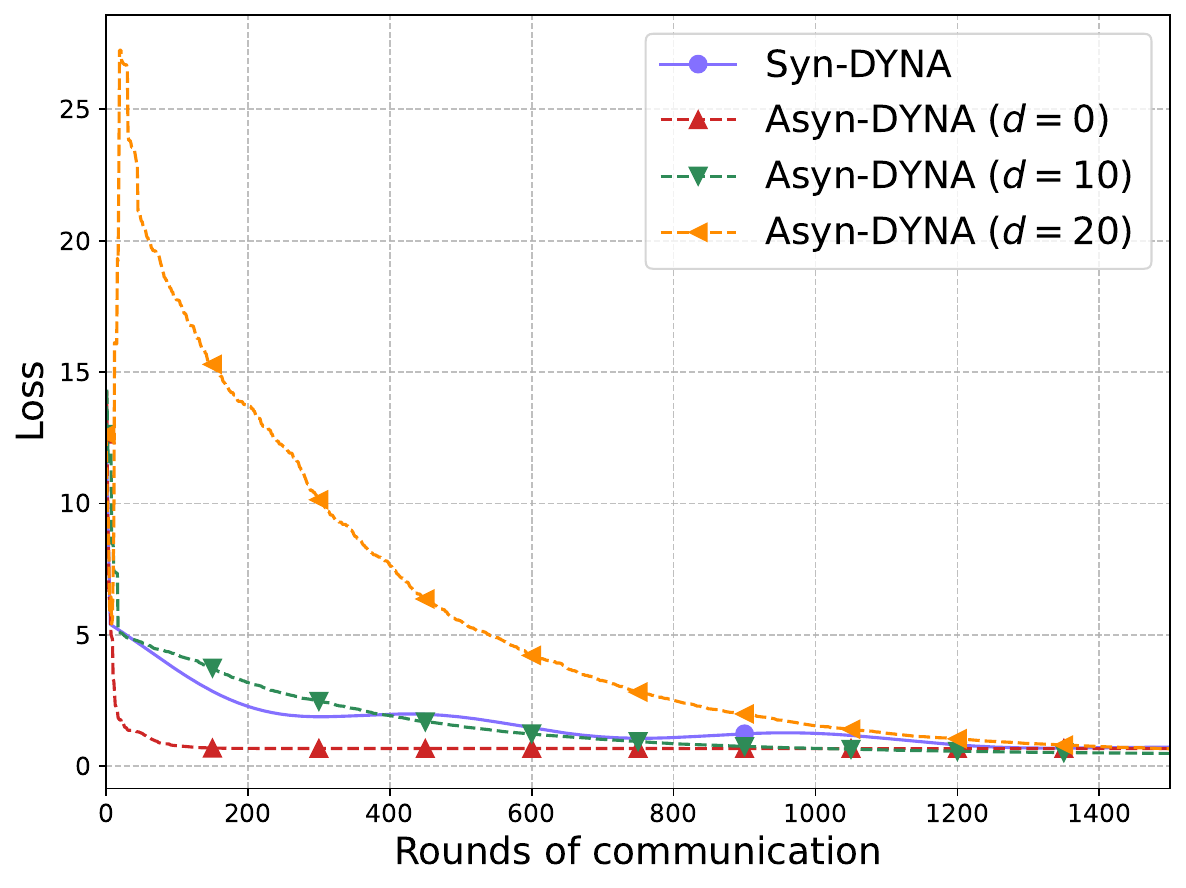}\label{Fig. 5-7-2}} \hfill
\subfloat[Convergence comparison over time.]{\includegraphics[width=1.15in,height=0.95in]{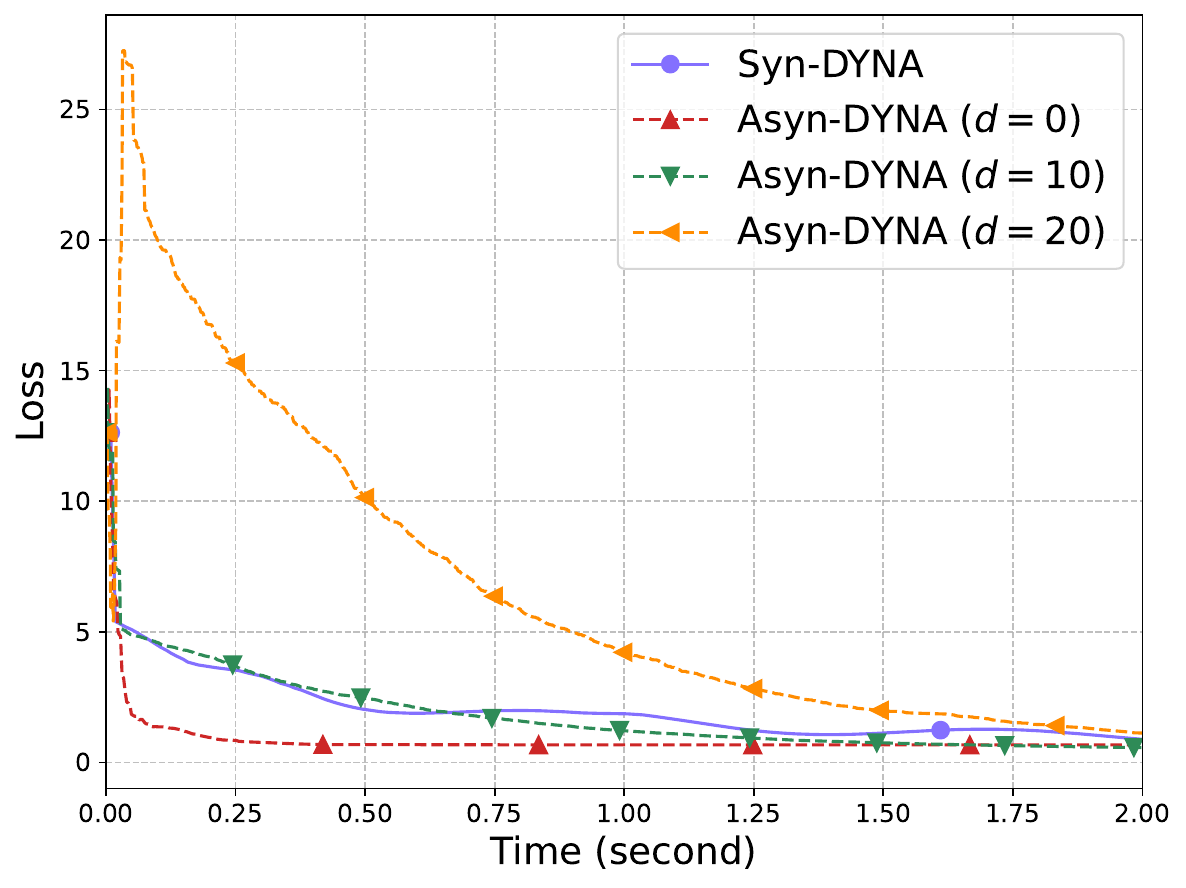}\label{Fig. 5-75-3}} \hfill
\caption{Convergence performance across different metrics.}
\label{Fig. 5-7}
\end{figure}

%\begin{figure*}[!h]
%\centering
%\subfloat[Convergence comparison over iterations.]{\includegraphics[width=2.3in,height=1.8in]{Figs/convergence_error_comparison_all.pdf}\label{Fig. 5-7-1}} \hfill
%\subfloat[Convergence comparison over communication.]{\includegraphics[width=2.3in,height=1.8in]{Figs/convergence_error_vs_commucation.pdf}\label{Fig. 5-7-2}} \hfill
%\subfloat[Convergence comparison over time.]{\includegraphics[width=2.3in,height=1.8in]{Figs/convergence_error_comparison_time.pdf}\label{Fig. 5-75-3}} \hfill
%\caption{Convergence performance across different metrics.}
%\label{Fig. 5-7}
%\end{figure*}
%异步和同步激活节点的细节  +  异步考虑了时间延迟为d=10的时候
To quantify the performance of the proposed algorithms under both synchronous and asynchronous schemes, we evaluate the evolution of traded power and optimal price signals during different periods, where the result related to the evolution of traded power has been deferred to Section \ref{S-1} in the supplementary material for length consideration.
%As illustrated in Fig. \ref{Fig. 5-1}: (a)-(d), the results illustrate the evolution of traded power for the \textit{Syn-DYNA} during different periods. Initially, the synchronous algorithm exhibits the traded power values fluctuate significantly as the agents update their variables to find optimal trading strategies. However, the traded power on all links converges to stable equilibrium values as iterations proceed. By contrast, Fig. \ref{Fig. 5-2} and Fig. \ref{Fig. 5-3} illustrate the evolution of traded power for \textit{Asyn-DYNA} in the cases of $d=0$ and $d=10$ respectively. When $d=0$, i.e., in the absence of time-varying communication delays, Fig. \ref{Fig. 5-2} shows that \textit{Asyn-DYNA} achieves the equilibrium state in less than 300 iterations, which is much faster than \textit{Syn-DYNA} as revealed by Fig. \ref{Fig. 5-1}. However, due to the randomness of asynchronous updates, Fig. \ref{Fig. 5-3} demonstrates that \textit{Asyn-DYNA} requires more iterations to converge when time-varying communication delays exist. Despite the performance gap, both algorithms successfully converge to a stable state. This steady state indicates that the prosumers ultimately reach a consensus on the optimal electricity transaction amounts, thereby balancing supply and demand within the P2P transactive networks.
Figs. \ref{Fig. 5-4}-\ref{Fig. 5-6} demonstrate the corresponding optimal price signals associated with the traded power in the stable state of \textit{Syn-DYNA}, \textit{Asyn-DYNA} ($d=0$), and \textit{Asyn-DYNA} ($d=10$). The results imply that the pricing strategies in the equilibrium point are consistent among the tested algorithms, despite their convergence speed gaps during iterations. To further investigate the convergence properties of \textit{Syn-DYNA} and \textit{Asyn-DYNA} in a more comprehensive way, we compare their convergence performance with respect to iterations, rounds of communication, and time (on the wall). Wherein, three different time-varying communication delays with upper bounds $d = 0,10,20$ are considered. From Fig. \ref{Fig. 5-7}: (a)-(c), it can be found that the convergence speed of \textit{Asyn-DYNA} is slowed down by larger time-varying communication delays with larger upper bounds in both cases. However, it is also shown that \textit{Asyn-DYNA} converges faster than \textit{Syn-DYNA} across all metrics in the absence of time-varying communication delays, i.e., $d=0$. Especially, Fig. \ref{Fig. 5-7}: (b)-(c) demonstrate that the gain in convergence speed of \textit{Asyn-DYNA} becomes more pronounced in terms of communication rounds and wall-clock time compared to iterations.

\section{Conclusion}\label{Section 6}

This paper studied a class of dynamic P2P energy management problems over asynchronous transactive networks. To handle the complex local constraints, we recasted the dynamic P2P energy management problem into a saddle-point problem formulation. Different to the \textit{ADMM}-based technical line, we adopt the operator-splitting theory to propose a synchronous decentralized dynamic energy management algorithm, namely \textit{Syn-DYNA}. To address the communication bottleneck caused by the synchronization clock, an asynchronous extension of \textit{Syn-DYNA} was developed by incorporating the random activation scheme and latest state tracking. Theoretical analysis established the linear convergence for \textit{Syn-DYNA} and asymptotic convergence for \textit{Asyn-DYNA}, respectively. Numerical experiments verify the effectiveness of the proposed algorithms and demonstrate the superior performance of \textit{Asyn-DYNA} in certain cases. Future work will focus on investigating the attack-robust version of \textit{Asyn-DYNA}.

\bibliographystyle{IEEEtran}
\bibliography{Asyn-DOEM}

\clearpage
\setcounter{page}{1}
\onecolumn

\section{Supplementary Material}\label{Section 7}
\subsection{Proof of Lemma \ref{L4-2}}\label{P-L4-2}
Recalling the update rule (\ref{E4-4}), we have
\begin{equation}\label{E7-1}
\begin{aligned}
 \left\| {{U^{k + 1}} - {U^ * }} \right\|_{{\mathbb{T}_s}}^2 - \left\| {{U^k} - {U^ * }} \right\|_{{\mathbb{T}_s}}^2 = & - \left\| {{U^{k + 1}} - {U^k}} \right\|_{{\mathbb{T}_s}}^2 + 2{\left( {{U^{k + 1}} - {U^ * }} \right)^ \top }{\mathbb{T}_s}\left( {{U^{k + 1}} - {U^k}} \right)\\
= & \left\| {{U^{k + 1}} - {U^k}} \right\|_{{\mathbb{T}_s}}^2 + 2{\left( {{U^k} - {U^ * }} \right)^ \top }{\mathbb{T}_s}\left( {{U^{k + 1}} - {U^k}} \right),
\end{aligned}
\end{equation}
where the first equality follows the weighted cosine law and the second equality applies the fact that ${U^k} - {U^*} = {U^k} - {U^{k + 1}} + {U^{k + 1}} - {U^*}$. In view of (\ref{E4-3-2}), since ${\mathbb{T}_M}$ is a trivially monotone operator, we have ${\mathbb{T}_s}\left( {{U^{k + 1}} - {U^k}} \right) = \left( {{\mathbb{T}_H} - {\mathbb{T}_M}} \right)\left( {{{\bar U}^k} - {U^k}} \right)$ such that
\begin{equation}\label{E7-3}
\begin{aligned}
\left\| {{U^{k + 1}} - {U^ * }} \right\|_{{\mathbb{T}_s}}^2 - \left\| {{U^k} - {U^ * }} \right\|_{{\mathbb{T}_s}}^2 = & \left\| {{U^{k + 1}} - {U^k}} \right\|_{{\mathbb{T}_s}}^2 + 2{\left( {{U^k} - {{\bar U}^k}} \right)^ \top }\left( {{\mathbb{T}_H} - {\mathbb{T}_M}} \right)\left( {{{\bar U}^k} - {U^k}} \right) \\
& + 2{\left( {{{\bar U}^k} - {U^ * }} \right)^ \top }\left( {{\mathbb{T}_H} - {\mathbb{T}_M}} \right)\left( {{{\bar U}^k} - {U^k}} \right)\\
= & \left\| {{U^{k + 1}} - {U^k}} \right\|_{{\mathbb{T}_s}}^2 + 2{\left( {{{\bar U}^k} - {U^ * }} \right)^ \top }\left( {{\mathbb{T}_H} - {\mathbb{T}_M}} \right)\left( {{{\bar U}^k} - {U^k}} \right) - 2\left\| {{{\bar U}^k} - {U^k}} \right\|^2_{{\mathbb{T}_p}},
\end{aligned}
\end{equation}
where ${\mathbb{T}_p}: = \left[ {\begin{array}{*{20}{c}}
  {\Lambda _\beta ^{ - 1}}&{\frac{1}{2}{\Psi ^ \top }} \\
  {\frac{1}{2}{\Psi ^ \top }}&{\Lambda _\alpha ^{ - 1}}
\end{array}} \right]$. We next handle the second term in the right-hand-side (RHS) of (\ref{E7-3}). In view of (\ref{E4-3-1}), we have
\begin{equation}\label{E7-4}
\begin{aligned}
{\left( {{{\bar U}^k} - {U^ * }} \right)^ \top }\left( {{\mathbb{T}_H} - {\mathbb{T}_M}} \right)\left( {{{\bar U}^k} - {U^k}} \right) = - {\left( {{{\bar U}^k} - {U^ * }} \right)^ \top }\left( {{\mathbb{T}_C}{U^k} + {\mathbb{T}_M}{{\bar U}^k} + {\mathbb{T}_A}{{\bar U}^k}} \right).
\end{aligned}
\end{equation}
Plugging (\ref{E7-4}) back into (\ref{E7-3}) obtains
\begin{equation}\label{E7-5}
\begin{aligned}
\left\| {{U^{k + 1}} - {U^ * }} \right\|_{{\mathbb{T}_s}}^2 - \left\| {{U^k} - {U^ * }} \right\|_{{\mathbb{T}_s}}^2 = & \left\| {{U^{k + 1}} - {U^k}} \right\|_{{\mathbb{T}_s}}^2 \!-\! 2{\left( {{{\bar U}^k} - {U^ * }} \right)^ \top }\left( {{\mathbb{T}_C}{U^k} \!+\! {\mathbb{T}_M}{{\bar U}^k} \!+\! {\mathbb{T}_A}{U^ * }} \right) - 2\left\| {{{\bar U}^k} - {U^k}} \right\|_{{\mathbb{T}_p}}^ \top  \\
& - 2{\left( {{{\bar U}^k} - {U^ * }} \right)^ \top }\left( {{\mathbb{T}_A}{{\bar U}^k} - {\mathbb{T}_A}{U^ * }} \right).
\end{aligned}
\end{equation}
Since ${\mathbb{T}_A}$ is a monotone operator, it is follows from  \cite{Combettes} that
\begin{equation}\label{E7-6}
{\left( {{{\bar U}^k} - {U^ * }} \right)^ \top }\left( {{\mathbb{T}_A}{{\bar U}^k} - {\mathbb{T}_A}{U^ * }} \right) \ge 0.
\end{equation}
According to Lemma \ref{L4-1}, we know ${\mathbb{T}_A}{U^ * } =  - {\mathbb{T}_M}{U^ * } - {\mathbb{T}_C}{U^ * }$ such that
\begin{equation}\label{E7-7}
\begin{aligned}
{\left( {{{\bar U}^k} - {U^ * }} \right)^ \top }\left( {{\mathbb{T}_C}{U^k} + {\mathbb{T}_M}{{\bar U}^k} + {\mathbb{T}_A}{U^ * }} \right) = {\left( {{{\bar U}^k} - {U^ * }} \right)^ \top }{\mathbb{T}_C}\left( {{U^k} - {U^ * }} \right) + {\left( {{{\bar U}^k} - {U^ * }} \right)^ \top }{\mathbb{T}_M}\left( {{{\bar U}^k} - {U^ * }} \right).
\end{aligned}
\end{equation}
To proceed, we tackle the first term in the RHS of (\ref{E7-7}) as follows:
\begin{equation}\label{E7-8}
\begin{aligned}
{\left( {{{\bar U}^k} - {U^ * }} \right)^ \top }{\mathbb{T}_C}\left( {{U^k} - {U^ * }} \right) = & {\left( {{{\bar x}^k} - {x^ * }} \right)^ \top }\left( {\nabla f\left( {{x^k}} \right) - \nabla f\left( {{x^ * }} \right)} \right) \ge   - \frac{1}{{{L_f}}}{\left\| {\nabla f\left( {{x^k}} \right) - \nabla f\left( {{x^ * }} \right)} \right\|^2} - \frac{{{L_f}}}{4}{\left\| {{x^k} - {{\bar x}^k}} \right\|^2}\\
 & + {\left( {{x^k} - {x^ * }} \right)^ \top }\left( {\nabla f\left( {{x^k}} \right) - \nabla f\left( {{x^ * }} \right)} \right) \ge  - \frac{{{L_f}}}{4}\left\| {{x^k} - {{\bar x}^k}} \right\|^2,
\end{aligned}
\end{equation}
where the second inequality is owing to the basic inequality and the last inequality applies the $L_f$-smoothness of ${\nabla f}$. Summarizing (\ref{E7-5})-(\ref{E7-8}) gives
\begin{equation}\label{E7-9}
\begin{aligned}
\left\| {{U^{k + 1}} - {U^ * }} \right\|_{{\mathbb{T}_s}}^2 - \left\| {{U^k} - {U^ * }} \right\|_{{\mathbb{T}_s}}^2 \le  \left\| {{U^{k + 1}} - {U^k}} \right\|_{{\mathbb{T}_s}}^2 - 2\left( {\left\| {{{\bar U}^k} - {U^k}} \right\|_{{\mathbb{T}_p}}^2 - \frac{{{L_f}}}{4}\left\| {{x^k} - {{\bar x}^k}} \right\|^2} \right).
\end{aligned}
\end{equation}
By revisiting the mathematical relationship established in equation (\ref{E4-3-2}), we can clearly deduce the fact that ${\left( {\mathbb{T}_s^{ - 1}\left( {{\mathbb{T}_H} - {\mathbb{T}_M}} \right)} \right)^{ - 1}}\left( {{U^{k + 1}} - {U^k}} \right) = {\bar U^k} - {U^k}$ with ${\left( {\mathbb{T}_s^{ - 1}\left( {{\mathbb{T}_H} - {\mathbb{T}_M}} \right)} \right)^{ - 1}} = \left[ {\begin{array}{*{20}{c}}
  {\mathbf{I}}&{- \Psi {\Lambda _\beta }} \\
  0&{\mathbf{I}}
\end{array}} \right]$, which implies
\begin{equation}\label{E7-10}
{\left( {{{\bar U}^k} \!-\! {U^k}} \right)^ \top }{\mathbb{T}_p}\left( {{{\bar U}^k} \!-\! {U^k}} \right) = {\left( {{U^{k + 1}} \!-\! {U^k}} \right)^ \top }{\mathbb{T}_{\tilde p}}\left( {{U^{k + 1}} \!-\! {U^k}} \right),
\end{equation}
where ${\mathbb{T}_{\tilde p}} := \left[ {\begin{array}{*{20}{c}}
  {\Lambda _\beta ^{ - 1}}&{ - \frac{1}{2}\Psi } \\
  { - \frac{1}{2}{\Psi ^ \top }}&{\Lambda _\alpha ^{ - 1}}
\end{array}} \right]$ is a symmetric matrix. Define ${\mathbb{T}_q} := \left[ {\begin{array}{*{20}{c}}
  {\mathbf{0}}&{\mathbf{0}} \\
  {\mathbf{0}}&{\frac{{{L_f}}}{4}{\mathbf{I}}}
\end{array}} \right]$ and let ${\mathbb{T}_{d}} = {\mathbb{T}_{\tilde p}} - {\mathbb{T}_q}$ such that the inequality (\ref{E7-9}) reduces to
\begin{equation}\label{E7-11}
\left\| {{U^{k + 1}} - {U^ * }} \right\|_{{\mathbb{T}_s}}^2 - \left\| {{U^k} - {U^ * }} \right\|_{{\mathbb{T}_s}}^2 \le - \left\| {{U^{k + 1}} - {U^k}} \right\|_{2{\mathbb{T}_{d}} - {\mathbb{T}_s}}^2.
\end{equation}
The proof is completed via choosing appropriate uncoordinated step-sizes to ensure that the matrix $2{\mathbb{T}_{d}} - {\mathbb{T}_s}$ is positive definite.

\subsection{Proof of Theorem \ref{T4-1}}\label{P-T4-1}
Based on the fundamental mathematical results that were explicitly established in Lemma \ref{L4-2}, we can now conclude that
\begin{equation}\label{E8-1}
\left\| {\mathbb{T}{U^k} - {U^ * }} \right\|_{{\mathbb{T}_s}}^2 \le \left\| {{U^k} - {U^ * }} \right\|_{{\mathbb{T}_s}}^2 - \left\| {\mathbb{T}{U^k} - {U^k}} \right\|_{2{\mathbb{T}_{d}} - {\mathbb{T}_s}}^2.
\end{equation}
We proceed to handle the first term in the RHS of (\ref{E8-1}). Owing to the fact that  $f$, $g$, and ${\mathcal{I}_{{\mathcal{S}^{\text{r}}}}}$ are Piecewise linear-quadratic functions, $ {\mathbb{T}_A}U + {\mathbb{T}_M}U + {\mathbb{T}_C}$ is globally metrically subregular such that there exists a constant $\eta > 0 $,
\begin{equation}\label{E8-2}
\begin{aligned}
 \left\| {{U^k} - {U^*}} \right\|_{{\mathbb{T}_s}}^2 \le & \left\| {{\mathbb{T}_s}} \right\|{\left( {{{\left\| {{{\bar U}^k} - {U^*}} \right\|}} + \left\| {{U^k} - \bar U^k} \right\|} \right)^2} \\ \le & \left\| {{\mathbb{T}_s}} \right\|{\left( {\eta \left( {\left\| {{\mathbb{T}_H} - {\mathbb{T}_M}} \right\| + {L_f}} \right)\left\| {{U^k} - {{\bar U}^k}} \right\| + \left\| {{U^k} - {{\bar U}^k}} \right\|} \right)^2}\\
\le & \left\| {{\mathbb{T}_s}} \right\|{\left( {1 + \eta \left( {\left\| {{\mathbb{T}_H} - {\mathbb{T}_M}} \right\| + {L_f}} \right)} \right)^2}{\left\| {{U^k} - {{\bar U}^k}} \right\|^2}.
\end{aligned}
\end{equation}
Since ${\mathbb{T}_s}{\left( {{\mathbb{T}_H} - {\mathbb{T}_M}} \right)^{ - 1}}\left( {\mathbb{T}{U^k} - {U^k}} \right) = {{{\bar U}^k} - {U^k}} $ holds true according to (\ref{E4-3-2}) and (\ref{E4-4}), we have
\begin{equation}\label{E8-3}
\begin{aligned}
 {\left\| {{U^k} - {{\bar U}^k}} \right\|^2} = & {\left\| {{\mathbb{T}_s}{{\left( {{\mathbb{T}_H} - {\mathbb{T}_M}} \right)}^{ - 1}}\left( {\mathbb{T}{U^k} - {U^k}} \right)} \right\|^2} \\
\le & {\left\| {{\mathbb{T}_s}{{\left( {{\mathbb{T}_H} - {\mathbb{T}_M}} \right)}^{ - 1}}} \right\|^2}\left\| {{{\left( {2{\mathbb{T}_{d}} - {\mathbb{T}_s}} \right)}^{ - 1}}} \right\|\left\| {\mathbb{T}{U^k} - {U^k}} \right\|_{2{\mathbb{T}_{d}} - {\mathbb{T}_s}}^2.
\end{aligned}
\end{equation}
Plugging (\ref{E8-3}) back into (\ref{E8-2}) reduces to
\begin{equation}\label{E8-4}
\left\| {{U^k} - {U^*}} \right\|_{{\mathbb{T}_S}}^2 \le \gamma \left\| {\mathbb{T}{U^k} - {U^k}} \right\|_{2{\mathbb{T}_{d}} - {\mathbb{T}_s}}^2,
\end{equation}
where $\gamma = \left\| {{\mathbb{T}_s}} \right\|\left\| {{{\left( {2{\mathbb{T}_{d}} - {\mathbb{T}_s}} \right)}^{ - 1}}} \right\|{\left\| {{\mathbb{T}_s}{{\left( {{\mathbb{T}_H} - {\mathbb{T}_M}} \right)}^{ - 1}}} \right\|^2}$ ${\left( {1 + \eta \left( {\left\| {{\mathbb{T}_H} - {\mathbb{T}_M}} \right\| + {L_f}} \right)} \right)^2}$. It can be verified that $\gamma > 1$ when $2{\mathbb{T}_{\tilde p}} - {\mathbb{T}_s}$ is positive definite, i.e., $0 < {\alpha _i} < 1/\left( {\left( {{L_f}/2} \right) + \left\| {\sum\nolimits_{j \in {\mathcal{N}_i}} {{\beta _{\left( {i,j} \right)}}\Omega _{ij}^ \top {\Omega _{ij}}} } \right\|} \right)$, $\forall i \in \mathcal{V}$. The proof is finished through substituting (\ref{E8-4}) back into (\ref{E8-1}).

\subsection{Proof of Lemma \ref{L4-2-2}}\label{P-L4-2-2}
For two arbitrary vectors $U$ and $Z$ satisfying (\ref{E4-3}), we have
\begin{equation}\label{E9-1}
\begin{aligned}
 \left\| {\mathbb{T}U - \mathbb{T}Z} \right\|_{{\mathbb{T}_s}}^2 - \left\| {U - Z} \right\|_{{\mathbb{T}_s}}^2 = & \left\| {\mathbb{T}U - \mathbb{T}Z} \right\|_{{\mathbb{T}_s}}^2 - 2{\left\langle {\mathbb{T}U - \mathbb{T}Z,U - Z} \right\rangle _{{\mathbb{T}_s}}} + \left\| {U - Z} \right\|_{{\mathbb{T}_s}}^2 - 2{\left\langle {U - Z,U - Z} \right\rangle _{{\mathbb{T}_s}}} \\
& + 2{\left\langle {\mathbb{T}U - \mathbb{T}Z,U - Z} \right\rangle _{{\mathbb{T}_s}}}\\
= & 2{\left\langle {U - Z,\left( {\mathbb{T}U - U} \right) - \left( {\mathbb{T}Z - Z} \right)} \right\rangle _{{\mathbb{T}_s}}} + \left\| {\left( {{\mathbf{Id}} - \mathbb{T}} \right)U - \left( {{\mathbf{Id}} - \mathbb{T}} \right)Z} \right\|_{{\mathbb{T}_s}}^2.
\end{aligned}
\end{equation}
We proceed to analyze ${\left\langle {U - Z,\left( {\mathbb{T}U - U} \right) - \left( {\mathbb{T}Z - Z} \right)} \right\rangle _{{\mathbb{T}_s}}}$ in the RHS of (\ref{E9-1}) as follows:
\begin{equation}\label{E9-2}
\begin{aligned}
 {\left\langle {U - Z,\left( {\mathbb{T}U - U} \right) - \left( {\mathbb{T}Z - Z} \right)} \right\rangle _{{\mathbb{T}_s}}}= & \left\langle {U - Z,\left( {{\mathbb{T}_H} - {\mathbb{T}_M}} \right)\left( {\left( {\bar U - U} \right) - \left( {\bar Z - Z} \right)} \right)} \right\rangle \\
= & \left\langle {\bar U - \bar Z,\left( {{\mathbb{T}_H} - {\mathbb{T}_M}} \right)\left( {\left( {\bar U - U} \right) - \left( {\bar Z - Z} \right)} \right)} \right\rangle - \left\| {\left( {\bar U - U} \right) - \left( {\bar Z - Z} \right)} \right\|_{{\mathbb{T}_p}}^2,
\end{aligned}
\end{equation}
where the first and second equalities apply (\ref{E4-3-2}) and the fact that ${U^ \top }\left( {{\mathbb{T}_H} - {\mathbb{T}_M}} \right)U = {U^ \top }{\mathbb{T}_p}U$, respectively. Therefore, according to (\ref{E4-3-1}), it follows that $\left( {{\mathbb{T}_H} - {\mathbb{T}_M}} \right)\left( {\bar U - U} \right) =  - {\mathbb{T}_M}\bar U - {\mathbb{T}_C}U - {\mathbb{T}_A}\bar U$, which implies
\begin{equation}\label{E9-3}
\begin{aligned}
\left\langle {\bar U - \bar Z,\left( {{\mathbb{T}_H} - {\mathbb{T}_M}} \right)\left( {\left( {\bar U - U} \right) - \left( {\bar Z - Z} \right)} \right)} \right\rangle = & - \left\langle {\bar U - \bar Z,{\mathbb{T}_A}\left( {\bar U - \bar Z} \right)} \right\rangle  - \left\langle {\bar U - \bar Z,{\mathbb{T}_M}\left( {\bar U - \bar Z} \right)} \right\rangle \\
& - \left\langle {\bar U - \bar Z,{\mathbb{T}_C}\left( {U - Z} \right)} \right\rangle.
\end{aligned}
\end{equation}
Since the operators ${\mathbb{T}_A}$ and ${\mathbb{T}_M}$ are maximally monotone and skew-symmetric such that $ - \left\langle {\bar U - \bar Z,{\mathbb{T}_A}\left( {\bar U - \bar Z} \right)} \right\rangle  \le  0$ and $\left\langle {\bar U - \bar Z,{\mathbb{T}_M}\left( {\bar U - \bar Z} \right)} \right\rangle  = 0$ holds true. Furthermore, since the objective function $f$ is $\mu$-strongly convex and $L_f$-smooth, it can be verified that
\begin{equation}\label{E9-4}
 - \left\langle {\bar U - \bar Z,{\mathbb{T}_C}\left( {U - Z} \right)} \right\rangle \le \left\| {\left( {\bar U - U} \right) - \left( {\bar Z - Z} \right)} \right\|_{{\mathbb{T}_{\tilde q}}}^2,
\end{equation}
where ${\mathbb{T}_{\tilde q}} = 2{\kappa _f} \cdot {\mathbb{T}_q}$. Plugging (\ref{E9-4})-(\ref{E9-5}) back into (\ref{E9-3}) and combining (\ref{E9-2}) reduces to
\begin{equation}\label{E9-5}
\begin{aligned}
 \left\| {\mathbb{T}U - \mathbb{T}Z} \right\|_{{\mathbb{T}_s}}^2 - \left\| {U - Z} \right\|_{{\mathbb{T}_s}}^2 \le & \left\| {\left( {{\mathbf{Id}} - \mathbb{T}} \right)U - \left( {{\mathbf{Id}} - \mathbb{T}} \right)Z} \right\|_{{\mathbb{T}_s}}^2 + \left\| {\left( {\bar U - U} \right) - \left( {\bar Z - Z} \right)} \right\|_{{\mathbb{T}_{\tilde q}}}^2\\
& - 2\left\| {\left( {\bar U - U} \right) - \left( {\bar Z - Z} \right)} \right\|_{{\mathbb{T}_p}}^2.
\end{aligned}
\end{equation}
Recall that ${{\mathbb{T}_{\tilde p}}} = \left[ {\begin{array}{*{20}{c}}
  {\Lambda _\beta ^{ - 1}}&{ - \frac{1}{2}\Psi } \\
  { - \frac{1}{2}{\Psi ^ \top }}&{\Lambda _\alpha ^{ - 1}}
\end{array}} \right]$ and $\mathbb{T}U{\kern 1pt}  - U = \mathbb{T}_s^{ - 1}\left( {{\mathbb{T}_H} - {\mathbb{T}_M}} \right)\left( {\bar U - U} \right)$, and we have
\begin{equation}\label{E9-6}
\left\| {\left( {\bar U - U} \right) - \left( {\bar Z - Z} \right)} \right\|_{{\mathbb{T}_p}}^2  = \left\| {\left( {\mathbb{T}U - U} \right) - \left( {\mathbb{T}Z - Z} \right)} \right\|_{{\mathbb{T}_{\tilde p}}}^2.
\end{equation}
Substituting (\ref{E9-6}) into (\ref{E9-5}) yields
\begin{equation}\label{E9-7}
\begin{aligned}
 \left\| {\mathbb{T}U - \mathbb{T}Z} \right\|_{{\mathbb{T}_s}}^2  \le &  \left\| {U - Z} \right\|_{{\mathbb{T}_s}}^2 - \left\| {\left( {\mathbb{T}U - U} \right) - \left( {\mathbb{T}Z - Z} \right)} \right\|_{2{\mathbb{T}_{\tilde p}} - {\mathbb{T}_{\tilde q}} - {\mathbb{T}_s}}^2.
 \end{aligned}
\end{equation}
The proof is completed by selecting appropriate uncoordinated step-sizes such that the matrix ${2{\mathbb{T}_{\tilde p}} - {\mathbb{T}_{\tilde q}} - {\mathbb{T}_s}}$ is positive semi-definite.

\subsection{Proof of Theorem \ref{T4-2-1}}\label{P-T4-2-1}
By revisiting the asynchronous update scheme that was previously defined in equation (\ref{E4-2-1}), we are now able to derive
\begin{equation}\label{E10-1}
\begin{aligned}
\mathbb{E}_k {\left\| {{U^{k + 1}} - {U^*}} \right\|_{{\mathbb{T}_s}}^2} = & \left\| {{U^k} - {U^*}} \right\|_{{\mathbb{T}_s}}^2 - 2{\theta _i}\mathbb{E}{\left\langle {{\mathbb{Z}_{{i_k}}}{{\hat U}^k},{U^k} - {U^*}} \right\rangle _{{\mathbb{T}_s}}} - \theta _i^2\mathbb{E}\left\| {{\mathbb{Z}_{{i_k}}}{{\hat U}^k}} \right\|_{{\mathbb{T}_s}}^2 \\
= & \left\| {{U^k} - {U^*}} \right\|_{{\mathbb{T}_s}}^2 - \frac{{2\tilde \theta }}{m}{\left\langle {\mathbb{Z}{{\hat U}^k},{U^k} - {U^*}} \right\rangle _{{\mathbb{T}_s}}} -\frac{{{{\tilde \theta }^2}}}{{{m^2}}}\sum\limits_{i = 1}^m {\frac{1}{{{P_i}}}\left\| {{\mathbb{Z}_i}{{\hat U}^k}} \right\|_{{\mathbb{T}_s}}^2}\\
\le & \left\| {{U^k} - {U^*}} \right\|_{{\mathbb{T}_s}}^2 - \frac{{2\tilde \theta }}{m}{\left\langle {\mathbb{Z}{{\hat U}^k},{U^k} - {U^*}} \right\rangle _{{\mathbb{T}_s}}} + \frac{{{\lambda _{\max }}\left( {{\mathbb{T}_s}} \right){{\tilde \theta }^2}}}{{{\lambda _{\min }}\left( {{\mathbb{T}_s}} \right){P_{\min }}{m^2}}}\left\| {\mathbb{Z}{{\hat U}^k}} \right\|_{{\mathbb{T}_s}}^2 \\
\mathop =  & \left\| {{U^k} - {U^*}} \right\|_{{\mathbb{T}_s}}^2 - \frac{{2\tilde \theta }}{m}{\left\langle {\mathbb{Z}{{\hat U}^k},{U^k} - {U^*}} \right\rangle _{{\mathbb{T}_s}}} +\frac{{{\kappa _s}}}{{{P_{\min }}{m^2}}}\left\| {{U^k} - {{\tilde U}^{k + 1}}} \right\|_{{\mathbb{T}_s}}^2,
\end{aligned}
\end{equation}
where the last equality follows (\ref{E4-2-4}). We proceed to analyze the term $\left\langle {\mathbb{Z}{{\hat U}^k},{U^k} - {U^*}} \right\rangle $ as follows:

%& \left\langle {\mathbb{Z}{{\hat U}^k},{U^k} - {U^*}} \right\rangle
\begin{equation}\label{E10-2}
\begin{aligned}
\left\langle {\mathbb{Z}{{\hat U}^k},{U^k} - {U^*}} \right\rangle = & \left\langle {\mathbb{Z}{{\hat U}^k},{{\hat U}^k} - \sum\limits_{l \in J\left( k \right)} {\left( {{U^l} - {U^{l + 1}}} \right)}  - {U^*}} \right\rangle \\
= & \left\langle {\mathbb{Z}{{\hat U}^k},{{\hat U}^k} - {U^*}} \right\rangle  - \frac{1}{{\tilde \theta }}\sum\limits_{l \in J\left( k \right)} {\left\langle {{U^k} - {{\tilde U}^{k + 1}},{U^l} - {U^{l + 1}}} \right\rangle } \\
\ge & - \frac{1}{{2\tilde \theta }}\sum\limits_{l \in J\left( k \right)} {\left( {\frac{1}{\xi }\left\| {{U^k} - {{\tilde U}^{k + 1}}} \right\|_{{\mathbb{T}_s}}^2 + \xi \left\| {{U^l} - {U^{l + 1}}} \right\|_{{\mathbb{T}_s}}^2} \right)}  + \left\langle {\mathbb{Z}{{\hat U}^k},{{\hat U}^k} - {U^*}} \right\rangle \\
= & - \frac{1}{{2\tilde \theta }}\sum\limits_{l \in J\left( k \right)} {\left( {\frac{1}{\xi }\left\| {{U^k} - {{\tilde U}^{k + 1}}} \right\|_{{\mathbb{T}_s}}^2 + \xi \left\| {{U^l} - {U^{l + 1}}} \right\|_{{\mathbb{T}_s}}^2} \right)}  + \left\langle {\mathbb{Z}{{\hat U}^k} - \mathbb{Z}{U^*},{{\hat U}^k} - {U^*}} \right\rangle \\
\ge & - \frac{1}{{2\tilde \theta }}\sum\limits_{l \in J\left( k \right)} {\left( {\frac{1}{\xi }\left\| {{U^k} - {{\tilde U}^{k + 1}}} \right\|_{{\mathbb{T}_s}}^2 + \xi \left\| {{U^l} - {U^{l + 1}}} \right\|_{{\mathbb{T}_s}}^2} \right)} + \frac{1}{2}\left\| {\mathbb{Z}{{\hat U}^k} - \mathbb{Z}{U^*}} \right\|_{{\mathbb{T}_s}}^2 \\
\ge & - \frac{1}{{2\tilde \theta }} ( {\frac{{\left| {J\left( k \right)} \right|}}{\xi }\left\| {{U^k} - {{\tilde U}^{k + 1}}} \right\|_{{\mathbb{T}_s}}^2 + \xi \sum\limits_{l \in J\left( k \right)} {\left\| {{U^l} - {U^{l + 1}}} \right\|_{{\mathbb{T}_s}}^2} } ) + \frac{1}{{2{{\tilde \theta }^2}}}\left\| {{{\tilde U}^{k + 1}} - {U^k}} \right\|_{{\mathbb{T}_s}}^2,
\end{aligned}
\end{equation}
where the first inequality applies Young's inequality, the third equality uses the fact that $\mathbb{Z}{U^*} = \left( {{\mathbf{Id}} - \mathbb{T}} \right){U^*} = {\mathbf{0}}$, and the penultimate inequality is owing to Lemma \ref{L4-2-2} and  \cite[Lemma 3]{Wu2018a}. Substituting (\ref{E10-2}) into (\ref{E10-1}) obtains
\begin{equation}\label{E10-3}
\begin{aligned}
 \mathbb{E}_k {\left\| {{U^{k + 1}} - {U^*}} \right\|_{{\mathbb{T}_s}}^2} \le & \left\| {{U^k} - {U^*}} \right\|_{{\mathbb{T}_s}}^2 + \frac{\xi }{m}\sum\limits_{l \in J\left( k \right)} {\left\| {{U^l} - {U^{l + 1}}} \right\|_{{\mathbb{T}_s}}^2} \\
 & - \left( {\frac{1}{{m\tilde \theta }} - \frac{{\left| {J\left( k \right)} \right|}}{{m\xi }} - \frac{{{\kappa _s}}}{{{P_{\min }}{m^2}}}} \right)\left\| {{{\tilde U}^{k + 1}} - {U^k}} \right\|_{{\mathbb{T}_s}}^2.
\end{aligned}
\end{equation}
To proceed, we analyze the term $\sum\nolimits_{l \in J\left( k \right)} {\left\| {{U^l} - {U^{l + 1}}} \right\|_{{\mathbb{T}_s}}^2} $. Recalling the definitions of the well-defined operator $\Theta $ and two stacked vectors ${V^k}$ and ${V^*}$, it follows from \cite[Lemma 5]{Wu2018a} that
\begin{equation}\label{E10-4}
\begin{aligned}
 \left\| {{V^k}} \right\|_\Theta ^2 = & \left\| {{U^k}} \right\|_{{\mathbb{T}_s}}^2 +  \sqrt {\frac{{{P_{\min }}}}{{{\kappa _s}}}} \sum\limits_{l = k - d}^{k - 1} {\left( {l - \left( {k - d} \right) + 1} \right)} \left\| {{U^l} - {U^{l + 1}}} \right\|_{{\mathbb{T}_s}}^2.
\end{aligned}
\end{equation}
Based on the relation (\ref{E10-4}), we proceed to seed an upper bound for $\mathbb{E}_k {\left\| {{V^{k + 1}} - {V^ * }} \right\|_\Theta ^2} $ as follows:
\begin{equation}\label{E10-5}
\begin{aligned}
\mathbb{E}_k {\left\| {{V^{k + 1}} - {V^ * }} \right\|_\Theta ^2} = & \mathbb{E}_k {\left\| {{U^{k + 1}} - {U^ * }} \right\|_{{\mathbb{T}_s}}^2}  + d\sqrt {\frac{{{P_{\min }}}}{{{\kappa _s}}}} \mathbb{E}_k {\left\| {{U^k} - {U^{k + 1}}} \right\|_{{\mathbb{T}_s}}^2} \\
 & + \sqrt {\frac{{{P_{\min }}}}{{{\kappa _s}}}} \sum\limits_{l = k + 1 - d}^{k - 1} {\left( {l - \left( {k - d} \right)} \right)\left\| {{U^l} - {U^{l + 1}}} \right\|_{{\mathbb{T}_s}}^2}.
\end{aligned}
\end{equation}
Fixing $\xi  = m\sqrt {{P_{\min }}/{\kappa _s}} $ and substituting (\ref{E10-3}) into (\ref{E10-5}) reduces to

\begin{equation}\label{E10-6}
\begin{aligned}
 \mathbb{E}_k {\left\| {{V^{k + 1}} - {V^ * }} \right\|_\Theta ^2} \le &   \left\| {{U^k} - {U^*}} \right\|_{{\mathbb{T}_s}}^2 + \frac{\xi }{m}\sum\limits_{l \in J\left( k \right)} {\left\| {{U^l} - {U^{l + 1}}} \right\|_{{\mathbb{T}_s}}^2} +  \sqrt {\frac{{{P_{\min }}}}{{{\kappa _s}}}} \sum\limits_{l = k + 1 - d}^{k - 1} {\left( {l - \left( {k - d} \right)} \right)\left\| {{U^l} - {U^{l + 1}}} \right\|_{{\mathbb{T}_s}}^2} \\
 & - \frac{1}{m}\left( {\frac{1}{{\tilde \theta }} - 2\frac{d}{m}\sqrt {\frac{{{\kappa _s}}}{{{P_{\min }}}}}  - \frac{{{\kappa _s}}}{{{P_{\min }}m}}} \right)\left\| {{{\tilde U}^{k + 1}} - {U^k}} \right\|_{{\mathbb{T}_s}}^2 \\
 \le & \left\| {{U^k} - {U^*}} \right\|_{{\mathbb{T}_s}}^2 + \sqrt {\frac{{{P_{\min }}}}{{{\kappa _s}}}} \sum\limits_{l = k - d}^{k - 1} {\left\| {{U^l} - {U^{l + 1}}} \right\|_{{\mathbb{T}_s}}^2} \\
 & + \sqrt {\frac{{{P_{\min }}}}{{{\kappa _s}}}} \sum\limits_{l = k - d}^{k - 1} {\left( {l - \left( {k - d} \right)} \right)\left\| {{U^l} - {U^{l + 1}}} \right\|_{{\mathbb{T}_s}}^2}  - \frac{1}{m}\left( {\frac{1}{{\tilde \theta }} - 2\frac{d}{m}\sqrt {\frac{{{\kappa _s}}}{{{P_{\min }}}}}  - \frac{{{\kappa _s}}}{{{P_{\min }}m}}} \right)\left\| {{{\tilde U}^{k + 1}} - {U^k}} \right\|_{{\mathbb{T}_s}}^2\\
\le & \left\| {{V^k} - {V^ * }} \right\|_\Theta ^2 - \frac{1}{m}\left( {\frac{1}{{\tilde \theta }} - 2\frac{d}{m}\sqrt {\frac{{{\kappa _s}}}{{{P_{\min }}}}}  - \frac{{{\kappa _s}}}{{{P_{\min }}m}}} \right) \times \left\| {{{\tilde U}^{k + 1}} - {U^k}} \right\|_{{\mathbb{T}_s}}^2.
\end{aligned}
\end{equation}
The proof is completed via choosing uncoordinated step-sizes to ensure that the sequence ${\left\{ {\left\| {{U^k} - {U^*}} \right\|_{{\mathbb{T}_s}}^2} \right\}_{k \ge 0}}$ is stochastic quasi-Fej\'{e}r monotone and then applying the Opial lemma  \cite[Lemma 26]{Wu2018a} to prove that the sequence ${\left\{ {{U^k}} \right\}_{k \ge 0}}$ converges to some optimal point ${U^*} \in {\mathcal{S}^*}$.

\subsection{Simulation Results Regarding Evolution of Traded Power}\label{S-1}
\begin{figure}[!htp]
\centering
\subfloat[$t=1$]{\includegraphics[width=1.75in,height=1.50in]{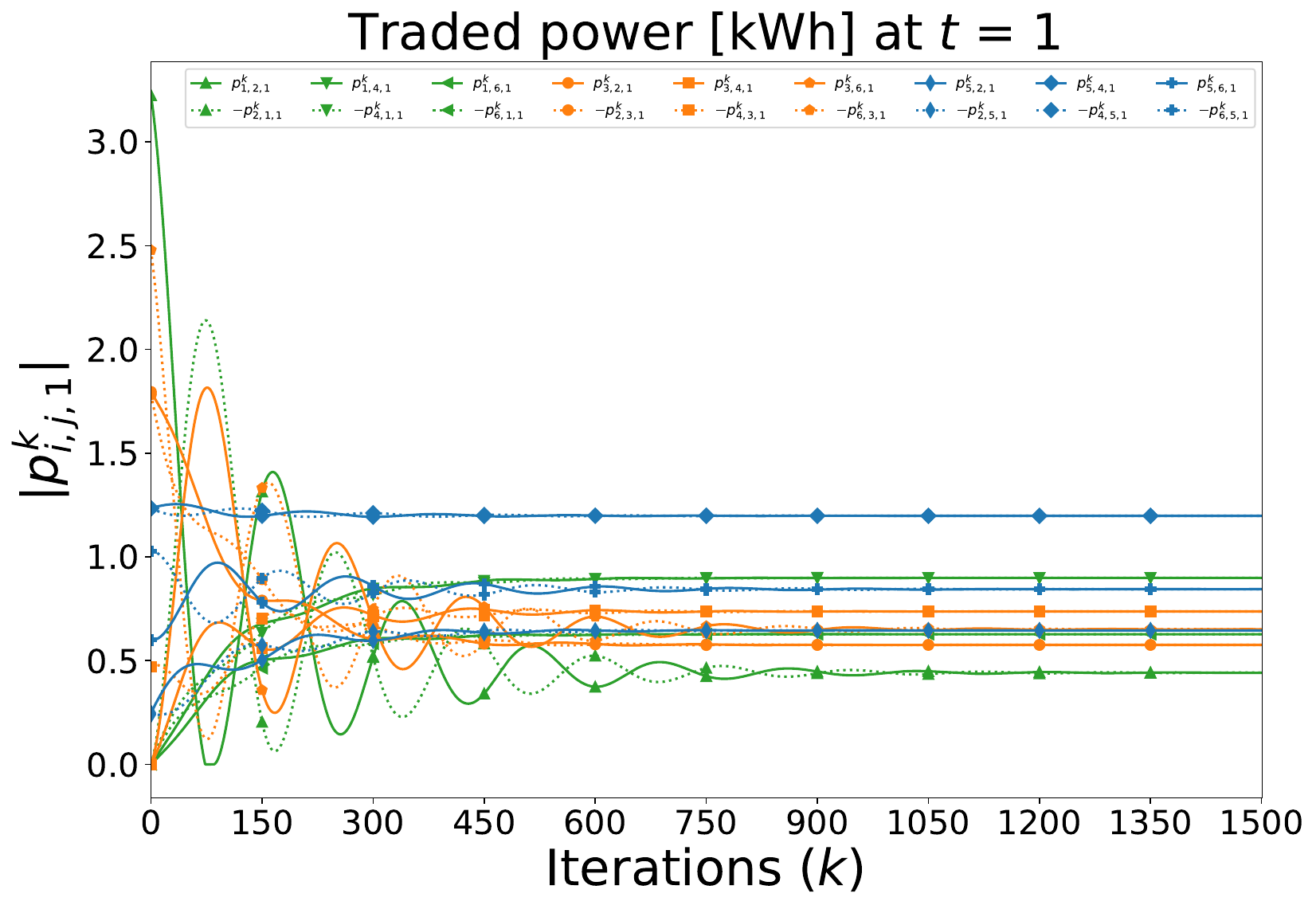}\label{Fig. 5-1-1}} \hfill
\subfloat[$t=2$]{\includegraphics[width=1.75in,height=1.50in]{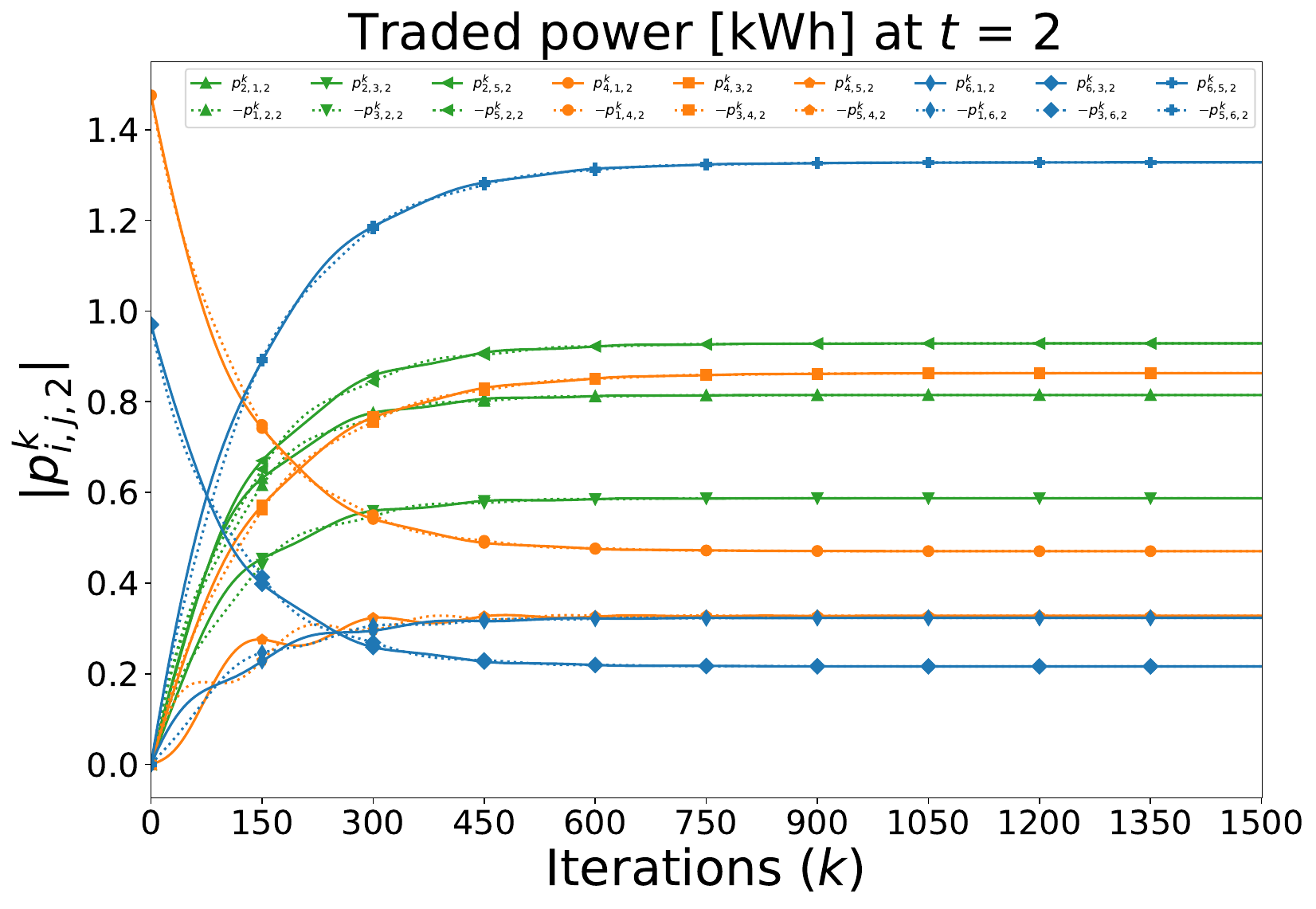}\label{Fig. 5-1-2}} \hfill
\subfloat[$t=3$]{\includegraphics[width=1.75in,height=1.50in]{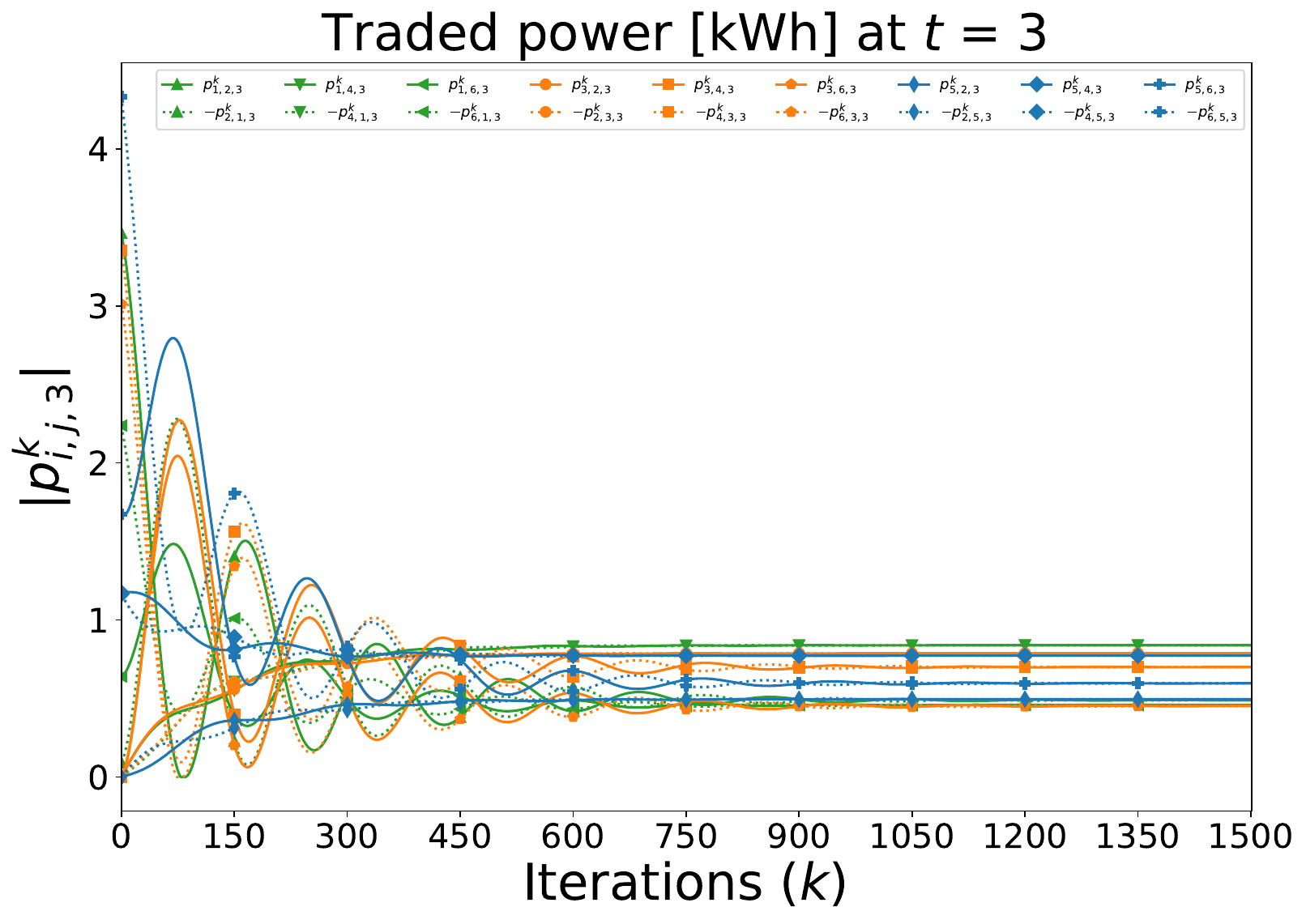}\label{Fig. 5-1-3}} \hfill
\subfloat[$t=4$]{\includegraphics[width=1.75in,height=1.50in]{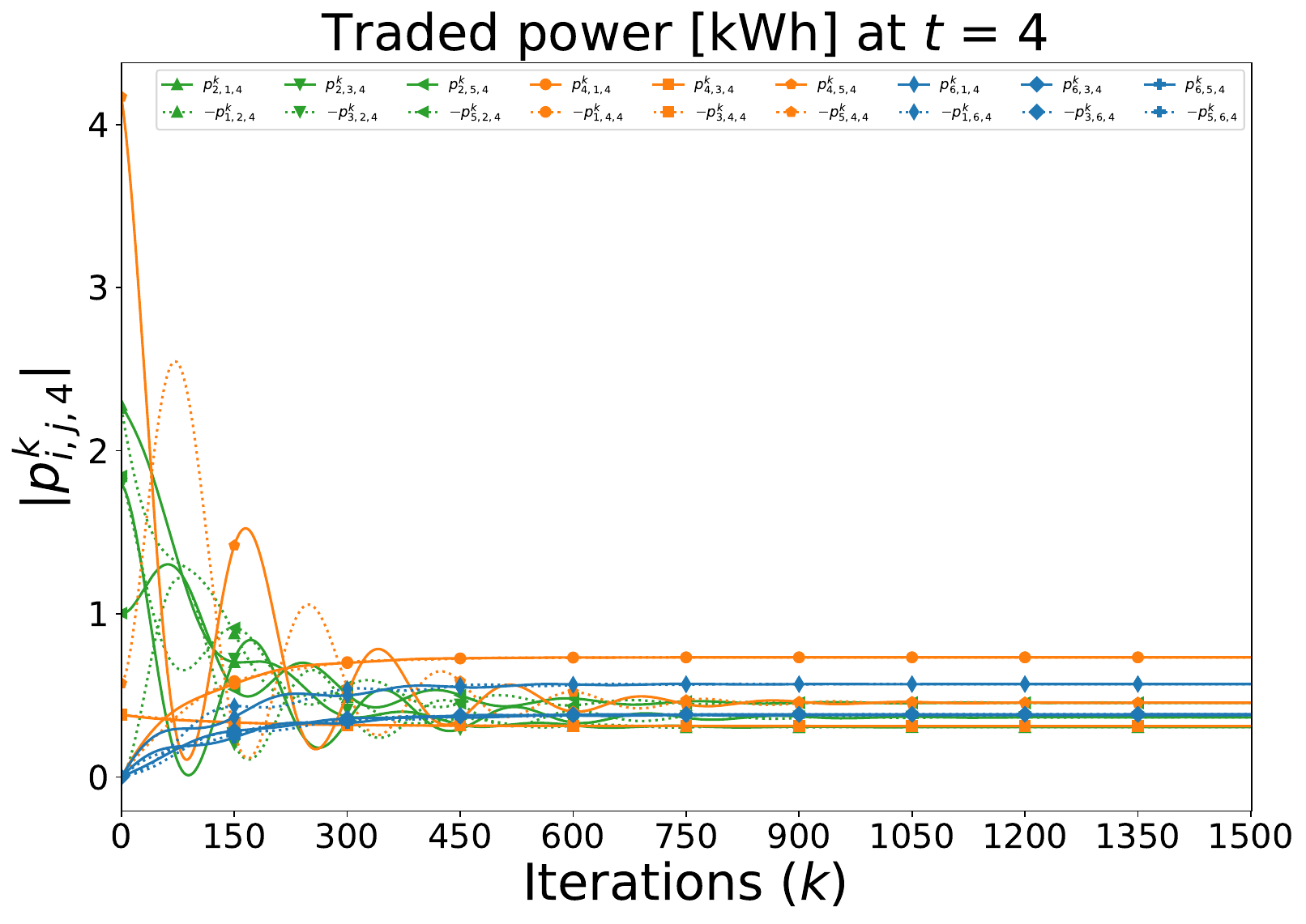}\label{Fig. 5-1-4}} \hfill
\caption{Evolution of traded power at different periods for \textit{Syn-DYNA} over a Syn-P2P transactive network.}
\label{Fig. 5-1}
\end{figure}

Fig. \ref{Fig. 5-1}: (a)-(d) show the evolution of traded power for the \textit{Syn-DYNA} during different periods. Initially, the synchronous algorithm exhibits the traded power values fluctuate significantly as the agents update their variables to find optimal trading strategies.

\begin{figure}[!htp]
\centering
\subfloat[$t=1$]{\includegraphics[width=1.75in,height=1.50in]{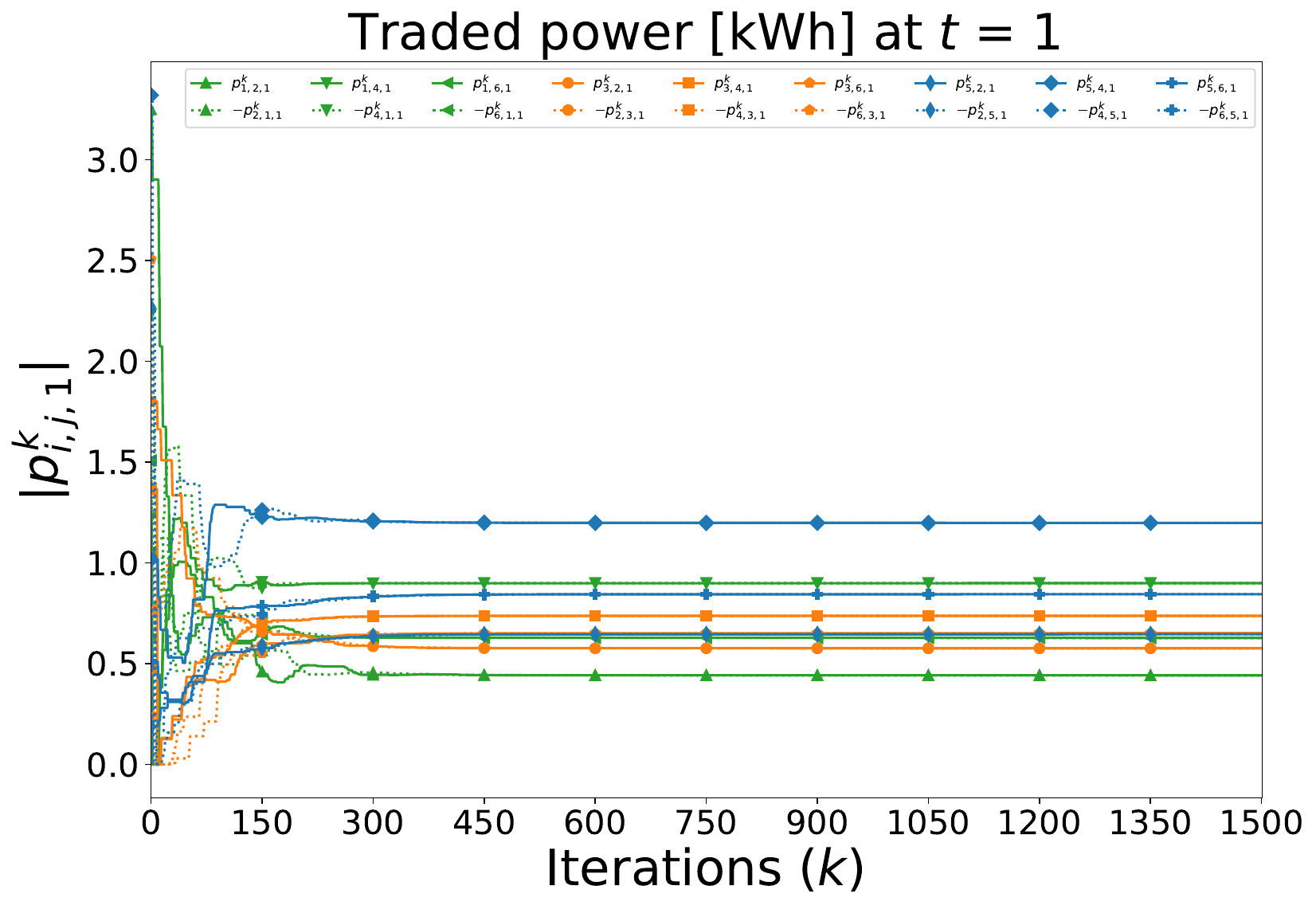}\label{Fig. 5-2-1}} \hfill
\subfloat[$t=2$]{\includegraphics[width=1.75in,height=1.50in]{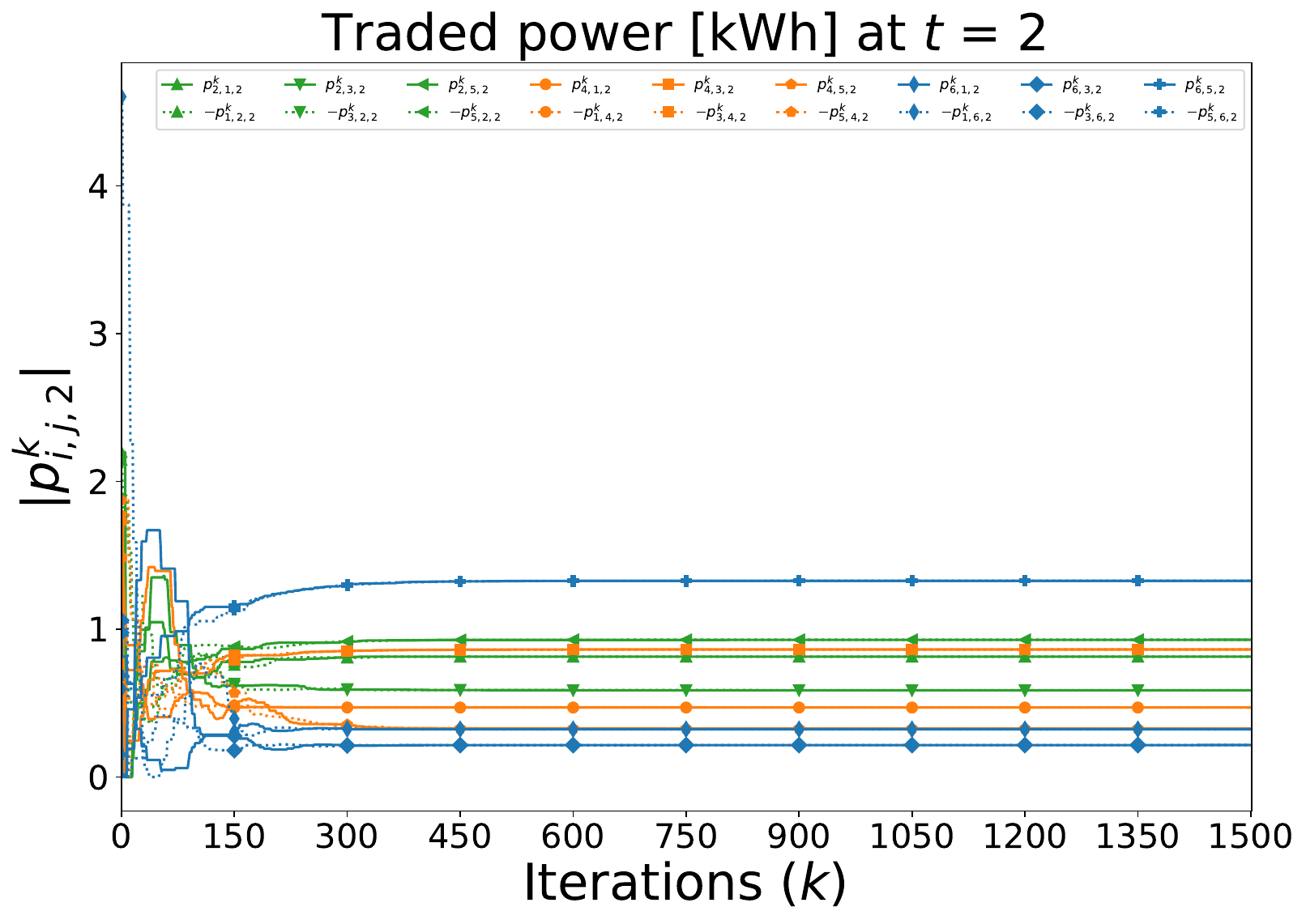}\label{Fig. 5-2-2}} \hfill
\subfloat[$t=3$]{\includegraphics[width=1.75in,height=1.50in]{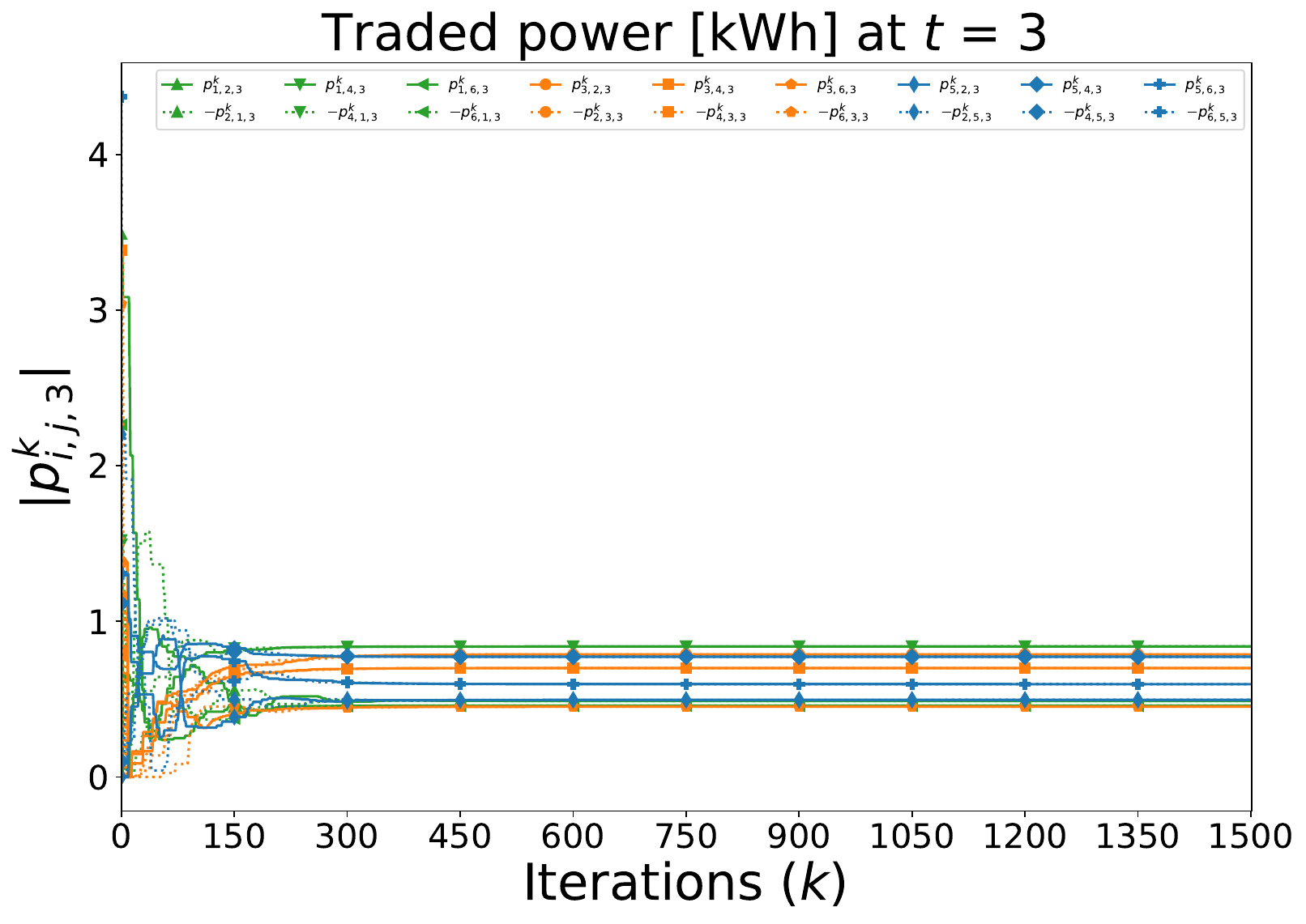}\label{Fig. 5-2-3}} \hfill
\subfloat[$t=4$]{\includegraphics[width=1.75in,height=1.50in]{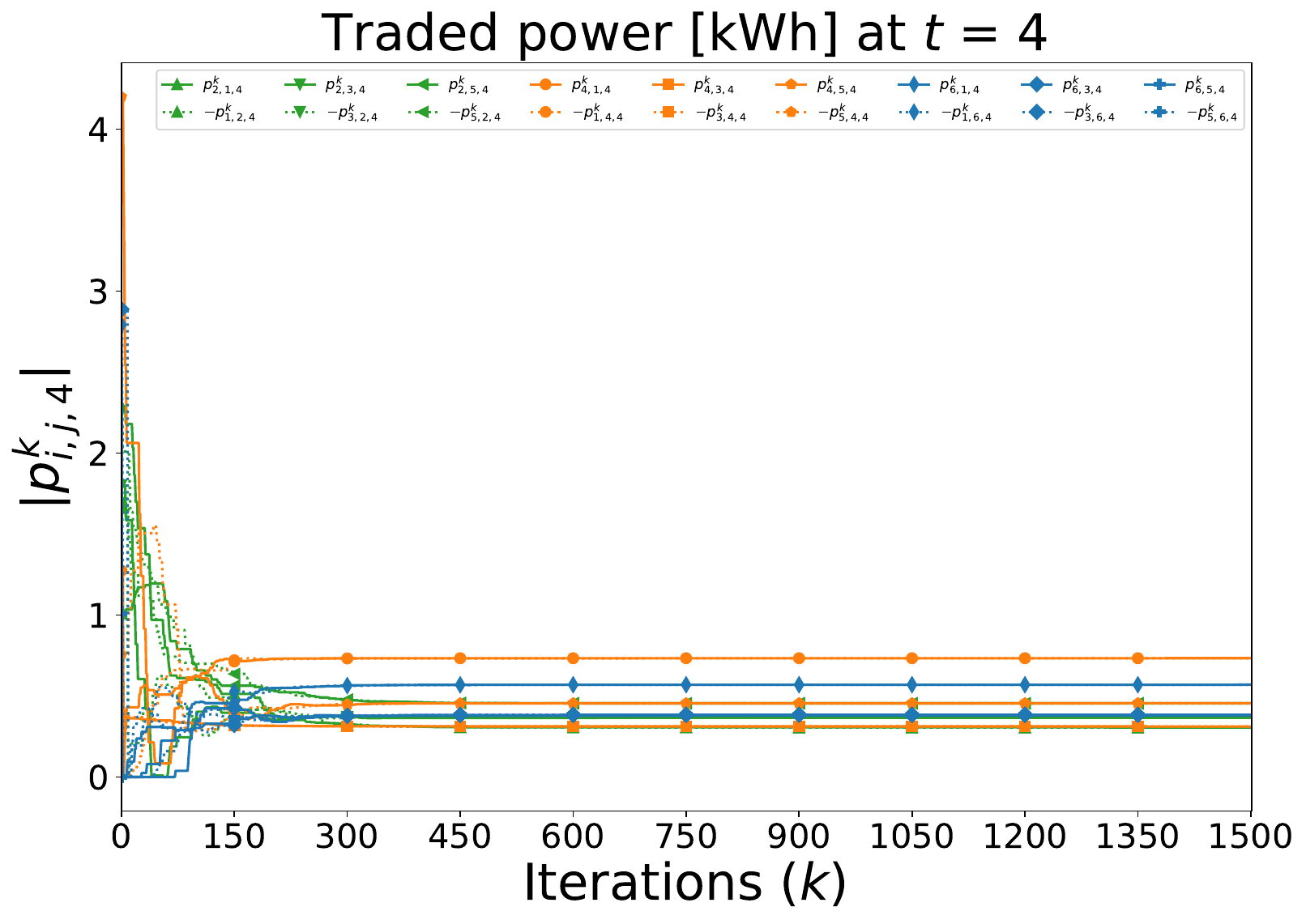}\label{Fig. 5-2-4}} \hfill
\caption{Evolution of traded power at different periods for \textit{Asyn-DYNA} over an Asyn-P2P transactive network with $d=0$.}
\label{Fig. 5-2}
\end{figure}

\begin{figure}[!htp]
\centering
\subfloat[$t=1$]{\includegraphics[width=1.75in,height=1.50in]{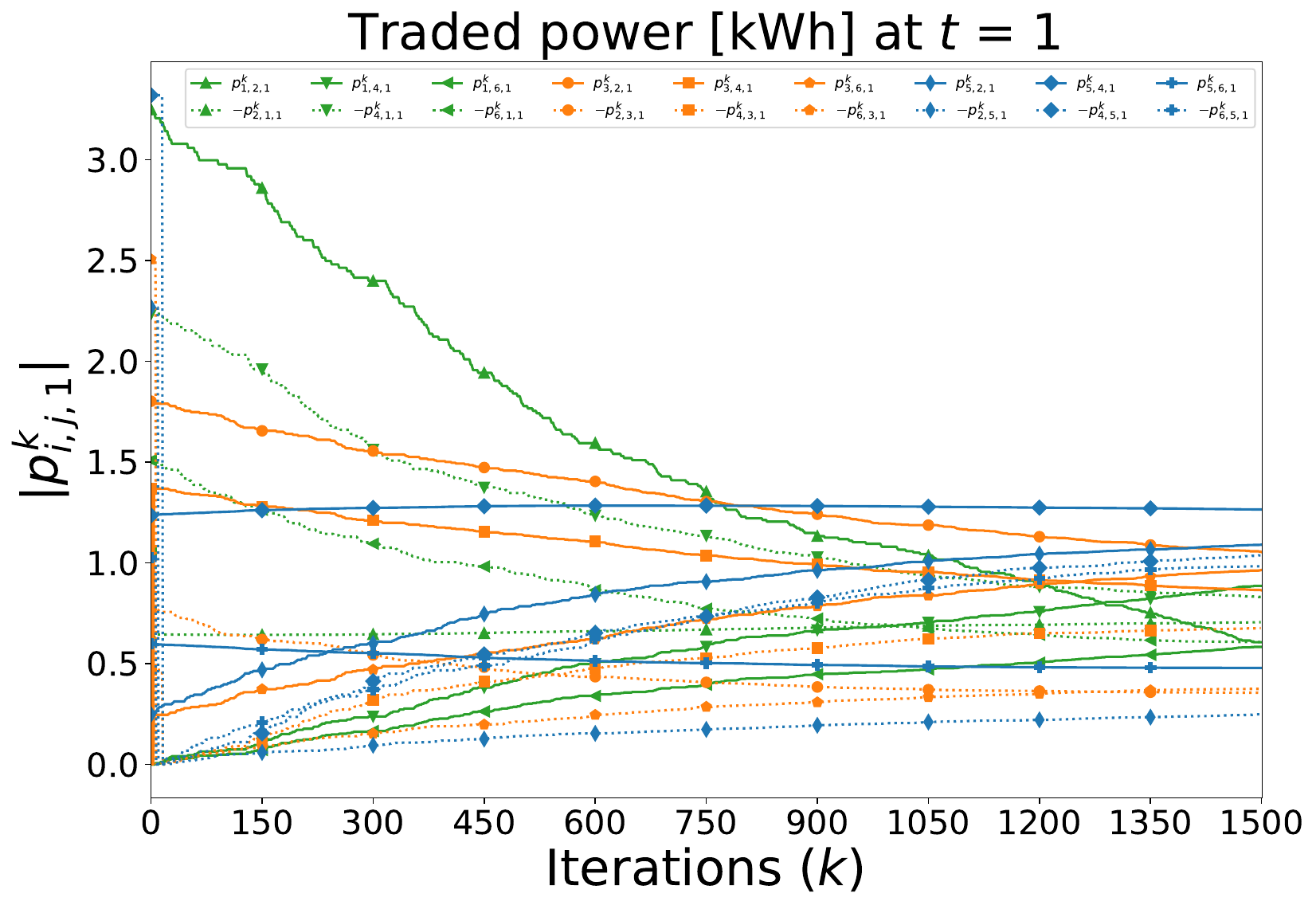}\label{Fig. 5-3-1}} \hfill
\subfloat[$t=2$]{\includegraphics[width=1.75in,height=1.50in]{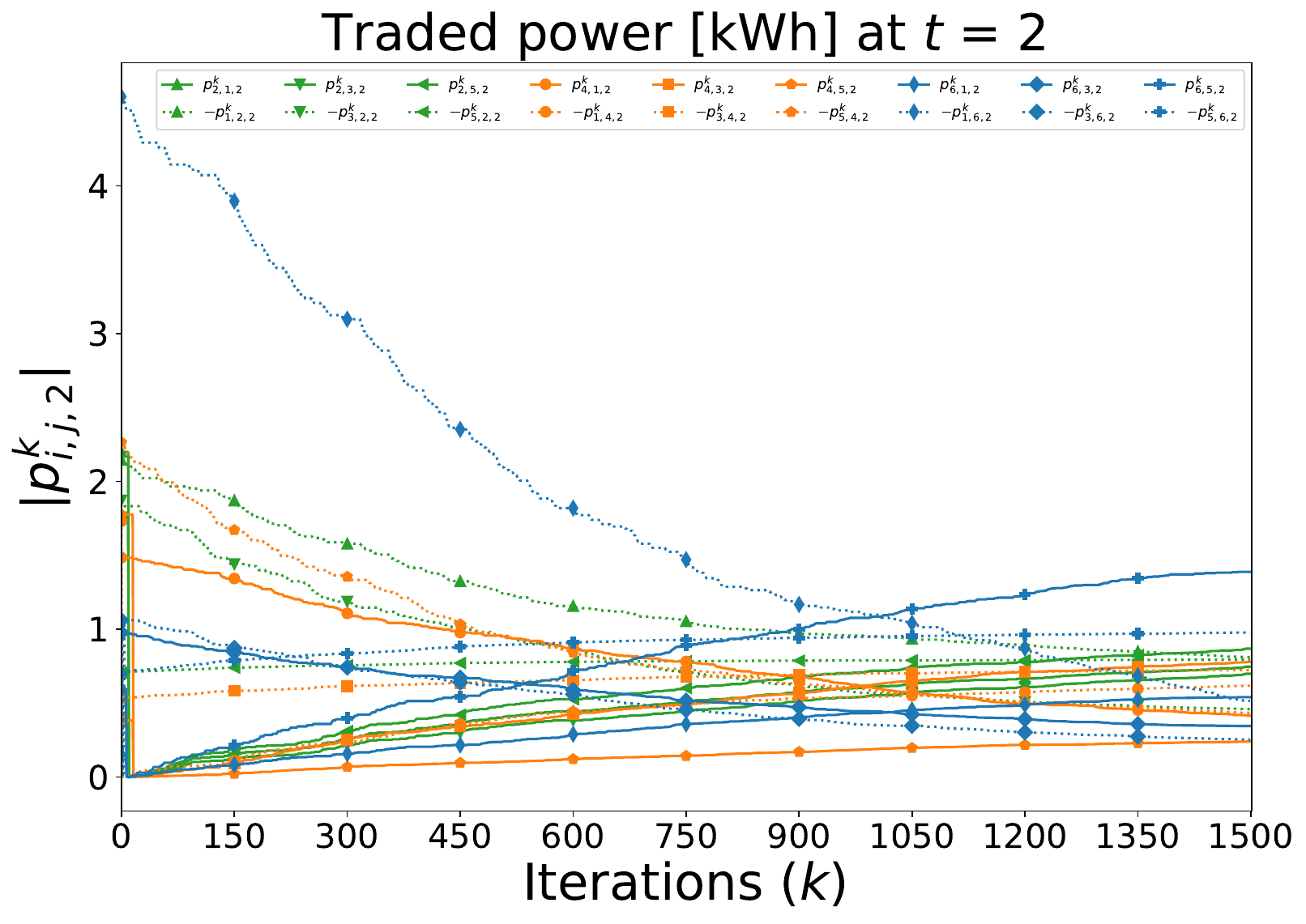}\label{Fig. 5-3-2}} \hfill
\subfloat[$t=3$]{\includegraphics[width=1.75in,height=1.50in]{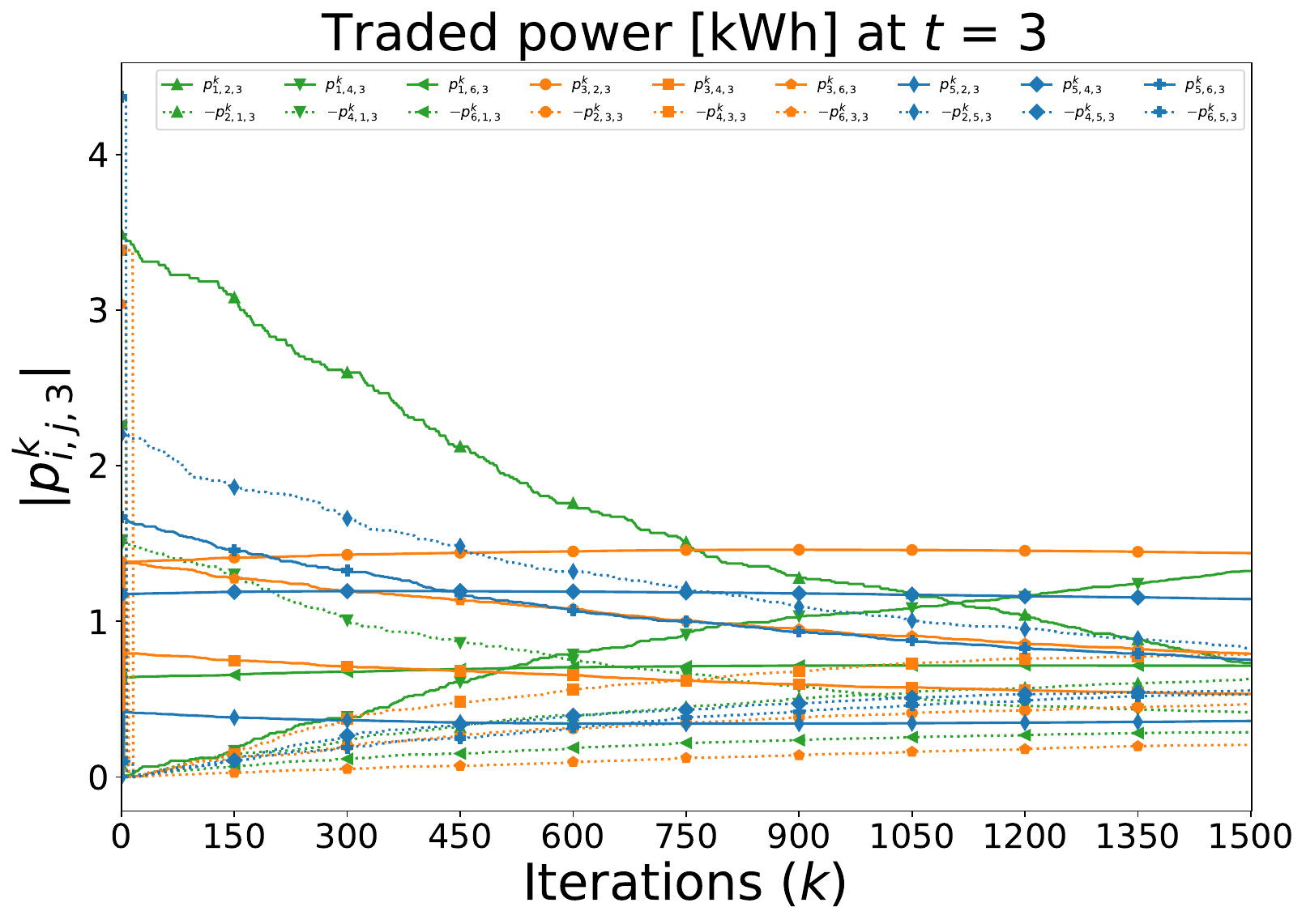}\label{Fig. 5-3-3}} \hfill
\subfloat[$t=4$]{\includegraphics[width=1.75in,height=1.50in]{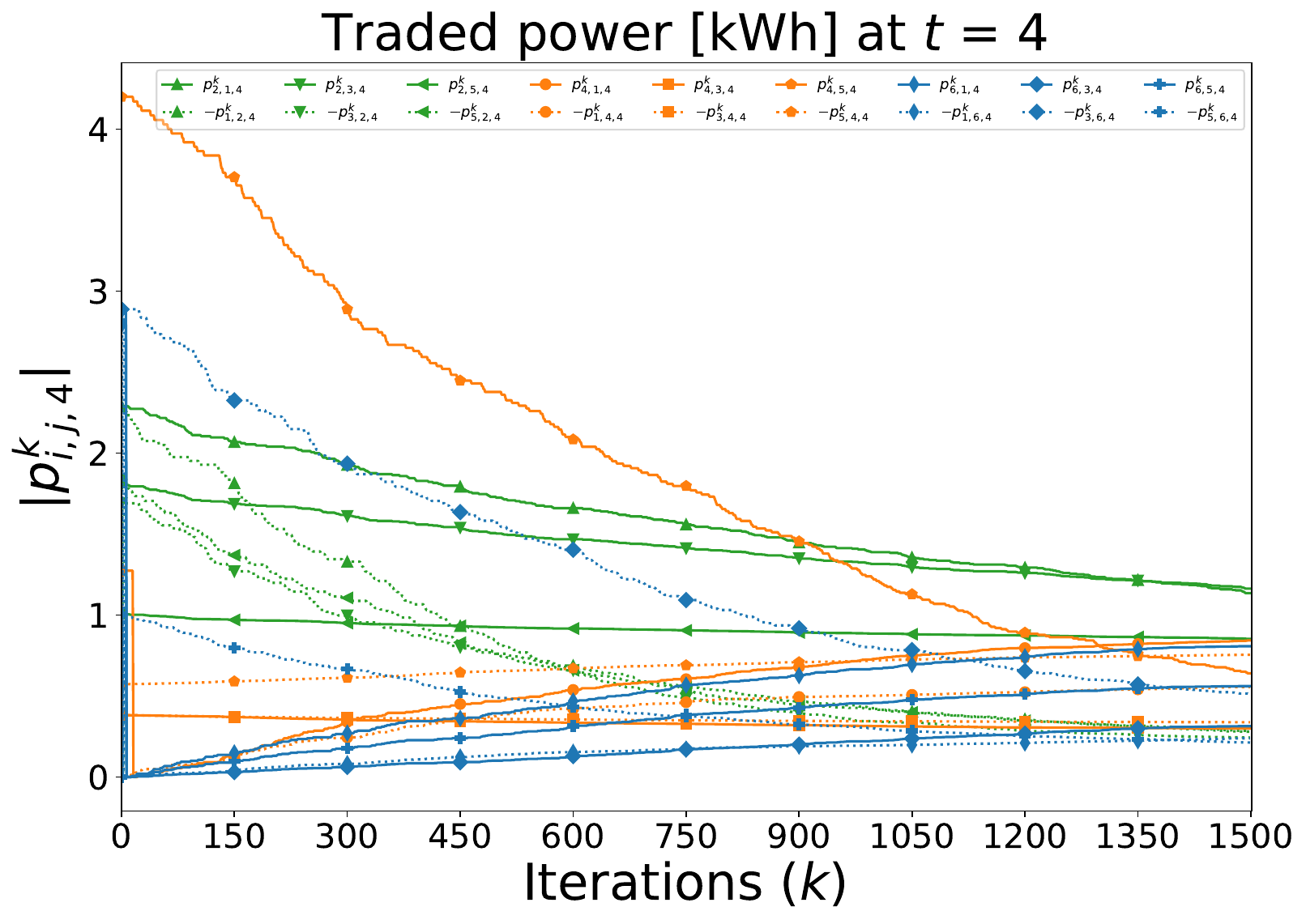}\label{Fig. 5-3-4}} \hfill
\caption{Evolution of traded power at different periods for \textit{Asyn-DYNA} over an Asyn-P2P transactive network with $d=10$.}
\label{Fig. 5-3}
\end{figure}
However, the traded power on all links converges to stable equilibrium values as iterations proceed.
By contrast, Fig. \ref{Fig. 5-2} and Fig. \ref{Fig. 5-3} illustrate the evolution of traded power for \textit{Asyn-DYNA} in the cases of $d=0$ and $d=10$, respectively. When $d=0$, i.e., in the absence of time-varying communication delays, Fig. \ref{Fig. 5-2} shows that \textit{Asyn-DYNA} achieves the equilibrium state in less than 300 iterations, which is much faster than \textit{Syn-DYNA} as revealed by Fig. \ref{Fig. 5-1}. However, due to the randomness of asynchronous updates, Fig. \ref{Fig. 5-3} demonstrates that \textit{Asyn-DYNA} requires more iterations to converge when time-varying communication delays exist. Despite the performance gap, both algorithms successfully converge to a stable state. This steady state indicates that the prosumers ultimately reach a consensus on the optimal electricity transaction amounts, thereby balancing supply and demand within the P2P transactive networks.

\end{document}